\documentclass[iop]{emulateapj}
\newcommand{\tc}[1]{#1}

\begin{document}

\slugcomment{Accepted to the Astrophysical Journal}

\shorttitle{3-D atmospheric circulation of eccentric hot Jupiters}
\shortauthors{Kataria et al.}

\title{Three-dimensional atmospheric circulation of hot Jupiters on highly eccentric orbits}

\author{T. Kataria\altaffilmark{1}, A. P. Showman\altaffilmark{1}, N. K. Lewis\altaffilmark{1,2,3}, J. J. Fortney\altaffilmark{4}, M. S. Marley\altaffilmark{5} and R. S. Freedman\altaffilmark{5,6}}

\email{tkataria@lpl.arizona.edu}
\altaffiltext{1}{Department of Planetary Sciences and Lunar and Planetary Laboratory, The University of Arizona,
Tucson, AZ 85721}
\altaffiltext{2}{Department of Earth, Atmospheric and Planetary
Sciences, Massachusetts Institute of Technology, Cambridge, MA 02139,
USA}
\altaffiltext{3}{Sagan Postdoctoral Fellow}
\altaffiltext{4}{Department of Astronomy \& Astrophysics, University of California, Santa Cruz, CA 95064}
\altaffiltext{5}{NASA Ames Research Center 245-3, Moffett Field, CA 94035}
\altaffiltext{6}{SETI Institute, 189 Bernardo Ave \#100, Mountain View, CA 94043}

\begin{abstract}
Of the over 800 exoplanets detected to date, over half are on non-circular orbits, with eccentricities as high as 0.93.  Such orbits lead to time-variable stellar heating, which has major implications for the planet's atmospheric dynamical regime. However, little is known about the fundamental dynamical regime of such planetary atmospheres, and how it may influence the observations of these planets. Therefore, we present a systematic study of hot Jupiters on highly eccentric orbits using the SPARC/MITgcm, a model which couples a three-dimensional general circulation model (the MITgcm) with a plane-parallel, two-stream, non-grey radiative transfer model.  In our study, we vary the eccentricity and orbit-average stellar flux over a wide range.  We demonstrate that the eccentric hot Jupiter regime is qualitatively similar to that of planets on circular orbits; the planets possess a superrotating equatorial jet and exhibit large day-night temperature variations.  As in Showman and Polvani (2011), we show that the day-night heating variations induce momentum fluxes equatorward to maintain the superrotating jet througout its orbit.  We find that as the eccentricity and/or stellar flux is increased (corresponding to shorter orbital periods), the superrotating jet strengthens and narrows, due to a smaller Rossby deformation radius.  For the cases that are most distant and slowly rotating, we find hints of a regime shift, with no eastward flow at lower pressures.  For a select number of model integrations, we generate full-orbit lightcurves and find that the timing of transit and secondary eclipse viewed from Earth with respect to periapse and apoapse can greatly affect what we see in infrared (IR) lightcurves; the peak in IR flux can lead or lag secondary eclipse depending on the geometry.  For those planets that have large temperature differences from dayside to nightside and rapid rotation rates, we find that the lightcurves can exhibit ``ringing" as the planet's hottest region rotates in and out of view from Earth.  These results can be used to explain future observations of eccentric transiting exoplanets.

\end{abstract}


\section{Introduction}

Since the first planetary confirmations in the mid-1990s \citep{mayor1995,wolszczan1994} the detection and characterization of
extrasolar planets continue to be major fields in astronomy and planetary science.  Over 700 planets have been detected from the ground and space, more than half of which are classified as ``hot Jupiters", Jovian-mass planets that orbit their parent stars at distances less than 0.1 AU.  A number of these hot Jupiters transit their host star along our line of sight, allowing us to observe them as they pass in front and behind their parent star \citep[e.g.][]{charb2008,swain2008,pont2008,knutson2007,knutson2008,knutson2009}. Using these observations, we can infer much about these planets' atmospheric composition, temperature structure, and circulation.

A fifth of these transiting exoplanets have eccentricities greater than 0.1, with values as large as 0.93 \citep[HD80606b,][]{naef2001}.  These eccentric planets are subject to highly time-variable heating which has a significant effect on the planet's atmospheric dynamics.  Among planets amenable to observational follow-up, HAT-P-2b, which has an eccentricity of 0.52, undergoes a factor of 9 variation in flux throughout its orbit.  HD 17156b ($e=0.67$) experiences a factor of 27 variation in stellar flux, and HD 80606b, with its large eccentricity ($e=0.93$), undergoes an impressive factor of 828 variation in flux.  Because they are transiting, we can probe their atmospheres as we can for planets on circular orbits.  However, while it is clear the strongly variable heating leads to a vastly different {\it forcing} regime than for exoplanets on circular orbits, it remains unknown whether this causes a fundamentally different {\it dynamical} regime: is the circulation quantitatively similar to that on circular hot Jupiters, or is it a completely new circulation regime?

Eccentric transiting exoplanets present unique challenges when one attempts to extract information about their atmospheres from observational data.  In particular, interpretation of flux maxima and minima in infrared lightcurves can be complicated by the convolution of spatial effects (for example, hot spots on the planet that rotate into and out of view along our line of sight) with temporal effects (planet getting colder/warmer at apoapse/periapse passage).  Langton and Laughlin (2008a,b) and Cowan and Agol (2011) address this problem as applied to particular targets, but use only a two-dimensional hydrodynamical model and one-dimensional semi-analytic model, respectively.  To fully capture these effects, a three-dimensional circulation model that self-consistently calculates heating and wind velocities is needed.

Hence, it is crucial to conduct a comprehensive study that establishes the dynamical regime, temperature structure, and observational implications of eccentric exoplanets.  We use a three-dimensional atmospheric circulation model coupled to a non-grey radiative transfer scheme to study eccentric hot Jupiters as a whole.  In Section 2 we will describe our model setup and integrations.  Section 3 describes the dynamical regime and its dependence on eccentricity and mean stellar flux.  Section 4 presents synthetic lightcurves and attempts to determine what can be learned about atmospheric circulations from remote measurements.  Finally, Section 5 concludes and compares results to known eccentric hot Jupiters.

\section{Model}
We adopt the Substellar and Planetary Atmospheric Radiation and Circulation (SPARC) model, which couples the MITgcm \citep{adcroft2004} with a two-stream implementation of the multi-stream, non-grey radiative transfer scheme developed by \cite{mm1999}.  A brief discussion of the model is included here; for a more complete description, see \cite{showman2009}.

The MITgcm solves the primitive equations, a simplification of the Navier-Stokes equations where the horizontal flow length scale exceeds vertical length scales; these equations are valid at large scales in stably stratified atmospheres.  Hot Jupiters generally satisfy these criteria, with horizontal and vertical length scales of $\sim\mathrm{10^4}$~km and $\sim$300 km, respectively, and highly stratified atmospheres. While the primitive equations do have limitations (for example, the inability to capture small-scale, weakly stratifed features such as fronts and storms) the large-scale flow (e.g. jets and waves) is captured.  The use of the primitive equations also present many computational advantages over the fully compressible equations.  The radiative transfer code solves the two-stream radiative transfer equations, and employs the correlated-$k$ method \citep[e.g.][]{mlawer1997,goody1989,fl1992,mm1999} to solve for upward/downward fluxes and heating/cooling rates through an inhomogeneous atmosphere.  This method retains most of the accuracy of full line-by-line calculations, while drastically increasing computational efficiency.  The correlated-$k$ method maps the absorption coefficients, denoted by $k(\nu)$ where $\nu$ is frequency, from spectral space ($k=k(\nu)$) to a space where $k$ varies with a variable $g$ ($k=k(g)$).  The inversion of this function, $g(k)$, is commonly referred to as the cumulative distribution,

\begin{equation}\label{cumdistr}
    g_{i}=\int_0^k \! f_{i}(k') \, \mathrm{d} k'
\end{equation}

\noindent where $f_{i}(k)$ is the distribution function for the absorption coefficient in the $i$th atmosphere layer in a spectral interval $\bigtriangleup\nu$.  In the space of cumulative probability, $g$ varies from 0 to 1 and represents the probability of the absorption coefficient being less than or equal to $k(g)$.  The total frequency space is split into a number of discrete frequency intervals (windows) and the absorption coefficients at each grid point in each window are sorted by frequency of occurrence.   From this, a $g$ distribution is derived for each window.  The radiative transfer is then carried out window by window, usually using a $g(k)$ that has been represented by a Gaussian division scheme in order to integrate the distribution over each window.  

%For each window, a monochromatic absorption coefficient, $K_i$, is determined.  These values are used to calculate the transmission, $T$, as a function of column mass, $u$, by the equation

For each window, $n$ different monochromatic absorption coefficients, $K_i$ (where $i$ varies from 1 to $n$), are computed at $n$ Gauss points in order to integrate over the $k(g)$ curve.  These values of $K_i$ are used to calculate the transmission, $T$, as a function of column mass, $u$, by the equation

%In this space, $g$ varies monotonically with $k$ and $k(g)$ can be easily computed.  The correlated-$k$ method splits $g$ into several subintervals each containing $10^4 - 10^5$ values of $k(g)$, and determining an equivalent monochromatic absorption coefficient, $K_i$, for each subinterval.  Each value of the monochromatic absorption coefficient, $K_i$, is used to calculate the transmission, $T$, as a function of column mass, $u$, by the equation

\begin{equation}\label{transmission}
T(u)=\sum_{i}^nW_{i}e^{-K_{i}u}
\end{equation}

\noindent where $W_{i}$ is the gauss weight for each fractional gauss point.  In this way, multiple opacity sources can be modeled.  This model accounts for gaseous (Rayleigh) scattering and pressure-induced absorption by $\rm H_2$ and He.  Scattering and absorption due to clouds, aerosols and dust can also be included, though for simplicity we do not include these effects here.  The fluxes within each spectral interval are weighted and summed to obtain the upwards and downwards fluxes for each bin at each atmospheric layer.  The sum of all intervals gives the total, wavelength-integrated fluxes for each layer.  We then calculate the heating rate by finite-differencing the fluxes and pressures between interfaces over and underlying a given dynamical level,

\begin{equation}\label{heating}
    q=g\frac{\partial F}{\partial p}
\end{equation}

\noindent Here $q$ is the heating rate, $g$ is the planet's gravity, and $\frac{\partial F}{\partial p}$ is the gradient of flux with pressure.
For each simulation, we utilize a cubed-sphere grid with a horizontal resolution of C32 (approximately equivalent to 64$\times$128 in latitude and longitude) and $\mathrm{N_{L}=}$40 or 76 pressure levels.  The lowermost $\mathrm{N_{L}-1}$  levels extend from a mean pressure of 200 bars at the bottom to 0.2 mbar at the top, evenly spaced in log pressure.  The top level extends from a pressure of 0.2 mbar to zero.

\tc{We implement a fourth-order Shapiro filter for temperature and momentum, with a timestep equal to double the dynamical timestep.  Models integrated with this Shapiro filter setup (but excluding any other large-scale drag) conserve total angular momentum to better than 0.1$\mathrm{\%}$, thereby demonstrating excellent conservation of angular momentum with our model configuration.}

\subsection{Updates to SPARC/MITgcm}
\subsubsection{Reduced number of frequency bins}
While the models of HD 189733b and HD 209458b in Showman et al. (2009) and the models of GJ 436b in Lewis et al. (2010) used 30 frequency bins to model circulation and heating on each planet, here the opacities are statistically weighted into 11 frequency bins to improve computational efficiency.  For both the 30- and 11-bin implementations, the radiative transfer is calculated at 8 values of $k(g)$ in each of the frequency bins.  Four of the values of $k(g)$ are sampled at $g<0.95$, meaning that four calculations are done for the weakest $95\%$ of the spectral lines.  The other four radiative transfer calculations are done for the strongest $5\%$ of the lines ($g>0.95$).  This allows us to calculate the radiative transfer for both strong and weak spectral lines.

\begin{deluxetable}{cc}
\tabletypesize{\scriptsize}
\tablecaption{Bounding wavelengths of the frequency bins used in the radiative transfer calculations. \label{opacitybins}}
\tablewidth{0pt}
\tablehead{
\colhead{Wavelength ($\mu$m)} & \colhead{Wavelength ($\mu$m)}
}
\startdata
324.68 & 20.00 \\
20.00 & 8.70 \\
8.70 & 4.40  \\
4.40 & 3.50  \\
3.50 & 2.50  \\
2.50 & 2.02  \\
2.02 & 1.32  \\
1.32 & 0.85  \\
0.85 & 0.61  \\
0.61 & 0.42  \\
0.42 & 0.26  
\enddata
\end{deluxetable}

We list the 11 new frequency bins in Table~\ref{opacitybins}.   Before implementing the updated scheme into our simulations, we conducted one-dimensional and three-dimensional tests to reproduce the results for HD 1897833b from \cite{showman2009}.  Once successful, we applied the new scheme to the model integrations described here. (Appendix A describes these validation tests in detail.)

\subsubsection{Updated collision-induced absorption due to $H_2-H_2$ and $H_2-He$ collisions}

The collision-induced absorption (CIA) by $\mathrm{H_2}$ molecules due to $\mathrm{H_2-H_2}$ and $\mathrm{H_2-He}$ collisions has recently been recomputed from first principles (Saumon et al. 2012 and references therein). The new formulation uses state-of-the-art quantum mechanical calculations of the potential energy surface and of the interaction-induced dipole moment surface for colliding $\mathrm{H_2-H_2}$ and $\mathrm{H_2-He}$ pairs.  This new calculation is more reliable at higher temperature and higher photon energies than the earlier calculations by Borysow and collaborators that radiative transfer groups have been using (http://www.astro.ku.dk\textbackslash$\sim$aborysow/) . We have included these new tabulations of $\mathrm{H_2}$ CIA opacity in our code.

\begin{figure}
%\epsscale{0.75}
\centering
\includegraphics[trim = 1.6in 3.3in 0.7in 3.3in, clip, width=0.58\textwidth]{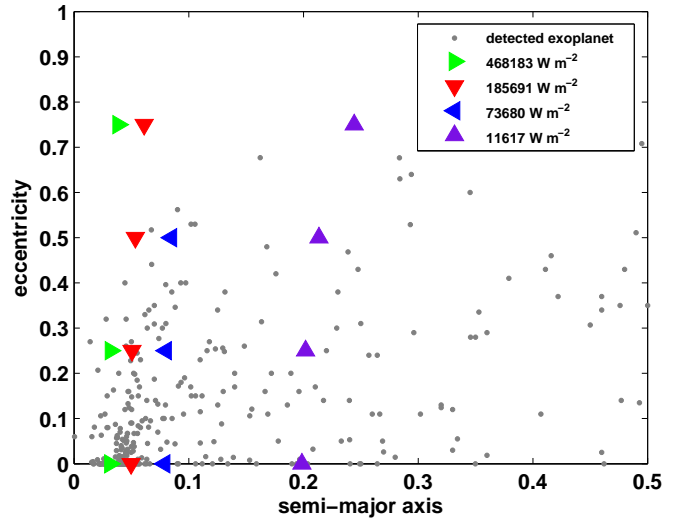}
\caption{Model integrations from this study plotted against a portion of the current exoplanet population (dots).  The green right-pointing triangles, red down-pointing triangles, blue left-pointing triangles and magenta up-pointing triangles correspond to average stellar fluxes of 468183, 185691, 73680 and 11617 $\mathrm{Wm^{-2}}$, respectively.  See Table \ref{simulationtable} and text for more details.}
\label{gridplot}
\end{figure}

\subsection{Grid simulations}
Table~\ref{simulationtable} lists our model integrations for this study.  For all models, we assume the radius and gravity of a generic hot Jupiter\footnote{HD189733b, $R_P=8.2396 \times 10^7~\mathrm{m}$, $\mathrm{g=21.4~m~s^{-2}}$.  Note that we could have easily chosen any other well-characterized hot Jupiter, such as HD 209458b.  However, the primary goal of this study is to illustrate the dynamical and observational effects of eccentricity and average stellar flux.  The precise planetary radius and gravity are of secondary importance.}.  Figure~\ref{gridplot} shows the parameter space plotted as a function of semi-major axis and eccentricity, along with detected exoplanets with semi-major axes less than 0.5 AU.  In these simulations, we systematically vary both the average stellar flux and eccentricity.  Orbital eccentricities of 0.0 to 0.75 and orbit-averaged stellar fluxes, $\langle F \rangle$ of 468183, 185691, 73680 and 11617 $\mathrm{W m^{-2}}$ were explored.  These values correspond to equilibrium temperatures ($T_{eq}$) of approximately 1199, 951, 755 and 476 K, respectively.  For these given values of $e$ and $\langle F \rangle$, the orbital semi-major axis, $a$, is calculated using the equation

\begin{equation}
\label{aucalc}
\langle F \rangle=\frac{L}{4\pi a^{2}(1-e^{2})^{1/2}}.
\end{equation}

\noindent where $L$ is the luminosity of HD 189733.  These chosen simulations fall well within the current exoplanet population.

\begin{deluxetable*}{cccccccccc}
\tabletypesize{\scriptsize}
\tablecaption{List of model integrations. \label{simulationtable}}
\tablewidth{0pt}
\tablehead{
\colhead{${\langle F \rangle}$ $(\mathrm{W~m^{-2}})$} & \colhead {$T_{eq}$ (K)} & \colhead{$a$~(AU)} & \colhead{$e$} &
\colhead{$r_p$ (AU)} & \colhead{$r_a$ (AU)} &
\colhead{$\mathrm{P_{orb}}$~(s)} & \colhead{$\mathrm{P_{rot}}$~(s)} & \colhead{$\mathrm{P_{orb}/P_{rot}}$} & \colhead{$\mathrm{\Omega~(s^{-1})}$}
}
\startdata
468183 & 1199 & 0.0313 & 0.00 & 0.0313 & 0.0313 & $1.917\times10^5$ & $1.917\times10^5$ & 1.0 & $3.27\times10^{-5}$ \\
468183 & 1199 & 0.0318 & 0.25 & 0.0239 & 0.0398 & $1.963\times10^5$ & $1.421\times10^5$ & 1.4 & $4.42\times10^{-5}$ \\
468183 & 1199 & 0.0385 & 0.75 & 0.00963 & 0.0674 & $2.615 \times10^5$ & $3.010\times10^4$ &  8.7 & $2.09\times10^{-4}$ \\
185691 & 951 & 0.0497 & 0.00 & 0.0497 & 0.0497 & $3.834\times10^5$ & $3.834\times10^5$ & 1.0 & $1.64\times10^{-5}$ \\
185691 & 951 & 0.0505 & 0.25 & 0.0379 & 0.0631 & $3.927\times10^5$ & $2.843\times10^5$ & 1.4 & $2.21\times10^{-5}$ \\
185691 & 951 & 0.0534 & 0.50 & 0.0267 & 0.0801 & $4.270\times10^5$ & $1.522\times10^5$ & 2.8 & $4.13\times10^{-5}$ \\
185691 & 951 & 0.0611 & 0.75 & 0.0153 & 0.1069 & $5.226\times10^5$ & $6.016\times10^4$ & 8.7 & $1.04\times10^{-4}$ \\
73680 & 755 & 0.0789 & 0.00 & 0.0789 & 0.0789 & $7.668\times10^5$ & $7.668\times10^5$ & 1.0 & $8.19\times10^{-6}$ \\
73680 & 755 & 0.0802 & 0.25 & 0.0602 & 0.1003 & $7.858\times10^5$ & $5.689\times10^5$ & 1.4 & $1.10\times10^{-5}$ \\
73680 & 755 & 0.0848 & 0.50 & 0.0424 & 0.1272 & $8.544\times10^5$ & $3.046\times10^5$ & 2.8 & $2.06\times10^{-5}$ \\
11617 & 476 & 0.1987 & 0.00 & 0.1987 & 0.1987 & $3.067\times10^6$ & $3.067\times10^6$ & 1.0 & $2.05\times10^{-5}$ \\
11617 & 476 & 0.2019 & 0.25 & 0.1514 & 0.2524 & $3.141\times10^6$ & $2.274\times10^6$ & 1.4 & $2.76\times10^{-5}$ \\
11617 & 476 & 0.2135 & 0.50 & 0.1068 & 0.3202 & $3.416\times10^6$ & $1.218\times10^6$ & 2.8 & $5.16\times10^{-6}$ \\
11617 & 476 & 0.2443 & 0.75 & 0.0611 & 0.4275 & $4.181\times10^6$ & $4.813\times10^5$ & 8.7 & $1.31\times10^{-5}$ 
\enddata
%\tablenotetext{a}{Average stellar flux, given in $\mathrm{W/m^2}$.}
%\tablenotetext{b}{Semi-major axis, given in AU.}
%\tablenotetext{c}{Orbital period, given in seconds.}
%\tablenotetext{d}{Rotational period, given in seconds.}
%\tablenotetext{e}{Ratio of the orbital period to rotational period.}
%\tablenotetext{f}{Planetary rotation rate, given in $s^{-1}$.}
\end{deluxetable*}

We assume the atmospheric composition to be 1$\times$ solar metallicity without TiO and VO; this composition has had success in explaining observations of HD 189733b.  As in Showman et al. (2009), opacities are determined assuming local chemical equilibrium (accounting for rainout of condensates) at each temperature and pressure over the 3D grid, assuming the elemental abundances of Lodders (2003).  We set $c_p$ and $\kappa$ to values appropriate for a $H_2$-dominated atmosphere ($1.3 \times 10^4~ \mathrm{J~kg^{-1}~K^{-1}}$ and 2/7, respectively).

For eccentric cases, we assume the planet is pseudo-synchronously rotating its host star---we assume the planet's tidal interactions with the star force the same side of the planet to approximately face the star every periapse passage.  We calculate this using the \cite{hut1981} formulation:

\begin{equation}\label{huteqn}
    P_{rot}=P_{orb}~\frac{(1+3e^2+\frac{3}{8}e^4)(1-e^2)^{3/2}}{1+\frac{15}{2}e^2+\frac{45}{8}e^4+\frac{5}{16}e^6}
\end{equation}

\begin{figure}
\centering
%\epsscale{0.75}
\includegraphics[trim = 1.4in 3.3in 0.7in 3.3in, clip, width=0.57\textwidth]{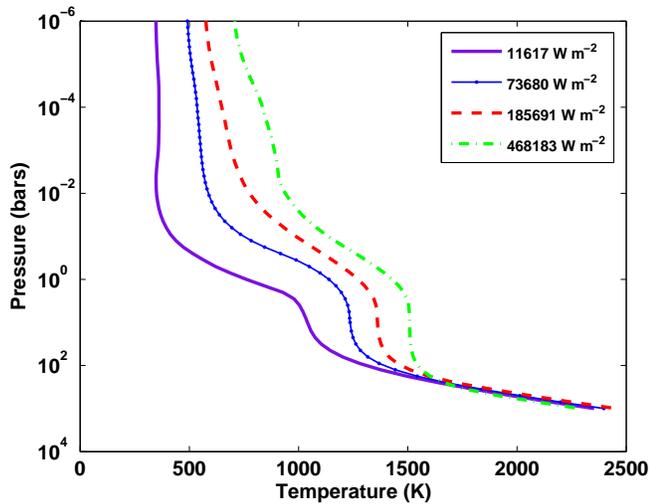}
\caption{Initial 1-D pressure-temperature profiles assigned to each column of the atmosphere at the start of each simulation, at the four different values of average stellar flux.  These profiles were calculated assuming radiative equilibrium, and use the same opacities as the full 3D model. }
\label{ptprofs}
\end{figure}

Pseudo-synchronization is a commonly-used assumption in the calculation of the rotation rate of eccentric exoplanets \citep[e.g.][]{deming2007,irwin2008,ll2008a,ll2008b,lewis2010}.  Nevertheless, the true dependence of rotation period on orbital period and eccentricity is very uncertain (e.g. Greenberg 2009), and other formulations besides Equation~\ref{huteqn} have also been suggested in the literature (e.g. Ivanov \& Papaloizou 2004).  We explore the sensitivity of the flow to the assumption of pseudo-synchronization in Section 3.4.

In each integration, we assume the winds to be initially zero, with each vertical atmospheric column assigned the same global-mean, radiative equilibrium pressure-temperature profile calculated using the same opacities as the main model (Figure~\ref{ptprofs}).  These temperature-pressure profiles are calculated for each value of average stellar flux (see above) using the methods described in \cite{fortney2005,fortney2008}.

\begin{figure}
%\epsscale{0.70}
\centering
\includegraphics[trim = 0.7in 2.7in 1.1in 2.4in, clip, width=0.5\textwidth]{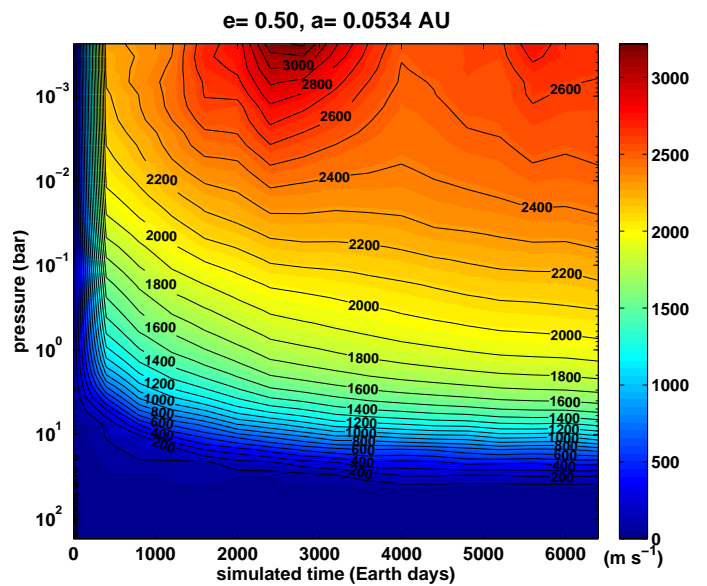}
\caption{Root-mean squared velocity plotted as function of pressure and simulation time for one of our model integrations.  The plot shows a case where $T_{eq}=951$ K, $e = 0.50$, and $a = 0.0534$ AU.  RMS velocity was calculated at each pressure level from wind speeds output every 100 days.The rapid decrease in velocity at the start of the run is merely an artifact at the start of the run.}  \label{vrms}
\end{figure}

Most model integrations were run until the winds reached a statistically steady configuration at upper levels.  Figure~\ref{vrms} shows the root-mean-squared (RMS) velocity plotted as a function of pressure and simulated time, for a model integrated for $>$6000 Earth days.  This behavior in RMS velocity is typical of most model integrations.  Weak drag was applied to the atmosphere, with a time constant of 100 days at 200 bars, with a drag coefficient that decreases linearly with pressure to zero at 10 bars.  Tests were conducted to explore the sensitivity of drag on the mean flow.  Extending the drag up to 1 bar affects the vertical and longitudinal extent of the equatorial jet, but not the qualitative nature of the circulation.  We reiterate that integrations without this drag scheme (but including our fourth-order Shapiro filter) conserve angular momentum to better than 0.1$\%$. 

\section{Results}

\subsection{Circulation regime}

\begin{figure}
\centering
%\epsscale{0.60}
\includegraphics[trim = 0.5in 2.7in 1.0in 2.3in, clip, width=0.45\textwidth]{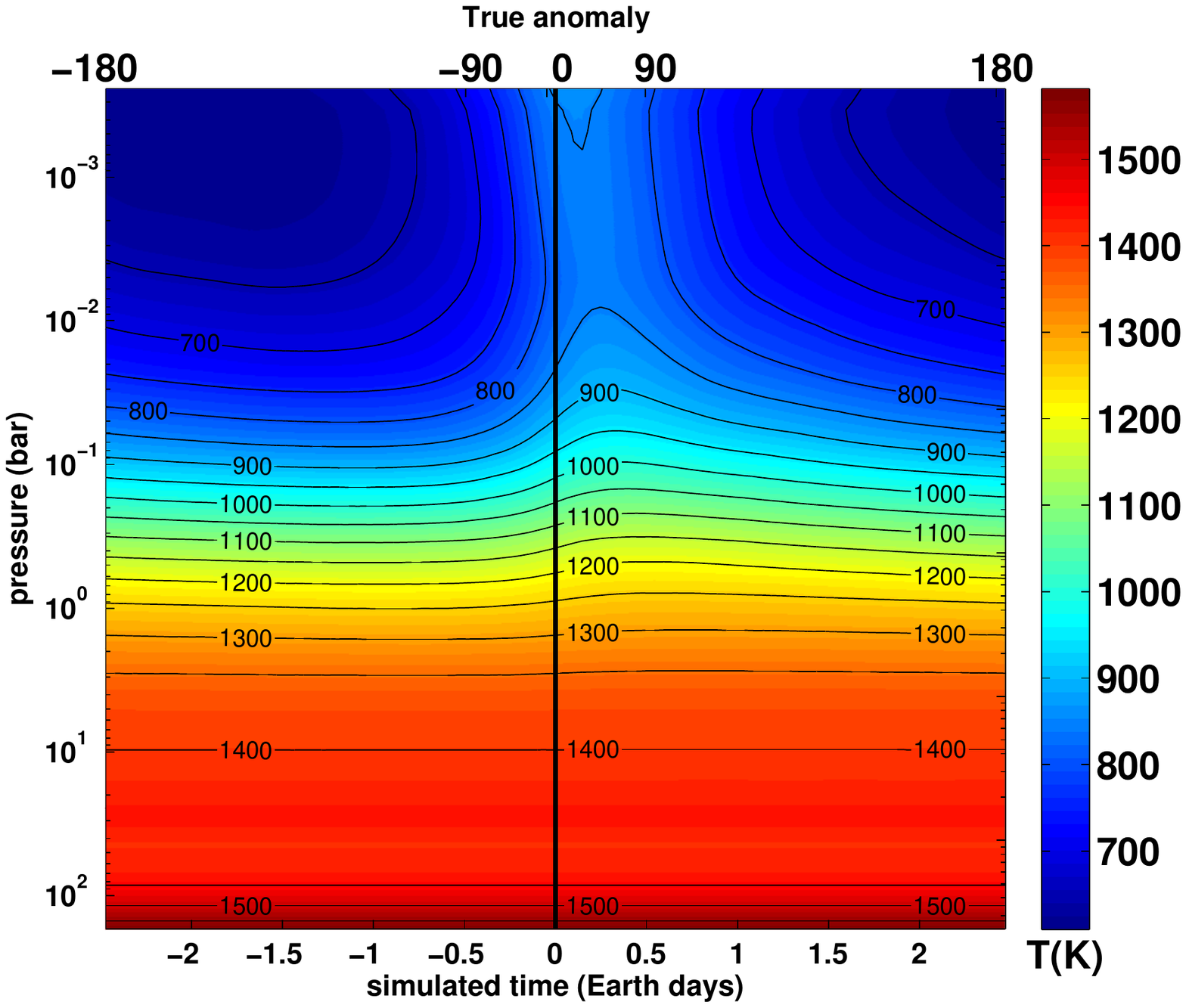}\\
\includegraphics[trim = 0.5in 2.7in 1.1in 2.5in, clip, width=0.45\textwidth]{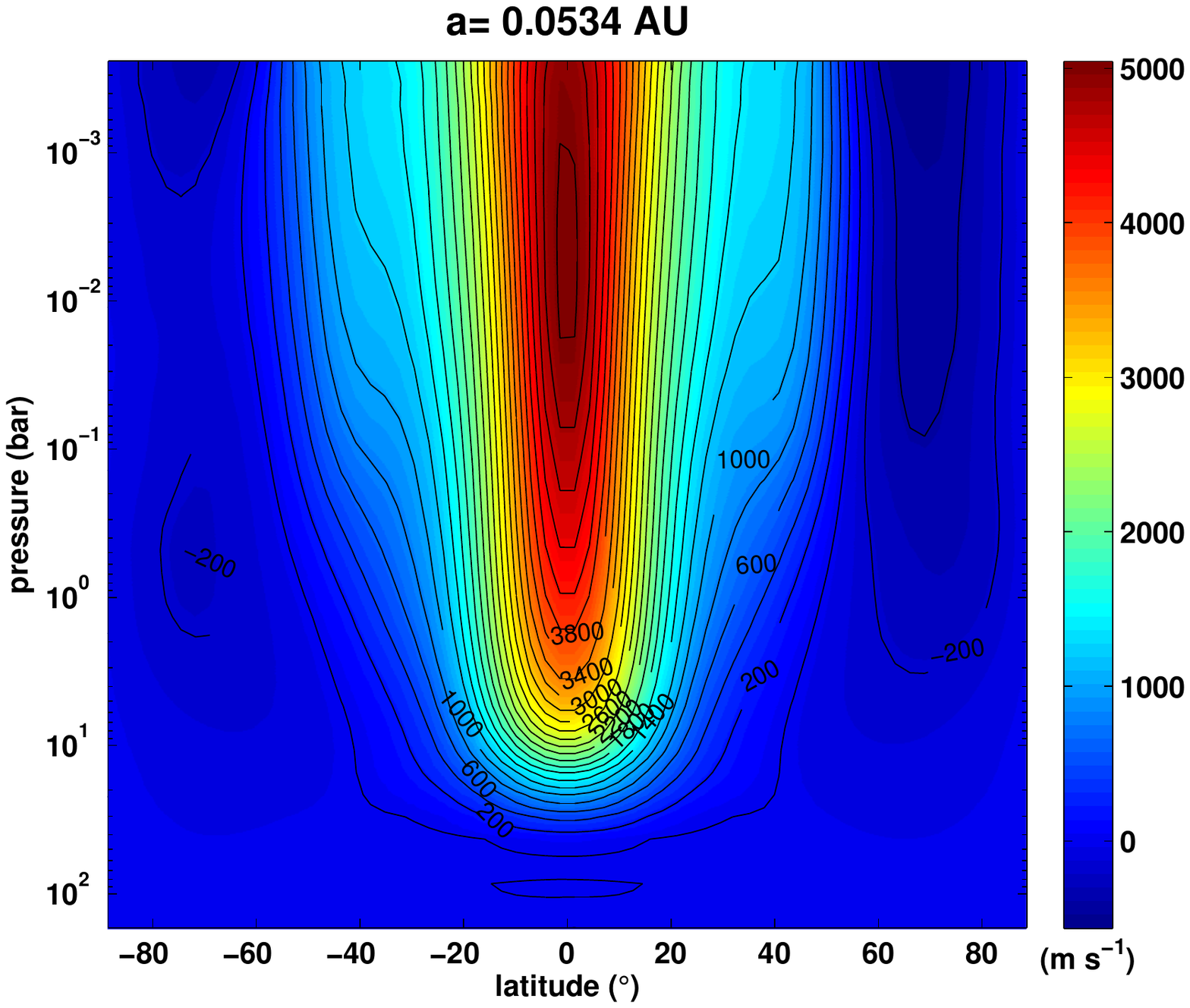}
\caption{Dynamical response of a model with $T_{eq}=951$ K, $e = 0.50$, and $a = 0.0534$ AU.  Top: Global, horizontally averaged temperature versus pressure and time throughout the full orbit; zero is defined as periapse passage.  The upper axis gives the true anomaly throughout the orbit.  Bottom: zonally-averaged zonal (east-west) winds versus pressure and latitude for the same model, time-averaged over a full orbit.}  \label{dynregime}
\end{figure}

To illustrate the circulation regime of an eccentric exoplanet, we present a nominal model where $e=0.50$, $a=0.0534$ AU, and $T_{eq}=951$ K,  representative of  the simulations (Figure~\ref{dynregime}).  The top panel shows the horizontally averaged global temperature as a function of pressure, time (in Earth days) and true anomaly, $f$, where $f=0^{\circ}$ at  periapse and $f=\pm180^{\circ}$ at apoapse.  Temperatures remain fairly constant at pressures exceeding 1 bar.  However, in the uppermost layers of the atmosphere, temperatures vary throughout the orbit, peaking $\sim$6 hours after periapse passage.  The bottom panel shows the zonal-mean zonal wind\footnote{Zonal wind is defined as east-west winds with positive (negative) values denoting eastward (westward) winds; a zonal average is an average in longitude.} of the same model time-averaged over a full orbit.  The flow exhibits a superrotating (eastward) equatorial jet which is the dominant feature of the flow, with peak winds of about 5000 $\mathrm{ms^{-1}}$.  Superrotation is commonly seen in models of hot Jupiters and hot Neptunes (e.g. Showman \& Guillot 2002; Cooper \& Showman 2005; Showman et al. 2008, 2009; Dobbs-Dixon \& Lin 2008; Dobbs-Dixon et al. 2010; Lewis et al. 2010; Heng et al. 2011; Perna et al. 2012; Rauscher \& Menou 2010, 2012), although not by all groups (Cho et al. 2008; Thrastorson and Cho 2010).

\tc{With peak winds exceeding 2-5 $\mathrm{km~s^{-1}}$, the equatorial jet is supersonic, as the sound speed for this class of planets is approximately 2 $\mathrm{km~s^{-1}}$.  As Heng (2012) shows, the presence of this supersonic flow generates shocks at the day-night terminator, upstream of the substellar point.  The SPARC/MITgcm does not properly account for the removal of kinetic energy as a result of these shocks.  The formation of shocks is expected to be shallow (i.e. at low pressure); in both of the model integrations that Heng (2012) presents, the Mach number falls below unity at pressures greater than 1 bar and hence shocks are not formed.  Nevertheless, the infrared photosphere and the peak jet speeds in our model occur at sufficiently low pressure that shocks could play some role, particularly when the orbital eccentricity is high.  Moreover, as a possible energy dissipation mechanism, shocks could have a non-local effect (i.e., they could cause an indirect effect on the flow even at pressures deeper than those where the shocks directly occur). Clearly, more work is warranted on the role of shocks in hot Jupiter atmospheres.}

\begin{figure*}
\centering
%\epsscale{0.90}
\includegraphics[trim = 1.35in 0.4in 0.7in 1.0in, clip, width=0.7\textwidth]{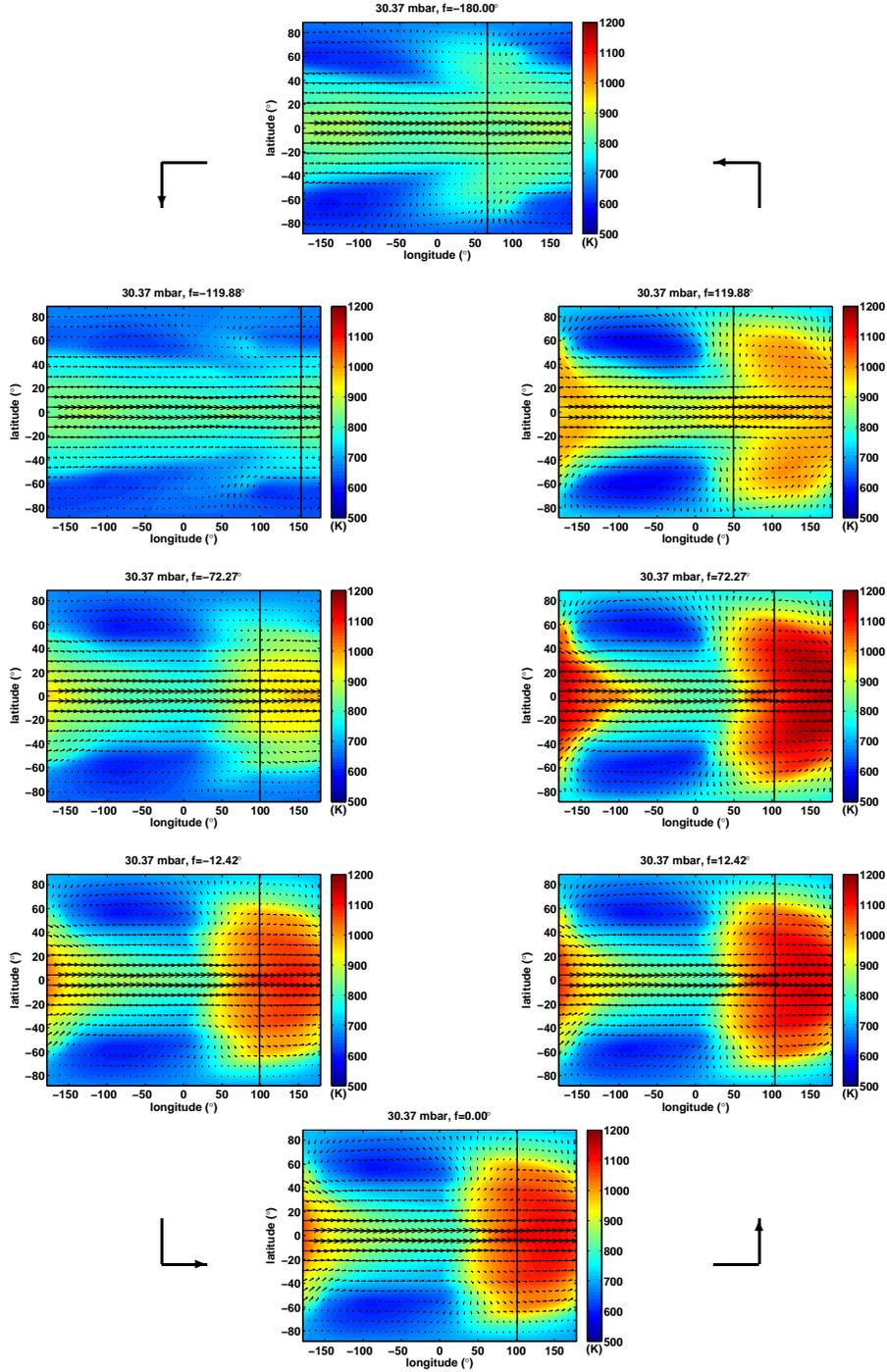}
%\plotone{fig5.ps}
\caption{Snapshots of the wind and temperature structure at 30 mbar of a planet with $T_{eq}=951$ K, $e = 0.50$, and $a = 0.0534$ AU.  To trace the behavior throughout one complete orbit, the different panels should be followeed counterclockwise from the the top panel.  Snapshots are taken from apoapse (true anomaly $f = -180.0^{\circ}$, top panel) to hours after periapse ($f=119^{\circ}$ , follow arrows) and back again to apoapse.  Each panel is plotted on the same temperature colorscale.  Snapshots were taken between 6103 and 6108 Earth days.  Note that peak temperatures lag periapse passage. Periapse corresponds to $f=0^{\circ}$ and apoapse to $180^{\circ}$.  The vertical line corresponds to substellar longitude.}
\label{snapshots}
\end{figure*}

\begin{figure*}
\centering
\includegraphics[trim = 1.4in 0.3in 0.7in 0.5in, clip, width=0.7\textwidth]{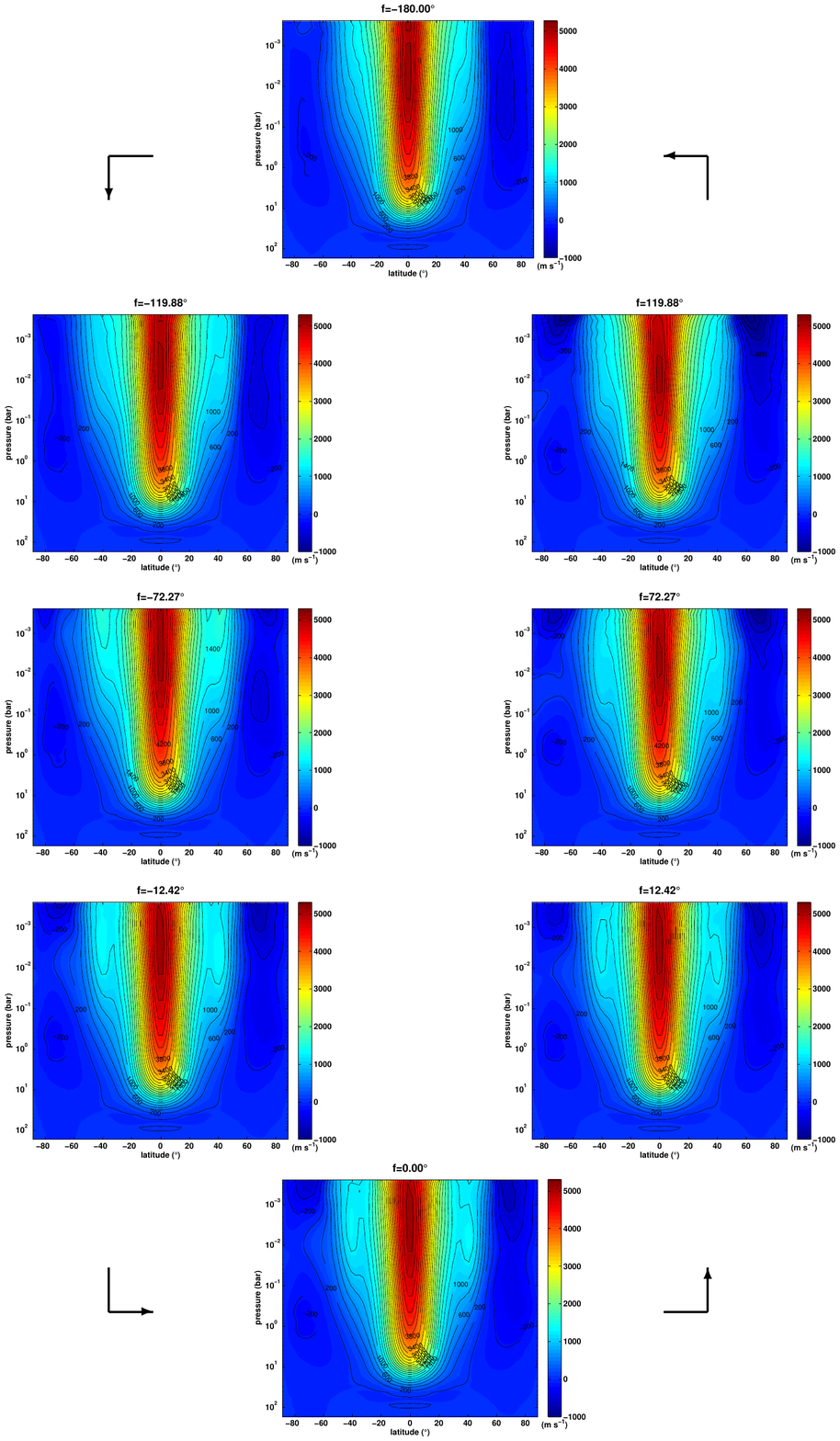}
%\epsscale{0.88}
%\plotone{fig6.ps}
\caption{Snapshots of the zonal-mean zonal wind for the same integration described in Figure~\ref{snapshots} ($T_{eq}=951$ K, $e = 0.50$, and $a = 0.0534$ AU).  Snapshots are again taken from apoapse ($f = -180.0^{\circ}$) to hours after periapse ($f=119^{\circ}$) and back again to apoapse.  Each panel is plotted on the same temperature colorscale.  The snapshots correspond to outputs from 6103-6108 Earth days.  Note that the equatorial superrotating jet is persistent throughout the orbit, with peak speeds exceeding 5000 $\mathrm{ms^{-1}}$.}
\label{snapshots_uz}
\end{figure*}

Snapshots of the wind and temperature profile throughout the orbit at a pressure level of 30 mbar (approximately the level of the infrared photosphere) show that the planet undergoes a flash heating event as the planet approaches periapse passage (Figure~\ref{snapshots}).  The eight panels show snapshots throughout one complete orbit from apoapse (top figure) to hours after periapse ($f=119^{\circ}$, follow arrows) and back again to apoapse, all with the same colorscale for temperature.  Like the circulation of HD 189733b in its nominally circular orbit \citep[see][]{showman2009}, the flow maintains the eastward equatorial superrotating jet throughout its orbit (Figure~\ref{snapshots_uz}).  As the planet approaches periapse (Figure~\ref{snapshots}, $f=-12^{\circ}$), temperatures rise to $\sim$1000 K at the equator, and a high-amplitude eddy structure develops with a pattern of temperature and wind vectors that tilt eastward toward the equator.  This so-called ``chevron"-shaped feature is prominent at periapse (bottom panel), extends over a wide range of latitude and longitude, and persists hours after periapse.  The chevron shape is also seen in our other integrations, and indicates that eddies transport momentum equatorward (see Section 3.6).

\begin{figure*}
%\epsscale{0.80}
\centering
\includegraphics[trim = 0.7in 2.7in 1.1in 2.6in, clip, width=0.36\textwidth]{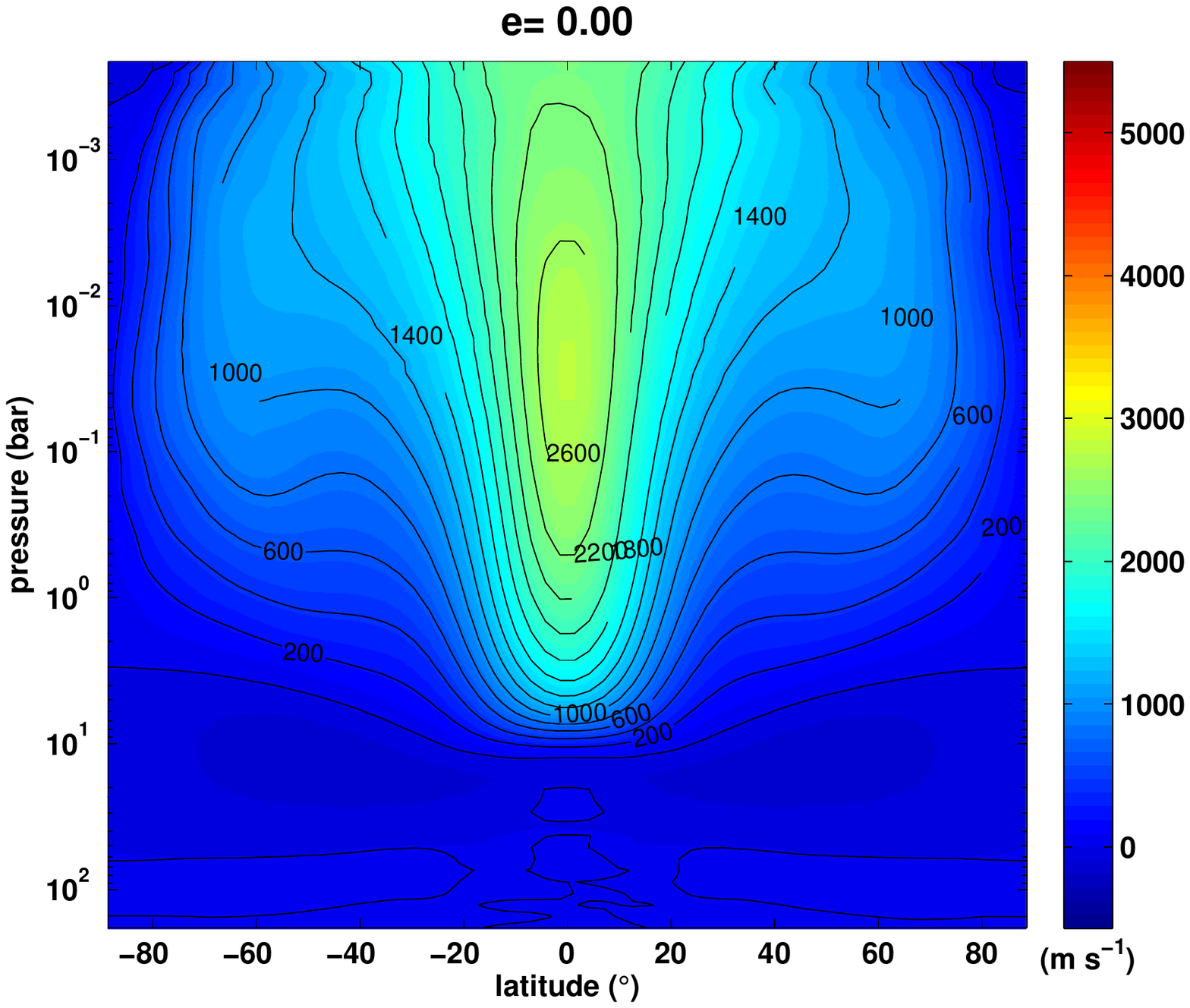}
\includegraphics[trim = 0.5in 2.8in 0.9in 2.7in, clip, width=0.41\textwidth]{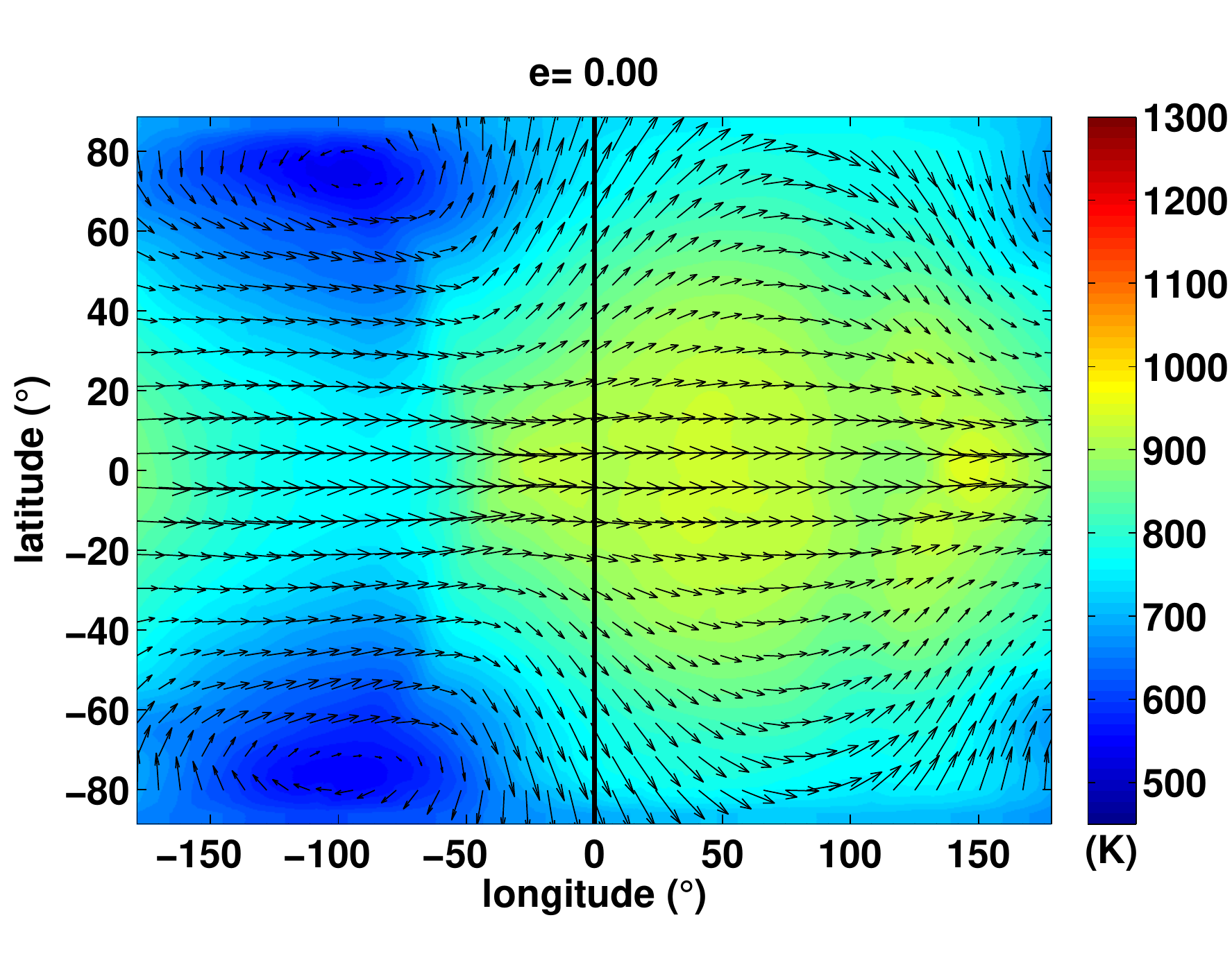}\\
\includegraphics[trim = 0.7in 2.7in 1.1in 2.6in, clip, width=0.36\textwidth]{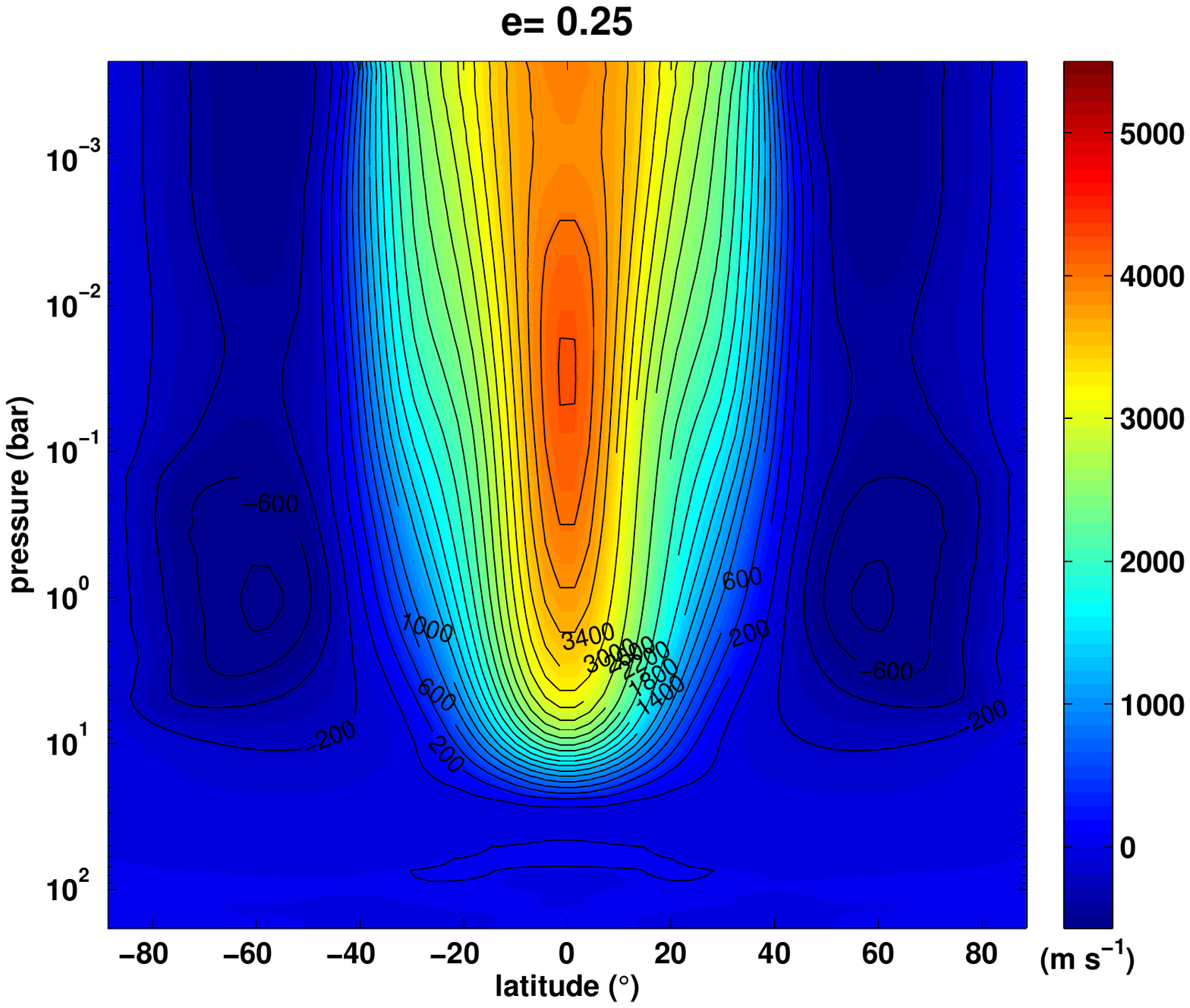}
\includegraphics[trim = 0.5in 2.8in 0.9in 2.7in, clip, width=0.41\textwidth]{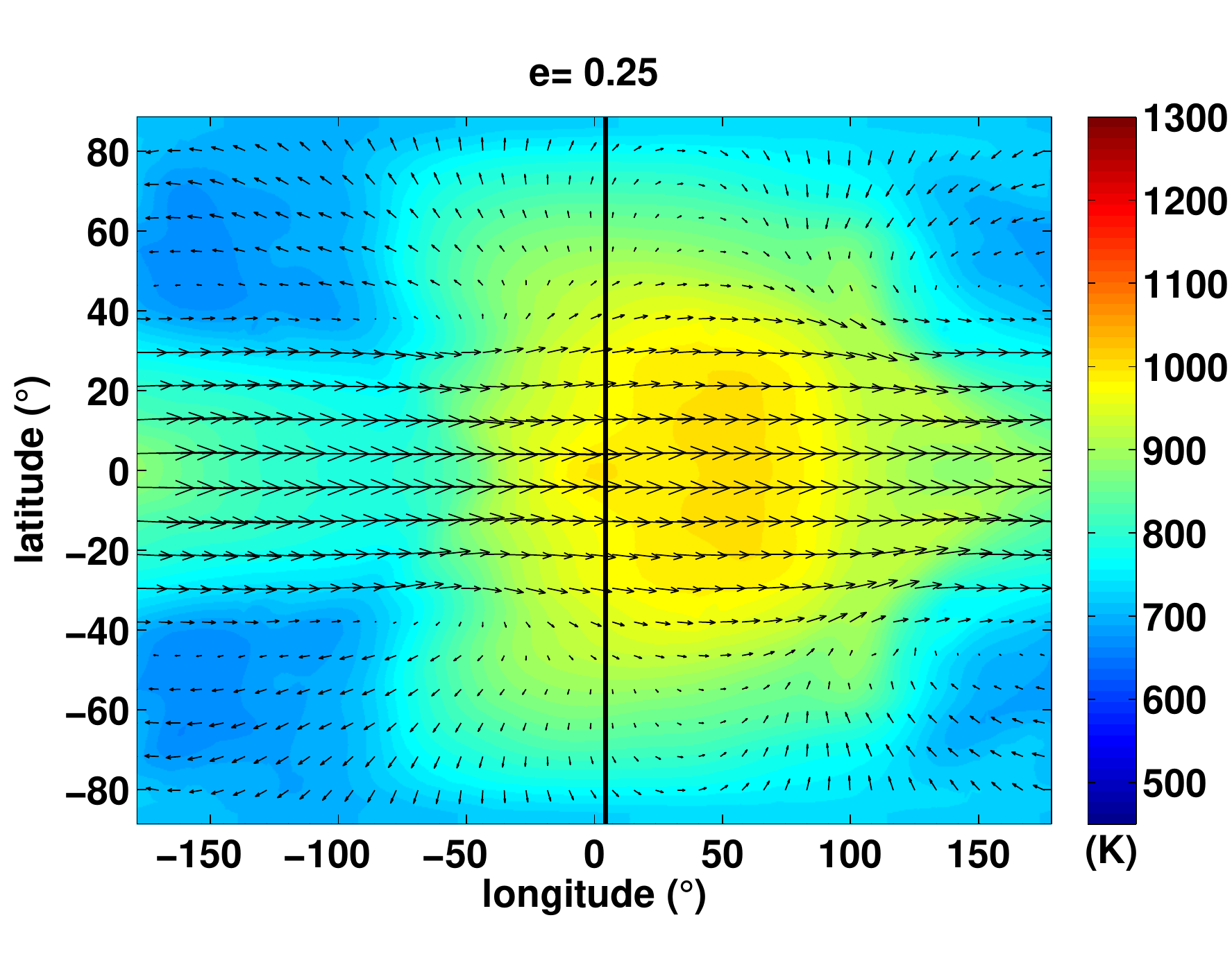}\\
\includegraphics[trim = 0.7in 2.7in 1.1in 2.6in, clip, width=0.36\textwidth]{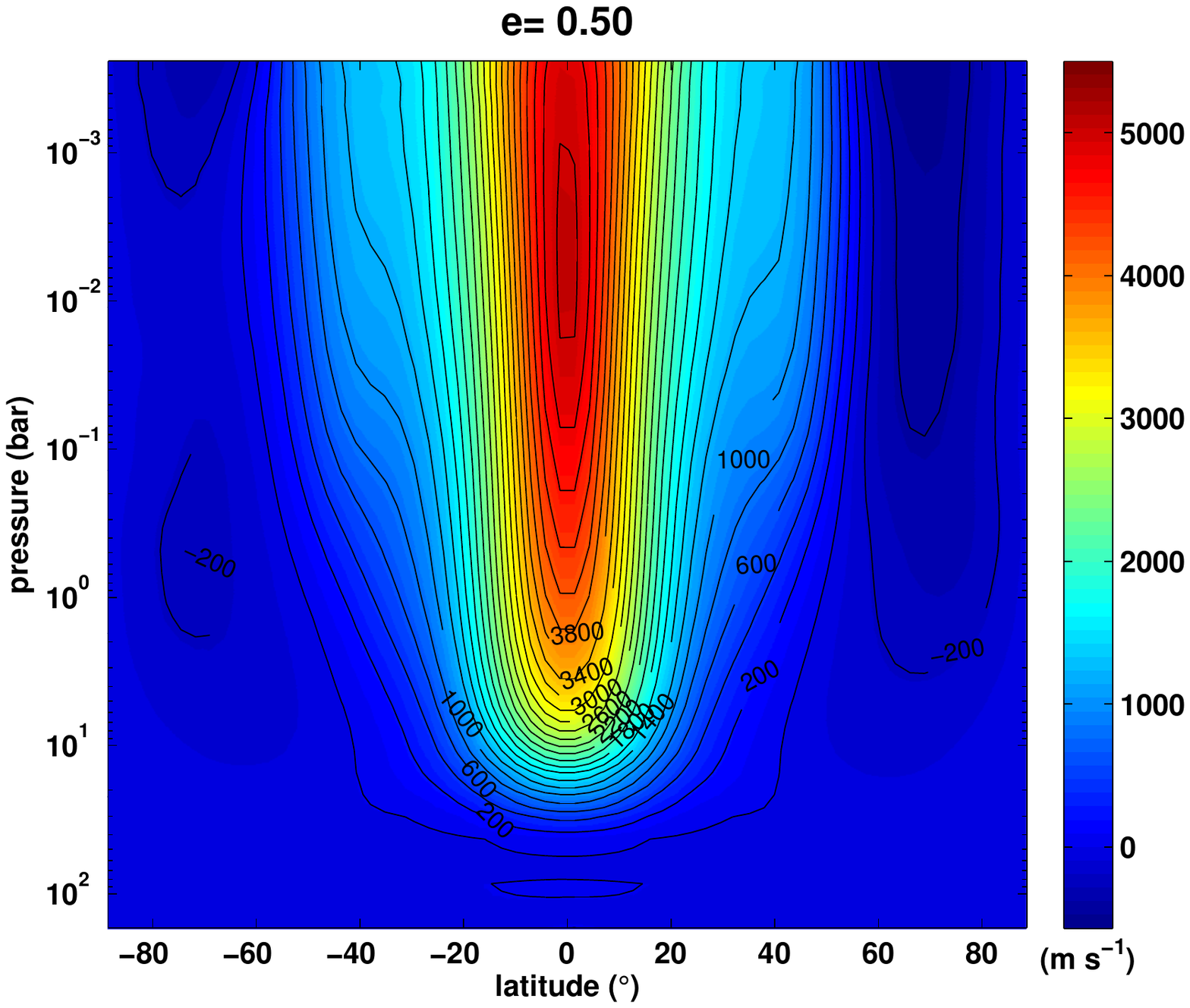}
\includegraphics[trim = 0.5in 2.8in 0.9in 2.7in, clip, width=0.41\textwidth]{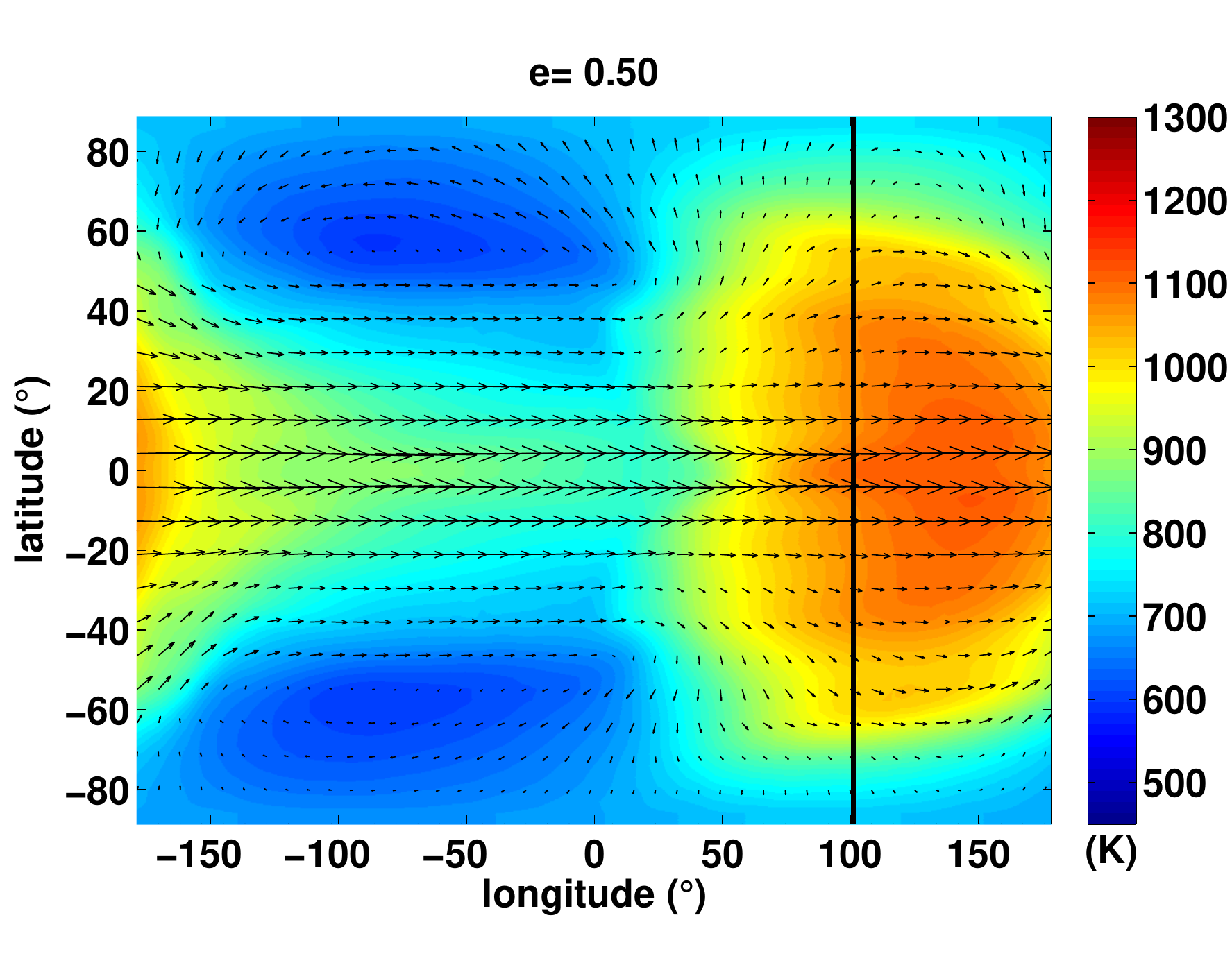}\\
\includegraphics[trim = 0.7in 2.7in 1.1in 2.6in, clip, width=0.36\textwidth]{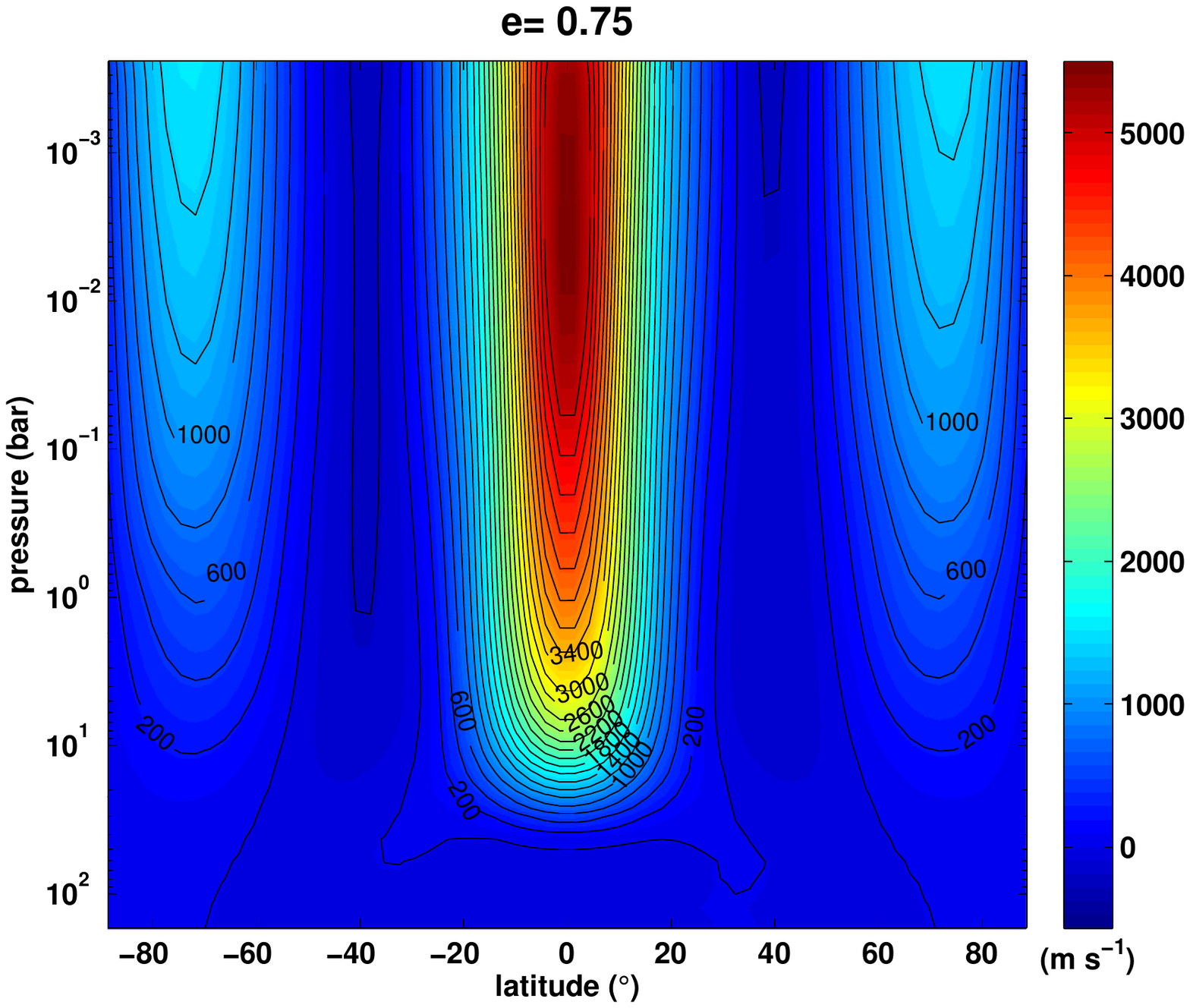}
\includegraphics[trim = 0.5in 2.8in 0.9in 2.7in, clip, width=0.41\textwidth]{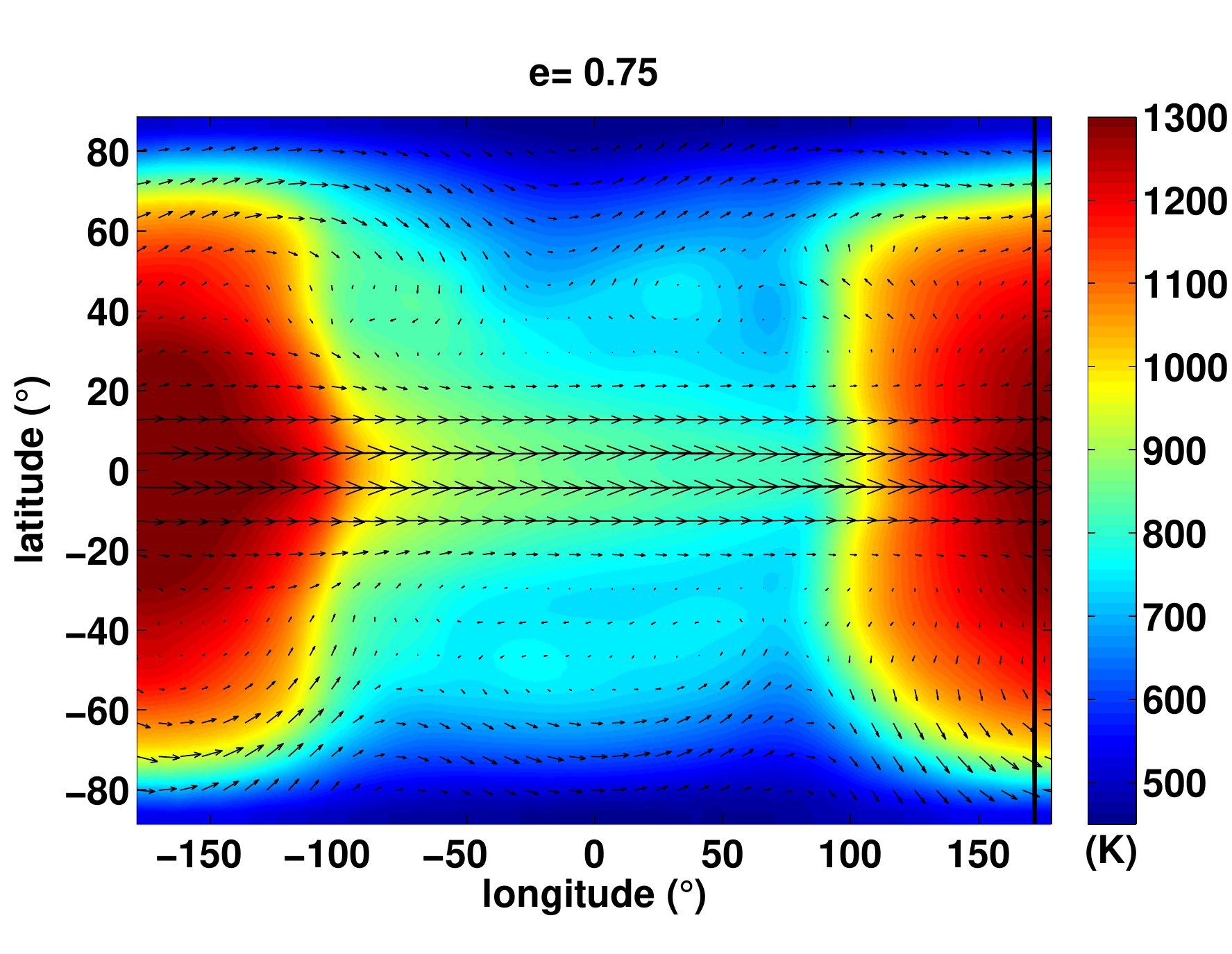}\\
\caption{Comparison of simulations with $T_{eq}=951$ K with increasing eccentricity, from $e=0.0$ (top row) to $e=0.75$ (bottom row).  Shown are plots of the zonal-mean zonal wind averaged over a full orbit (left column) and snapshots of the wind and temperature profiles at 30 mbar (right column). The snapshots of the wind and temperature profiles in the right column were taken at periapse passage.  The vertical bars in the right column denote the longitude of the substellar point.  The snapshots of wind and temperature correspond to model outputs of (from top to bottom) 4100, 5006, 6106 and 6904 Earth days.
}
\label{ecompare}
\end{figure*}

\begin{figure*}
\centering
%\epsscale{0.80}
%\includegraphics[trim = 0.0in 0.0in 0.0in 0.0in, clip, width=0.35\textwidth]{fig8_newgrid-014_uztimeavg.pdf}
%\includegraphics[trim = 0.0in 0.0in 0.0in 0.0in, clip, width=0.39\textwidth]{fig8_newgrid-014p_uvt.pdf}\\
\includegraphics[trim = 0.7in 2.7in 1.1in 2.6in, clip, width=0.36\textwidth]{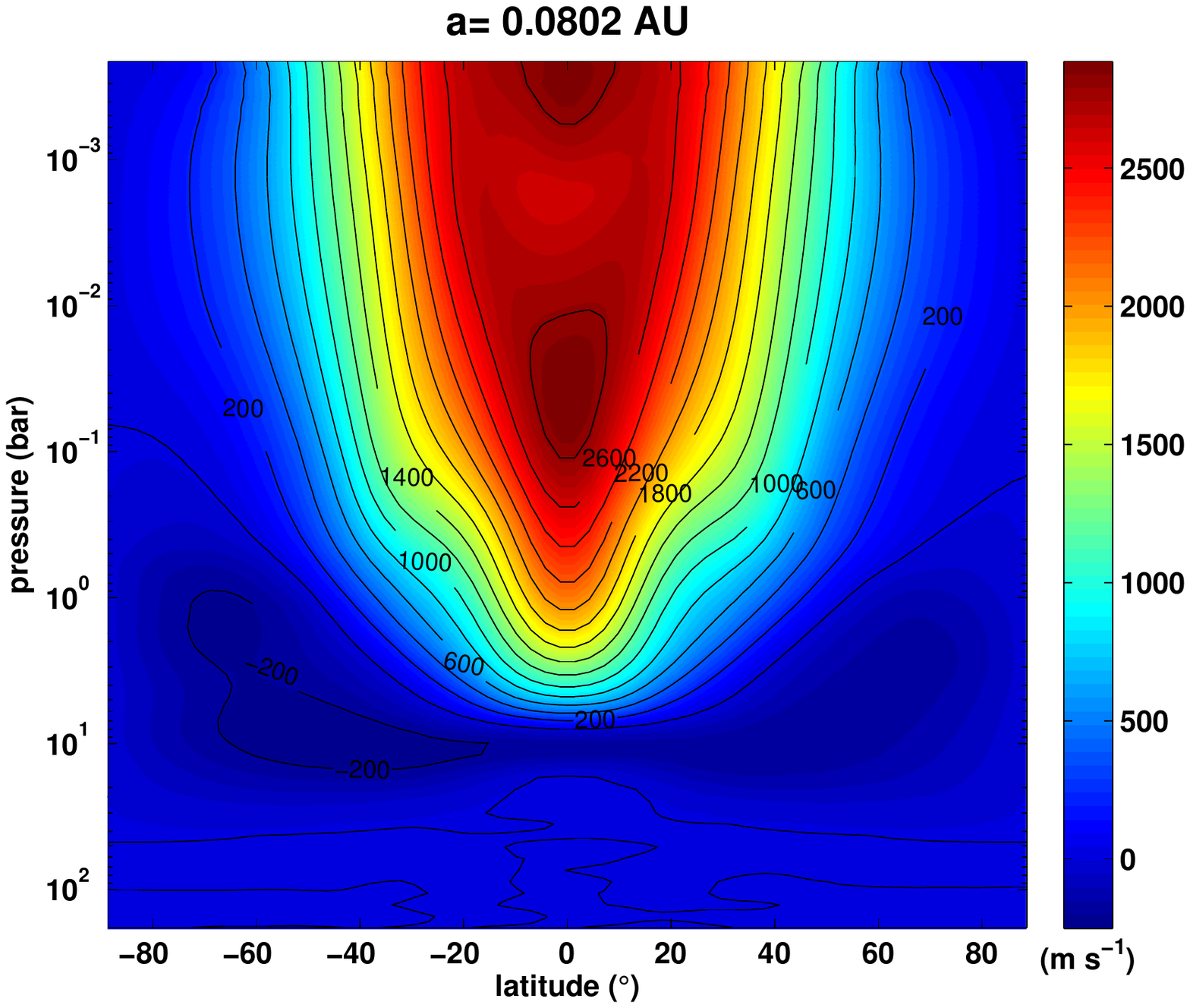}
\includegraphics[trim = 0.5in 2.8in 0.9in 2.7in, clip, width=0.41\textwidth]{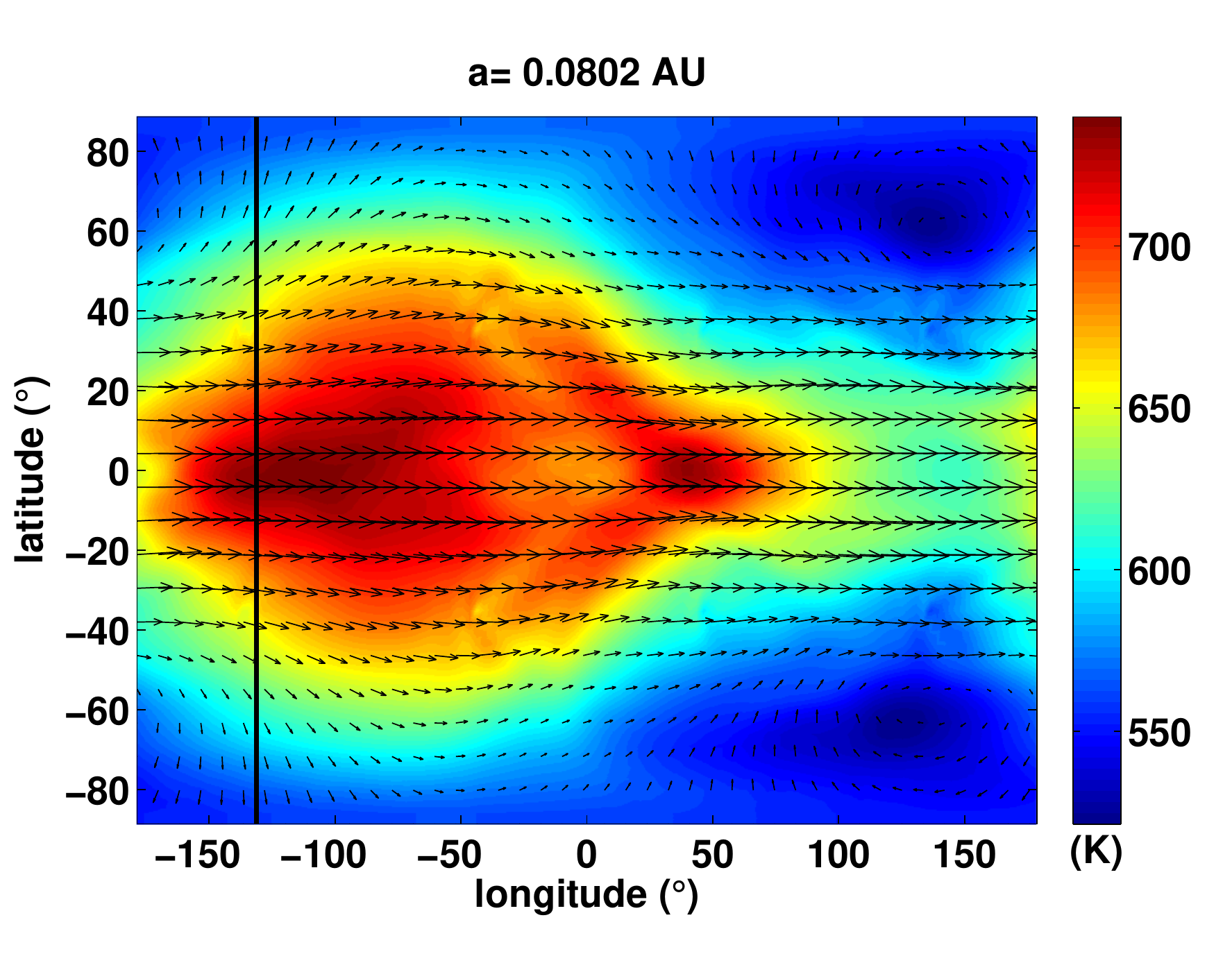}\\
\includegraphics[trim = 0.7in 2.7in 1.1in 2.6in, clip, width=0.36\textwidth]{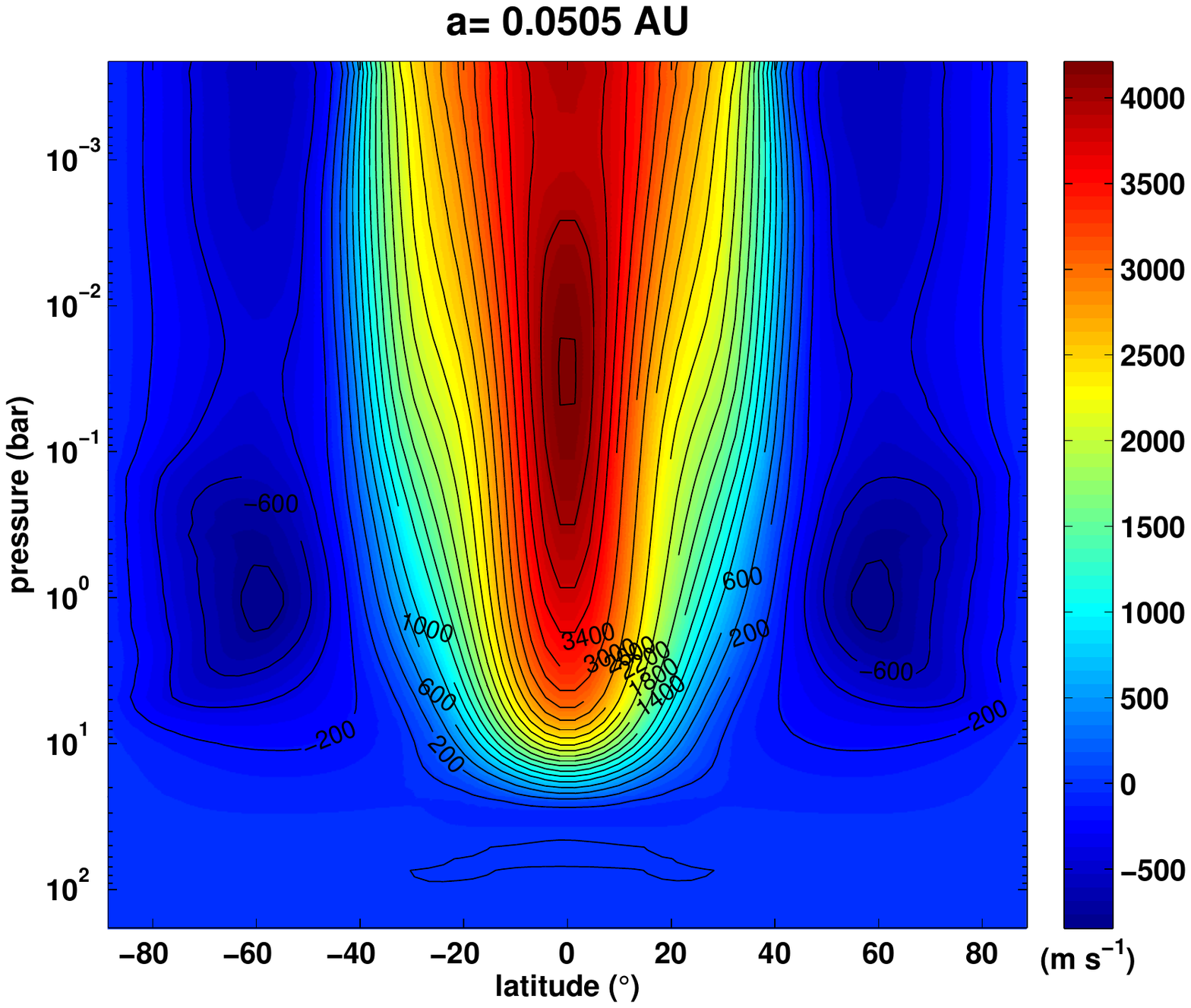}
\includegraphics[trim = 0.5in 2.8in 0.9in 2.7in, clip, width=0.41\textwidth]{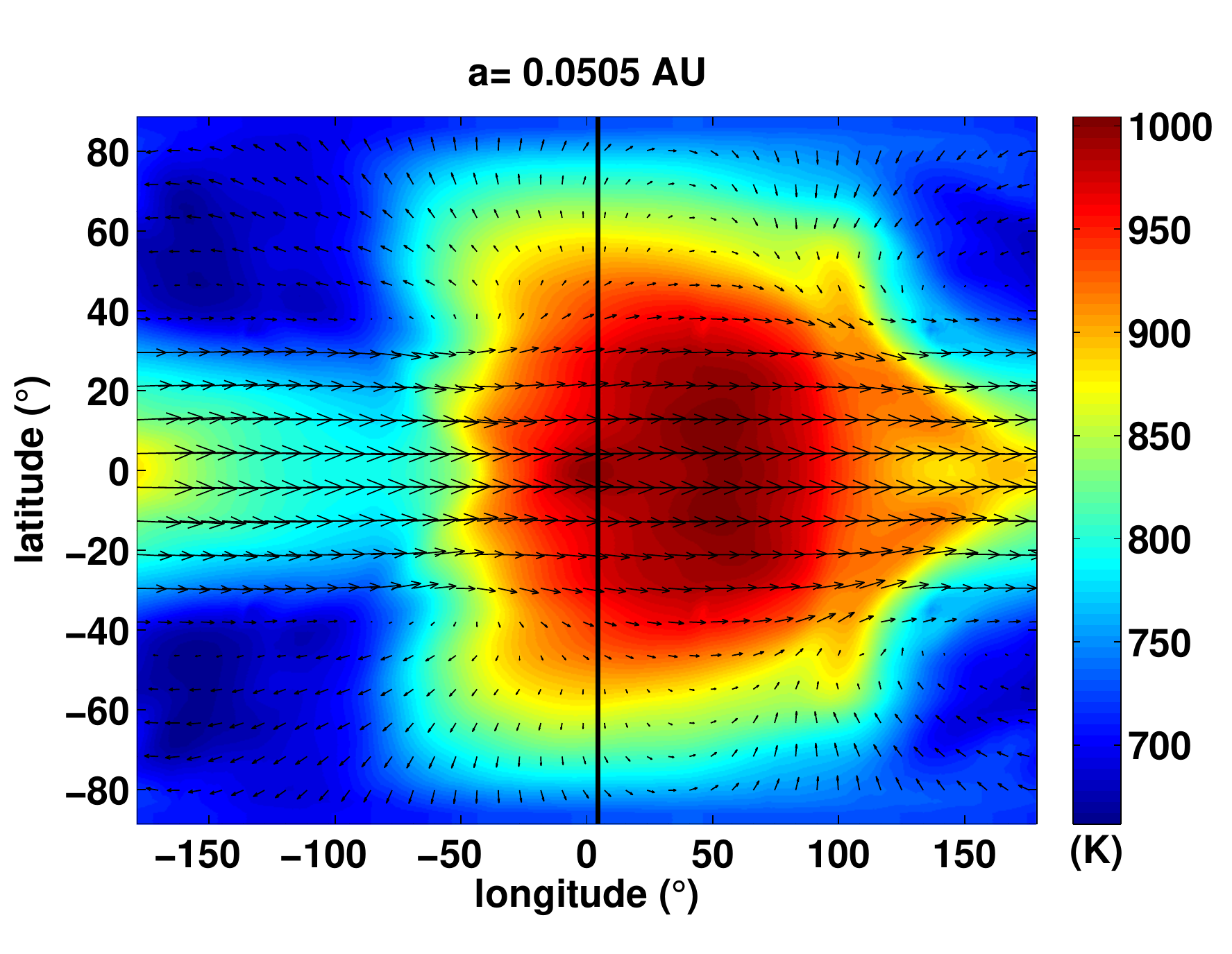}\\
\includegraphics[trim = 0.7in 2.7in 1.1in 2.6in, clip, width=0.36\textwidth]{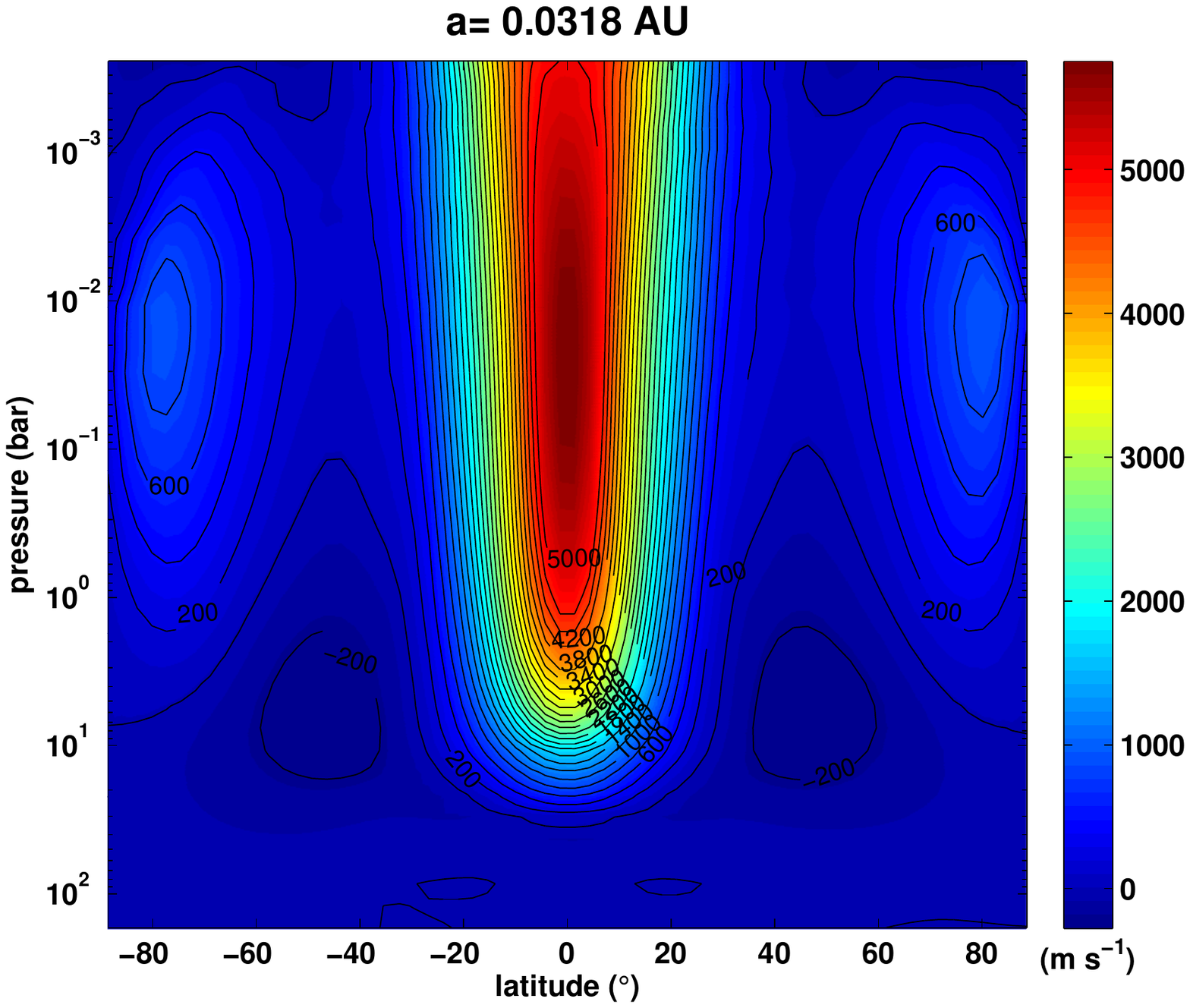}
\includegraphics[trim = 0.5in 2.8in 0.9in 2.7in, clip, width=0.41\textwidth]{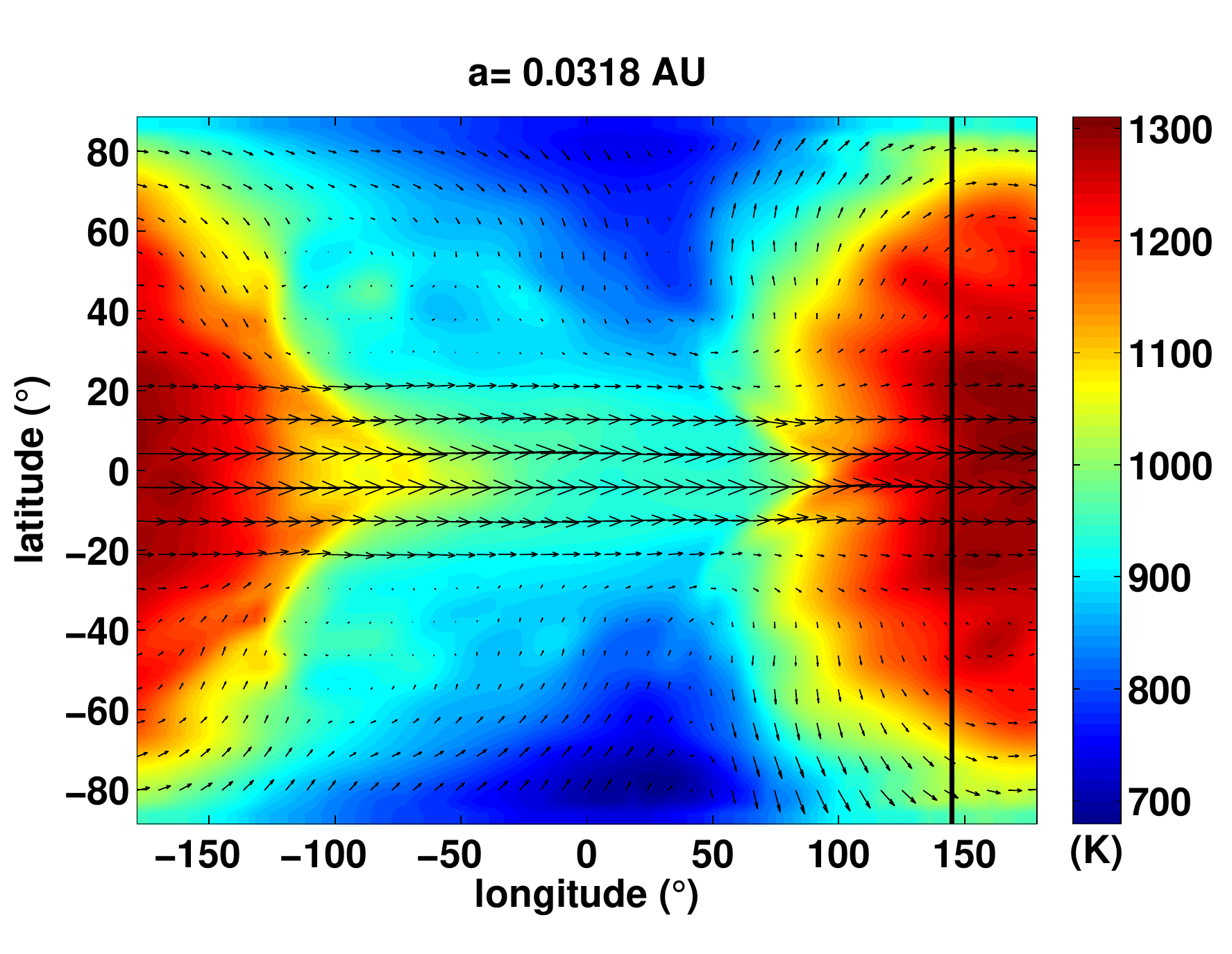}\\
\caption{Comparison of simulations with $e=0.25$ with decreasing average stellar flux, from bottom to top panels.  Shown are plots of the orbit-averaged zonal-mean zonal wind (left column) and the wind and temperature profiles at 30 mbar at periapse (right column).  Each plot has an independent colorscale.  The vertical bars in the right column denote the substellar longitude.  The snapshots of wind and temperature correspond to model outputs of (from top to bottom) 3706, 5006, and 2201 Earth days. }
\label{fluxcompare}
\end{figure*}

\subsection{Effect of eccentricity}
Our models show that, at constant orbit-mean stellar flux, the dayside temperatures, day-night temperature differences, and wind speeds at periapse passage all increase with increasing orbital eccentricity.  This is illustrated in Figure~\ref{ecompare}, which compares  three simulations with an equilibrium temperature of 951 K ($\langle F \rangle=185691~\mathrm{Wm^{-2}}$), that vary in eccentricity from 0.0 (top row) to 0.75 (bottom row).  Again, we plot each column on the same colorscale (zonal wind and temperature) for comparison.  As eccentricity increases, peak temperatures increase from 1000 K to 1300 K.  In addition, the variation in temperature from dayside to nightside increases; this strengthens the winds within the equatorial superrotating jet.  Peak winds reach $\sim$2500 $\mathrm{ms^{-1}}$ in the circular case, but increase to $\sim$5000 $\mathrm{ms^{-1}}$ in the high eccentricity case.

\begin{figure*}
\centering
\epsscale{0.80}
\includegraphics[trim = 1.0in 3.2in 0.9in 3.1in, clip, width=0.43\textwidth]{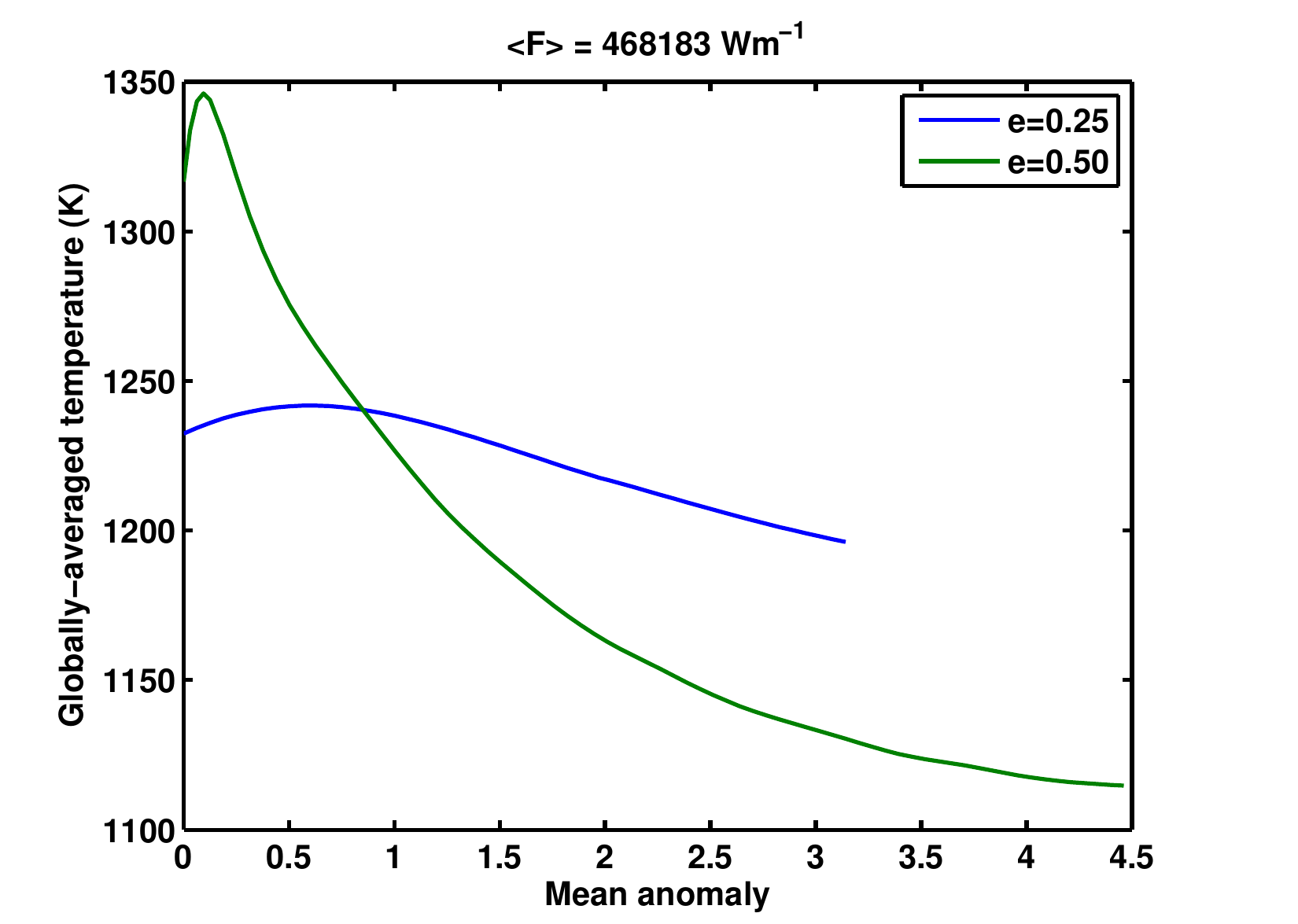}
\includegraphics[trim = 1.0in 3.2in 0.9in 3.1in, clip, width=0.43\textwidth]{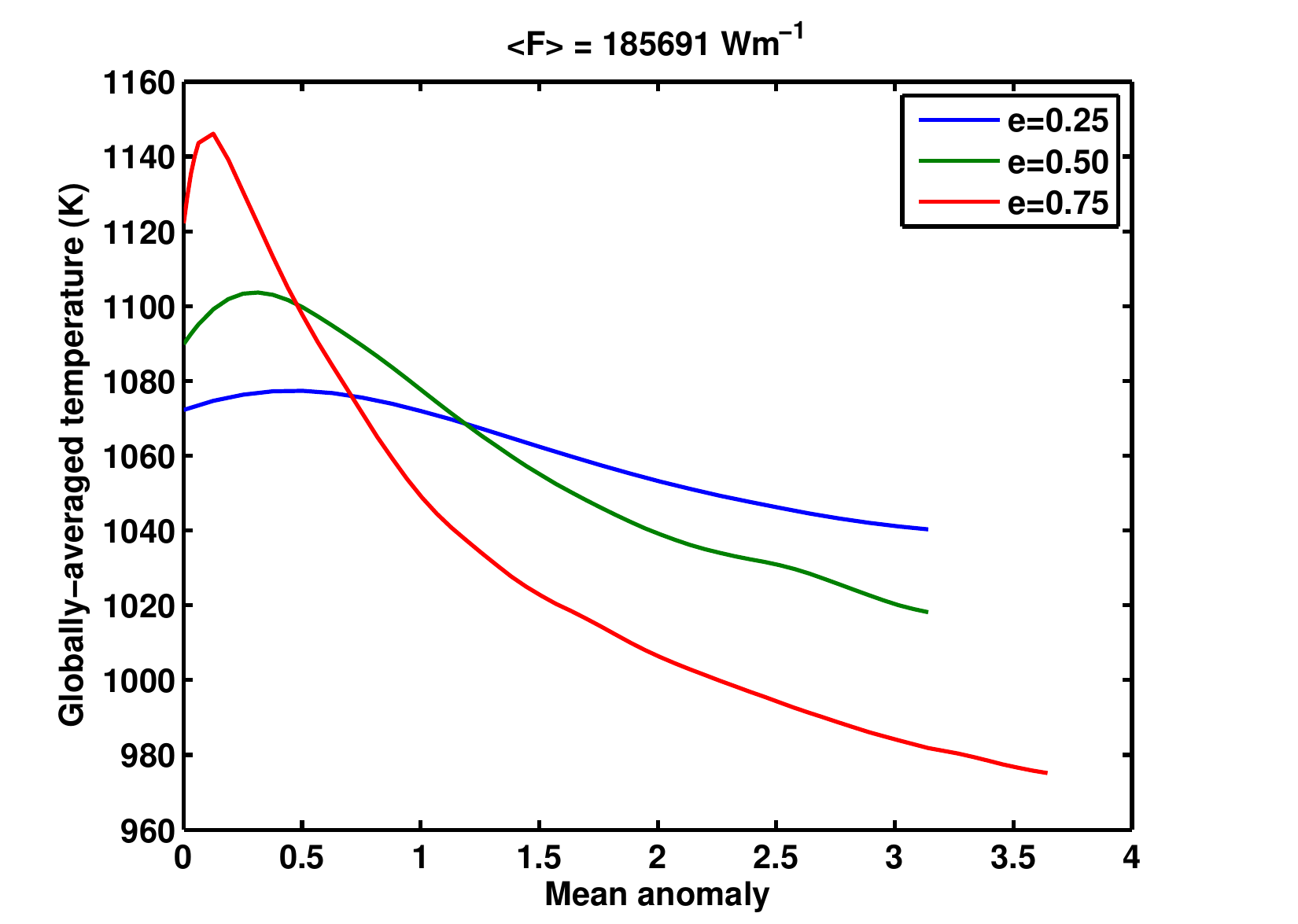}\\
\includegraphics[trim = 1.0in 3.2in 0.9in 3.1in, clip, width=0.43\textwidth]{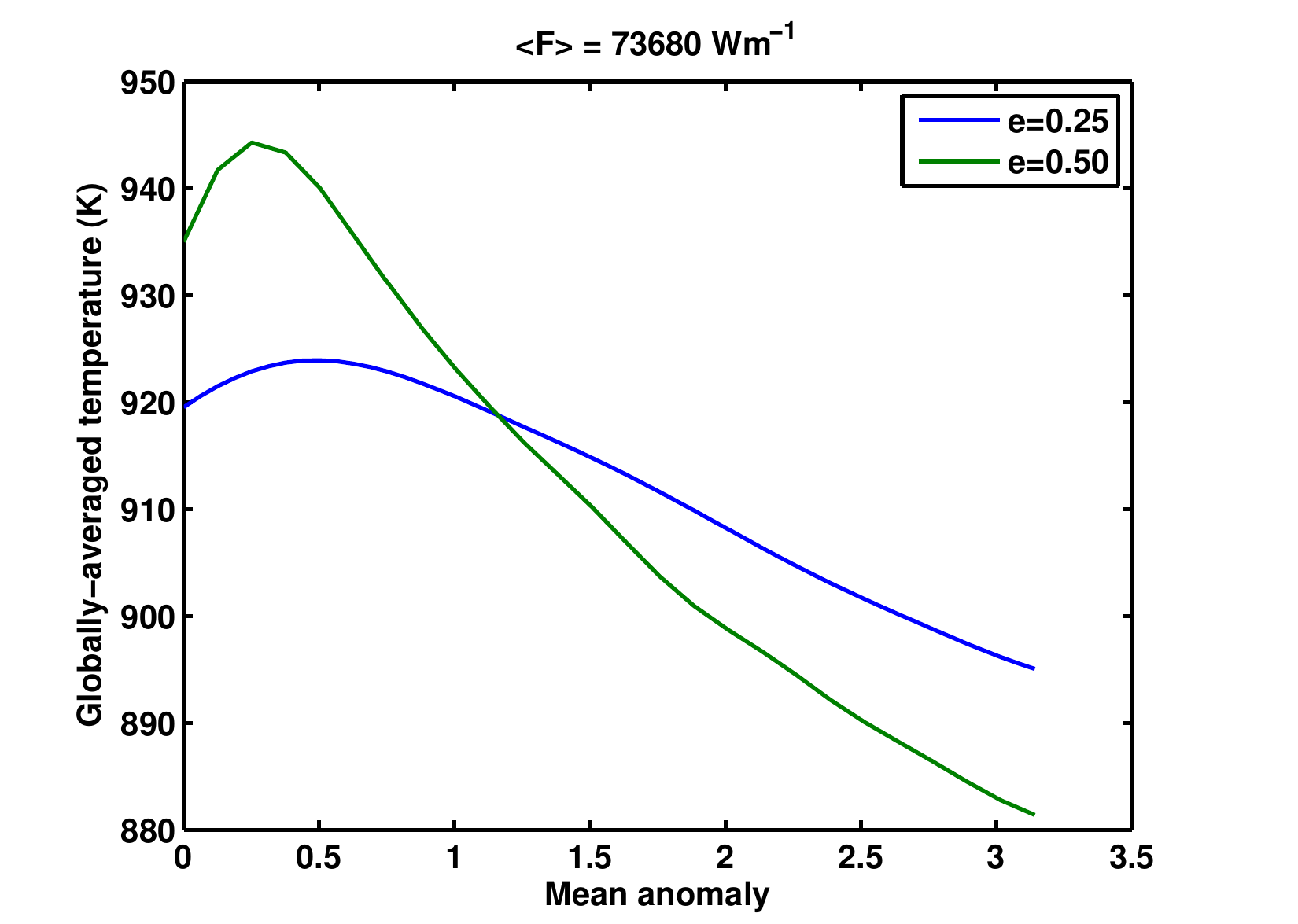}
\includegraphics[trim = 1.0in 3.2in 0.9in 3.1in, clip, width=0.43\textwidth]{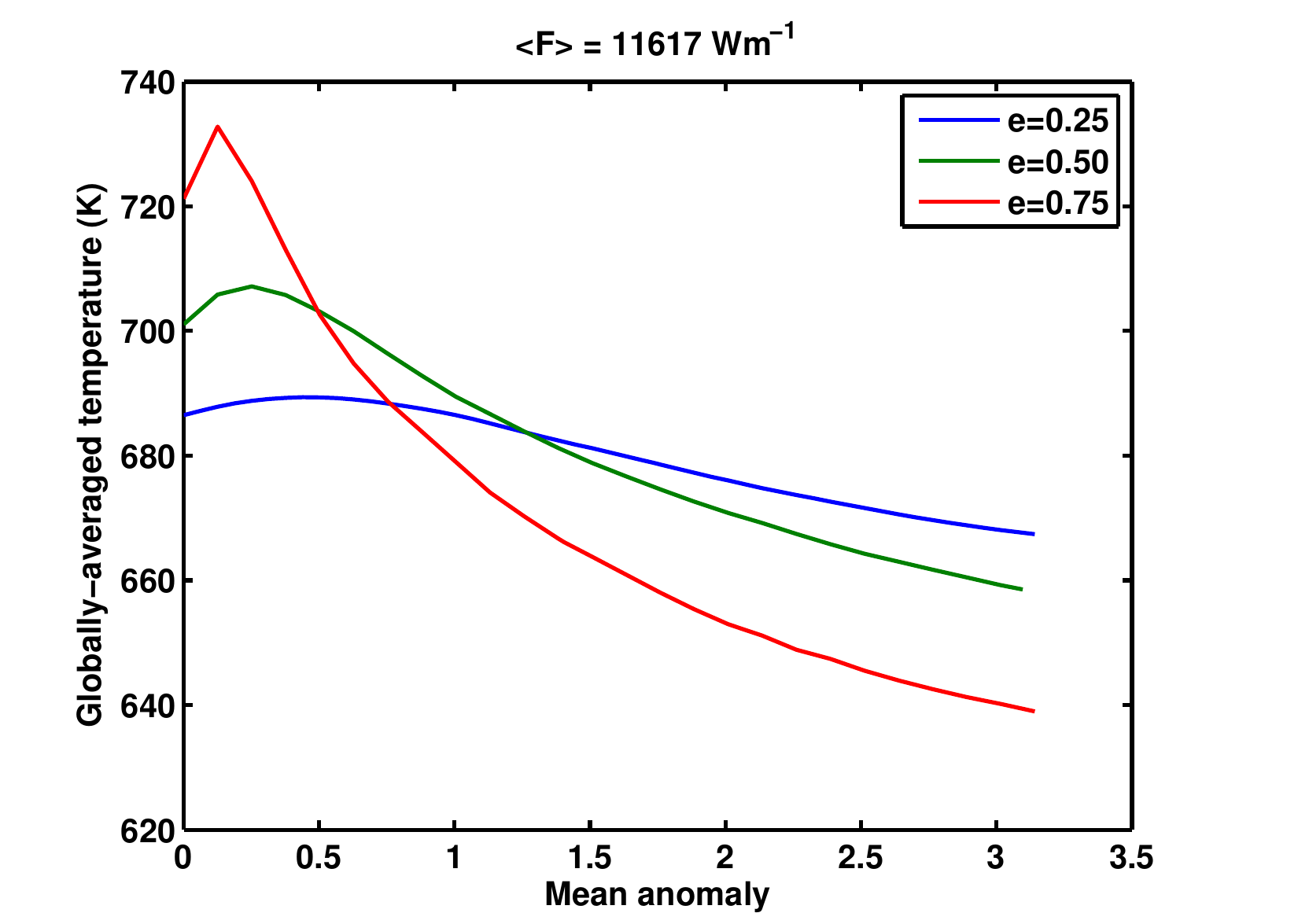}\\
\caption{Plots of global average temperature versus mean anomaly for each model integration.  Each plot contains profiles of a given average stellar flux (468183, 185691, 73680, and 11617 $\mathrm{Wm^{-1}}$), with blue, green and red representing profiles for $e=0.25, 0.50$ and $0.75$, respectively.}
\label{rows_tave1}
\end{figure*}

As the eccentricity increases, the equatorial jet also narrows in latitude.  In the case of highest eccentricity, the atmosphere develops westward midlatitude jets and eastward high-latitude jets.  The narrowing of the equatorial jet arises from the fact that jet width is confined to the Rossby radius of deformation (Showman \& Polvani 2011), the length scale over which the atmosphere adjusts to planetary-scale phenomena. The Rossby radius of deformation (Holton 2004) is defined at the equator as

\begin{equation}
\label{rossbycalc}
L_{R} = \sqrt{\frac{c}{\beta}}.
\end{equation}

\noindent where $c$ is the gravity wave speed, and $\beta$ is the derivative of the Coriolis parameter, $f$, with respect to latitudinal distance $y$, $\beta=\frac{df}{dy}=\frac{2\Omega_{rot}\cos{\phi}}{R_p}$, where $\Omega_{rot}$ is the planetary rotation rate, $\phi$ is the latitude and $R_p$ is the planetary radius.  At the equator, $\beta$ is simply $\frac{2\Omega_{rot}}{R_p}$.  Hence, at the equator $L_R$ is

\begin{equation}
L_R=\sqrt{\frac{cR_p}{2\Omega_{rot}}}.
\end{equation}

Therefore, $L_{R}\propto\frac{1}{\sqrt{\Omega_{rot}}}$, and an increased rotation rate leads to a narrower deformation radius.  From the \cite{hut1981} formulation, the combined effect of the larger orbital period and increase in eccentricity produces a shorter rotational period.  This, in turn, increases the rotation rate, which reduces the deformation radius.  For example, the ratio between $\Omega_{rot}$ for the $e$=0.0 case and the $e$=0.75 case is 6.3; hence, we would expect that the equatorial jet should be narrower by a factor of $\sqrt{6.3}\sim2.5$.  This is indeed the case, as the $e=0.75$ case has an average jet width of $\sim40^{\circ}$ latitude, while the $e=0.0$ case has an average jet width of $\sim100^{\circ}$ latitude.

\subsection{Effect of varying average stellar flux}
An increase in average stellar flux leads to increases in temperature and wind speeds, illustrated in Figure~\ref{fluxcompare}.  Each case shown has an eccentricity of 0.25; these maps have independent colorscales to better show atmospheric structure.  The planet that receives the highest average stellar flux throughout its orbit is hottest overall (bottom), with peak temperatures of 1300 K.  Furthermore, because the planet with the highest $\langle F \rangle$ orbits at the smallest orbital distance, it has a faster rotation rate and hence a narrower equatorial jet; this comparison makes the effect of rotation on the jet structure much more apparent.

%Interestingly, the wind and temperature structure of the farthest case ($a=0.2019$ AU, top row), is unlike the other three; the flow is fully westward at pressure levels above 1 mbar.  Peak temperatures reach only 390 K, and no strong day-night temperature variations occur.  This would suggest that a regime shift occurs between orbital distances of 0.08 and 0.20 AU.  The homogeneity in temperature is due to the planet's long radiative time constant and the slow rotation rate.  The planet effectively transports heat both latitudinally and longitudinally, thus leading to small temperature variations.  Exploring this regime shift in depth will be the subject of a future paper.

\subsection{Global average temperature vs. eccentricity and $\langle F \rangle$}

\begin{figure}
\centering
\epsscale{0.80}
\includegraphics[trim = 1.4in 3.2in 0.9in 3.1in, clip, width=0.55\textwidth]{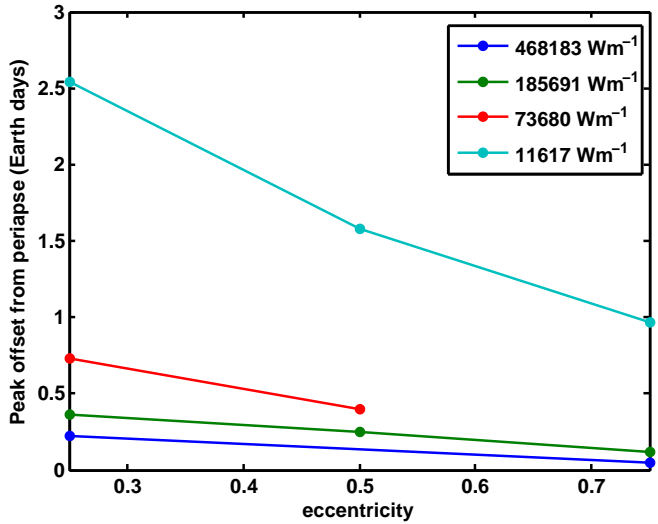}
\caption{Time after periapse passage at which the global-average temperature peaks (from Figure~\ref{rows_tave1}) versus eccentricity for all model integrations.  The blue, green, red and teal profiles correspond to average stellar fluxes of 468183, 185691, 73680, and 11617 $\mathrm{Wm^{-1}}$, respectively. }
\label{rows_tave2}
\end{figure}

Here we illustrate the variation in time of maximum planet-wide temperature with eccentricity and average stellar flux.  Figure \ref{rows_tave1} shows the globally-averaged temperature for each model integration, plotted as a function of mean anomaly\footnote{The mean anomaly, $M$, is defined as $\frac{2\pi}{P_{orb}}(t-\tau)$, where $\tau$ is the time of periapse passage.}, where $M=0$ corresponds to periapse passage.  With increasing eccentricity, the time at which the global-mean temperature reaches its peak value decreases.  This is expected, as a planet with a larger eccentricity will have a higher global-average temperature at periapse and hence shorter radiative time constant ($\tau_{rad}\propto{T^{-3}}$, see Showman et al. 2011).  Figure \ref{rows_tave2} plots the time of peak temperature from periapse versus eccentricity.  Here, the inverse relationship between eccentricity and the time of peak temperature is clear.

%Hence, $M$ linearly increases in time, and we can use it to compare each model integration without being limited by shortest period planets.

Figures \ref{rows_tave1} and \ref{rows_tave2} serve as a nice summary of the model integrations and their relation to observations, as temperature is one of the planetary properties that shapes lightcurves at secondary eclipse.  Observers can use these predictions to estimate the timing of peak flux (and global temperature) based on a given stellar flux and/eccentricity (see Section 4).

\begin{figure*}
\centering
%\epsscale{1.0}
\includegraphics[trim = 0.5in 2.7in 0.9in 2.6in, clip, width=0.43\textwidth]{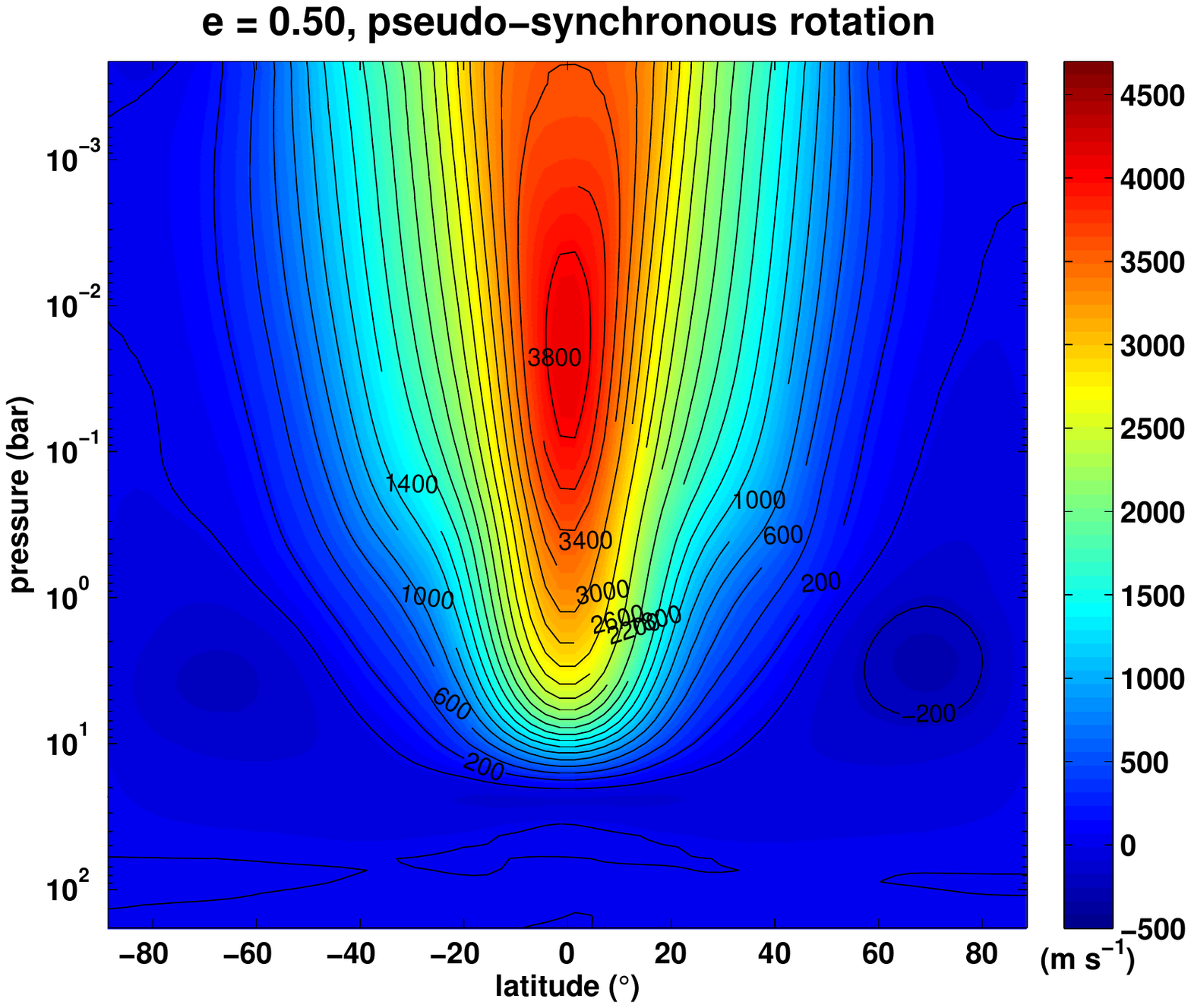}
\includegraphics[trim = 0.5in 2.7in 0.9in 2.6in, clip, width=0.43\textwidth]{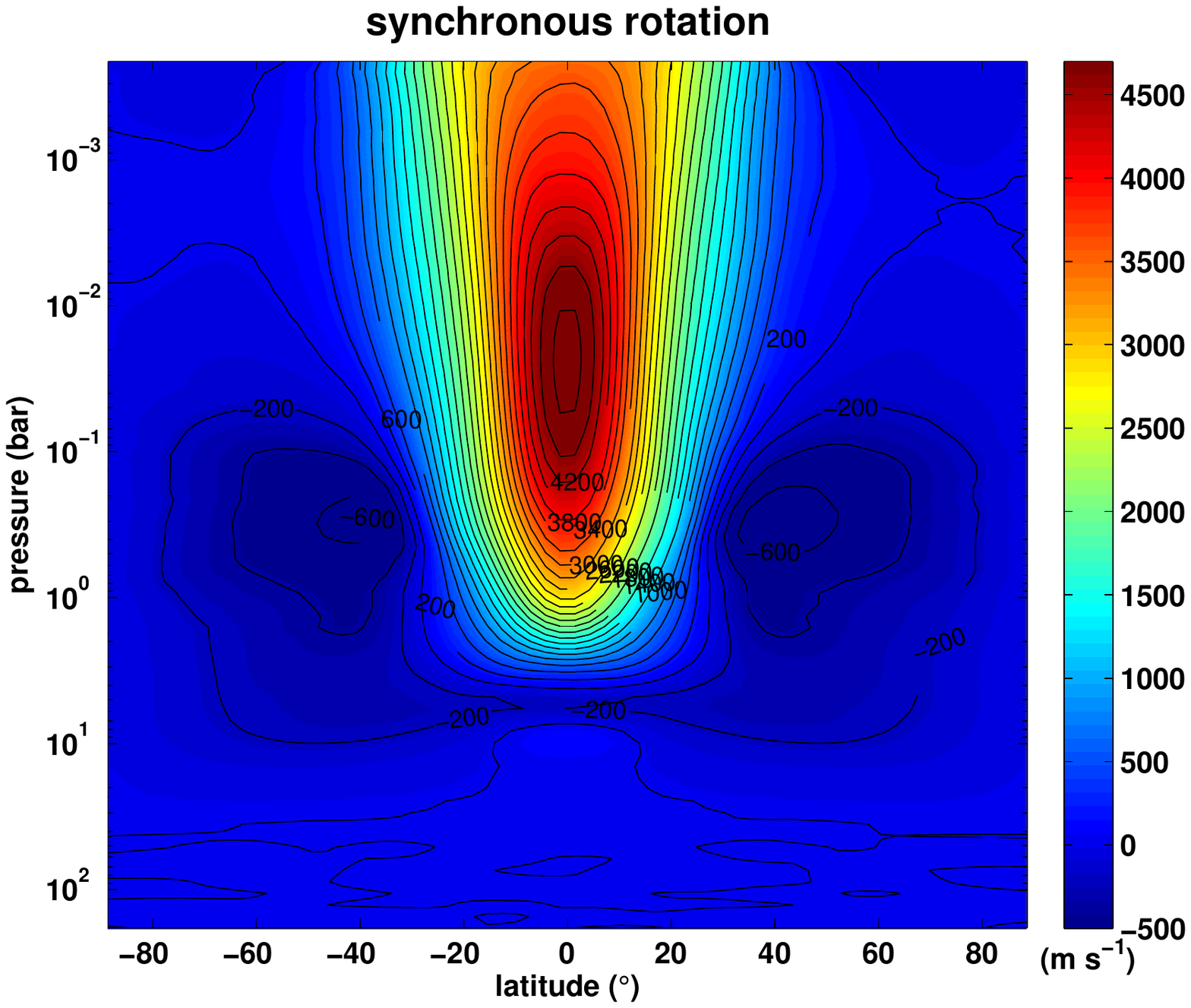} \\
\includegraphics[trim = 0.5in 2.8in 0.9in 2.7in, clip, width=0.43\textwidth]{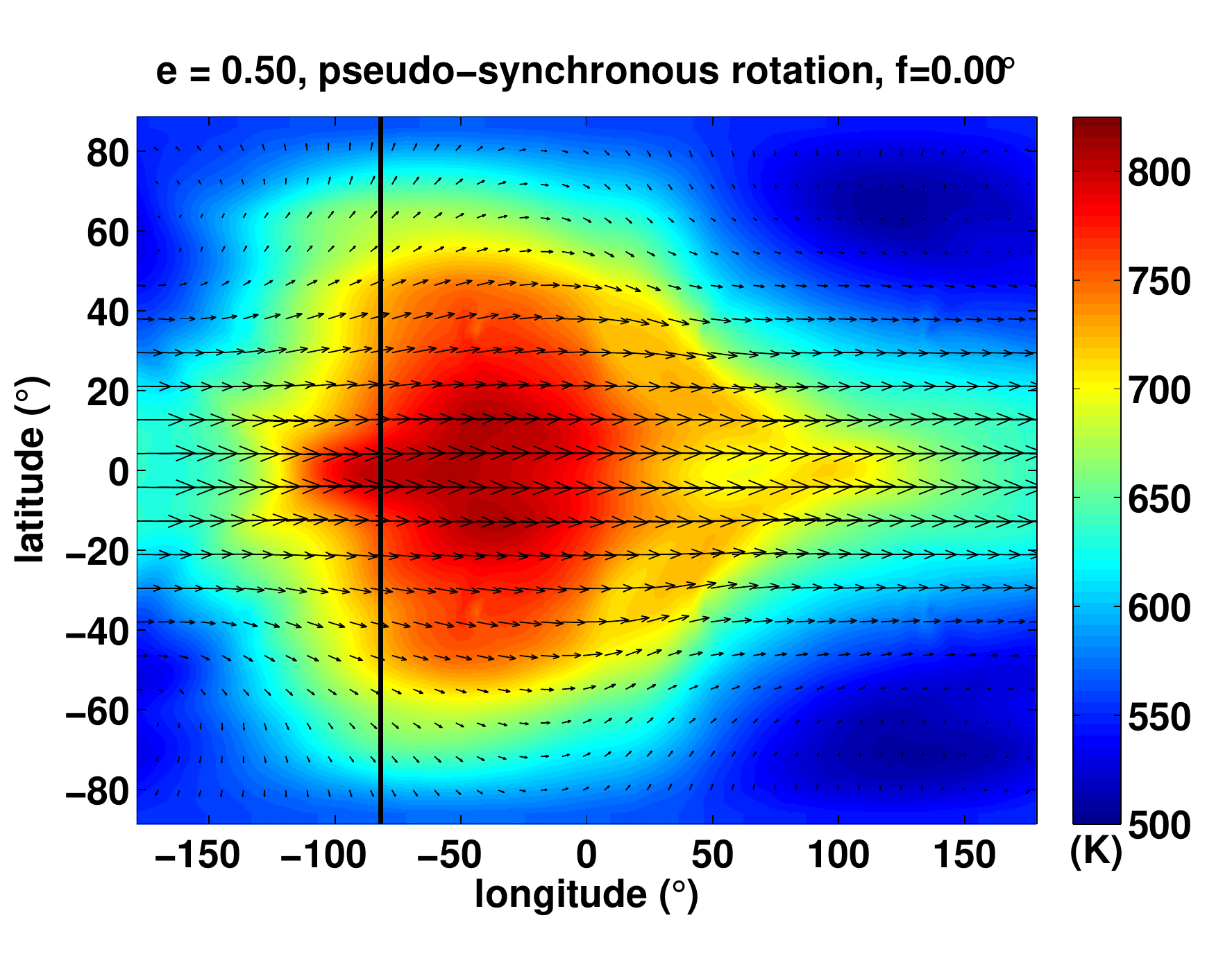}
\includegraphics[trim = 0.5in 2.8in 0.9in 2.7in, clip, width=0.43\textwidth]{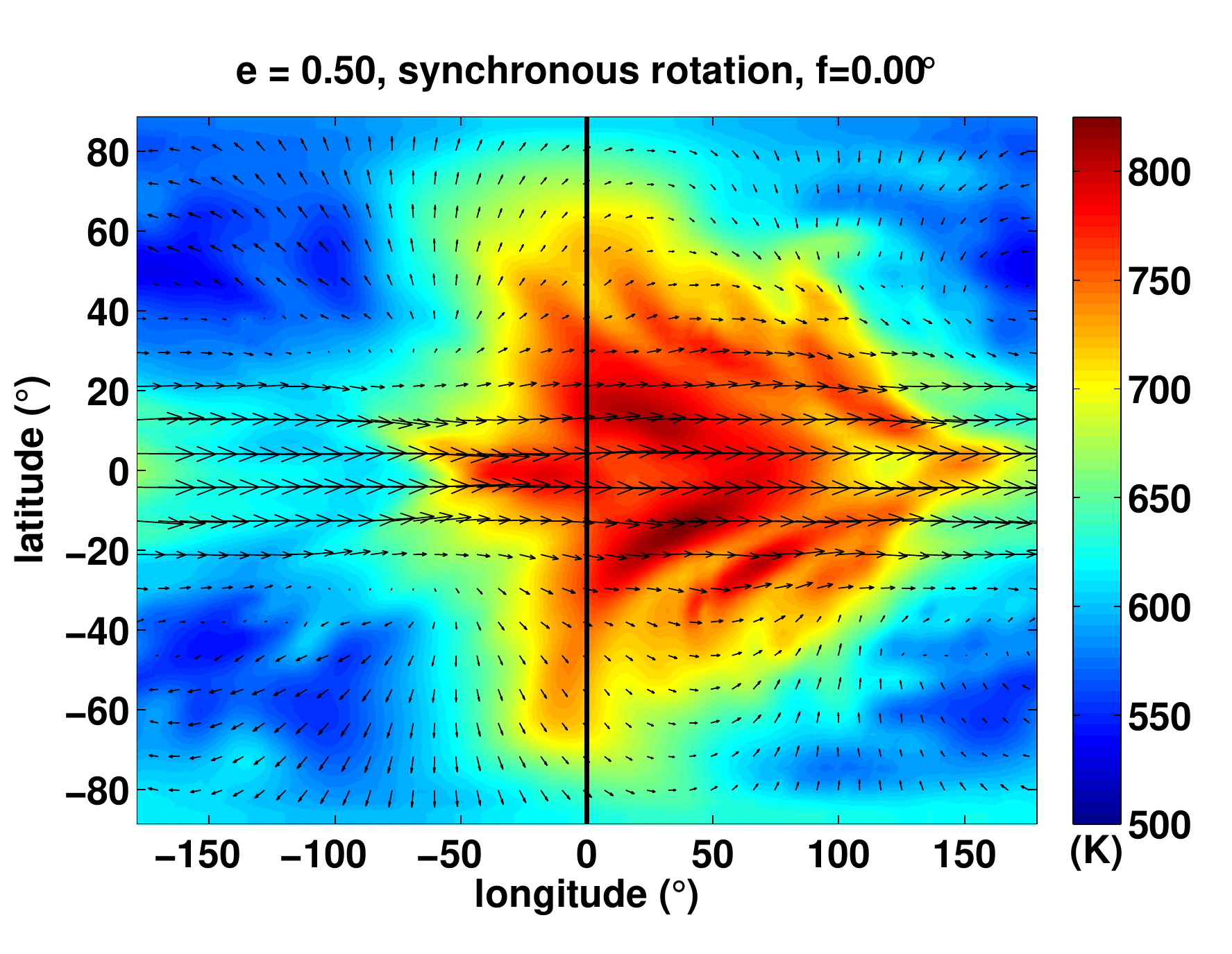} \\
\includegraphics[trim = 0.5in 2.8in 0.9in 2.7in, clip, width=0.43\textwidth]{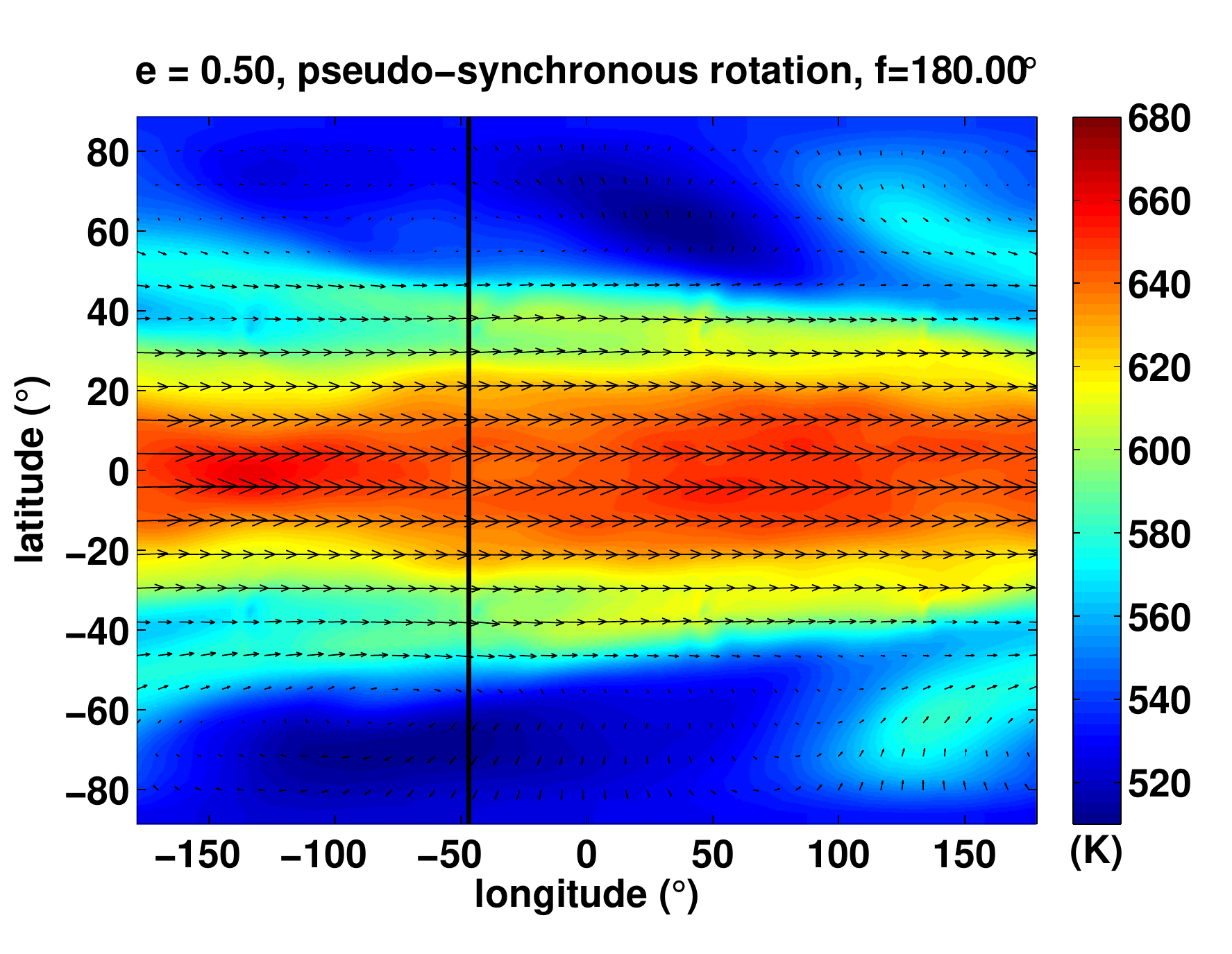}
\includegraphics[trim = 0.5in 2.8in 0.9in 2.7in, clip, width=0.43\textwidth]{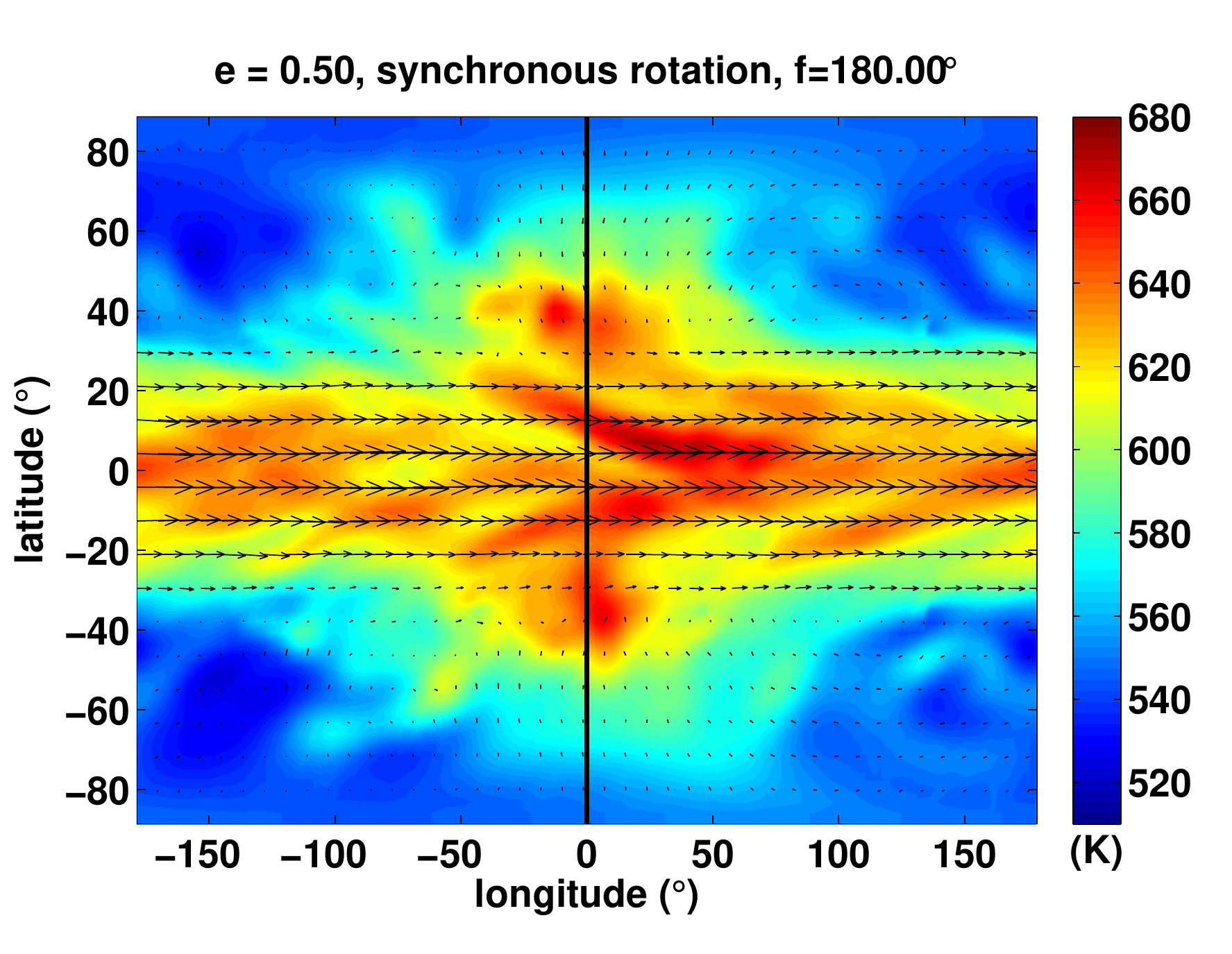} \\
\caption{Comparison of pseudo-synchronous ($\mathrm{T_{rot}}=3.046 \times 10^5$ s) versus synchronous ($\mathrm{T_{rot}}=8.544 \times 10^5$ s) rotation, for a case with $T_{eq}=755$ K, $e = 0.50$, and $a=0.0848$ AU.  Wind and temperature snapshots are taken at periapse (middle row) and apoapse (bottom row) at 5008 Earth days and 5013 Earth days, respectively.  The vertical bar denotes substellar longitude.}
\label{synchcompare1}
\end{figure*}

\subsection{Synchronous vs. pseudo-synchronous rotation}
Here we explore the sensitivity of the circulation to rotation rate by comparing models performed with the pseudo-synchronous rotation rate to otherwise identical models performed with the synchronous rotation rate.  This is illustrated in Figure~\ref{synchcompare1}.  Here, the rotation rate varies by almost a factor of 3.  Both cases still have an equatorial superrotating jet, and the overall atmospheric structure is similar.  However, there are subtle differences: in the synchronous case, the equatorial eastward winds and high-latitude westward winds are higher in magnitude (right column).  The synchronous case also exhibits more latitudinal and longitudinal temperature variation, particularly on the dayside; this is more apparent at apoapse (bottom row).  In the pseudo-synchronous case, the faster rotation rate serves to homogenize the temperature in longitude because the time for rotation to carry air parcels from day to night is shorter.  The temperature difference from equator-to-pole is also larger because the faster rotation rate inhibits equator-to-pole heat transport.  Thus, while it is unlikely that an eccentric planet synchronously rotates its star, the distinction can lead to notable differences in the mean flow.

These results can also be used to compare the \cite{hut1981} formulation, shown by the pseudo-synchronous case, to formulations by other groups.  As shown in Figure~\ref{synchcompare1}, if another prescription for rotation rate differed from \cite{hut1981} by a factor of 2-3 (as illustrated by the synchronous case), the qualitative picture would not look that different.  However, if the prescription differed by a factor of more than 3-4, the picture could substantially differ.  For example, the rotation rate calculated from \cite{ip2004}, Eq. 63, yields a value of $1.4 \times 10^6$ seconds, a factor of four larger than the \cite{hut1981} calculation.  So, results using this prescription would look very different.

Still, the rotation rate of eccentric exoplanets is a completely unconstrained problem. One must be careful, then, in choosing a formulation, noting its assumptions and limitations.

\begin{figure*}
\centering
\includegraphics[trim = 1.4in 0.2in 0.7in 0.5in, clip, width=0.7\textwidth]{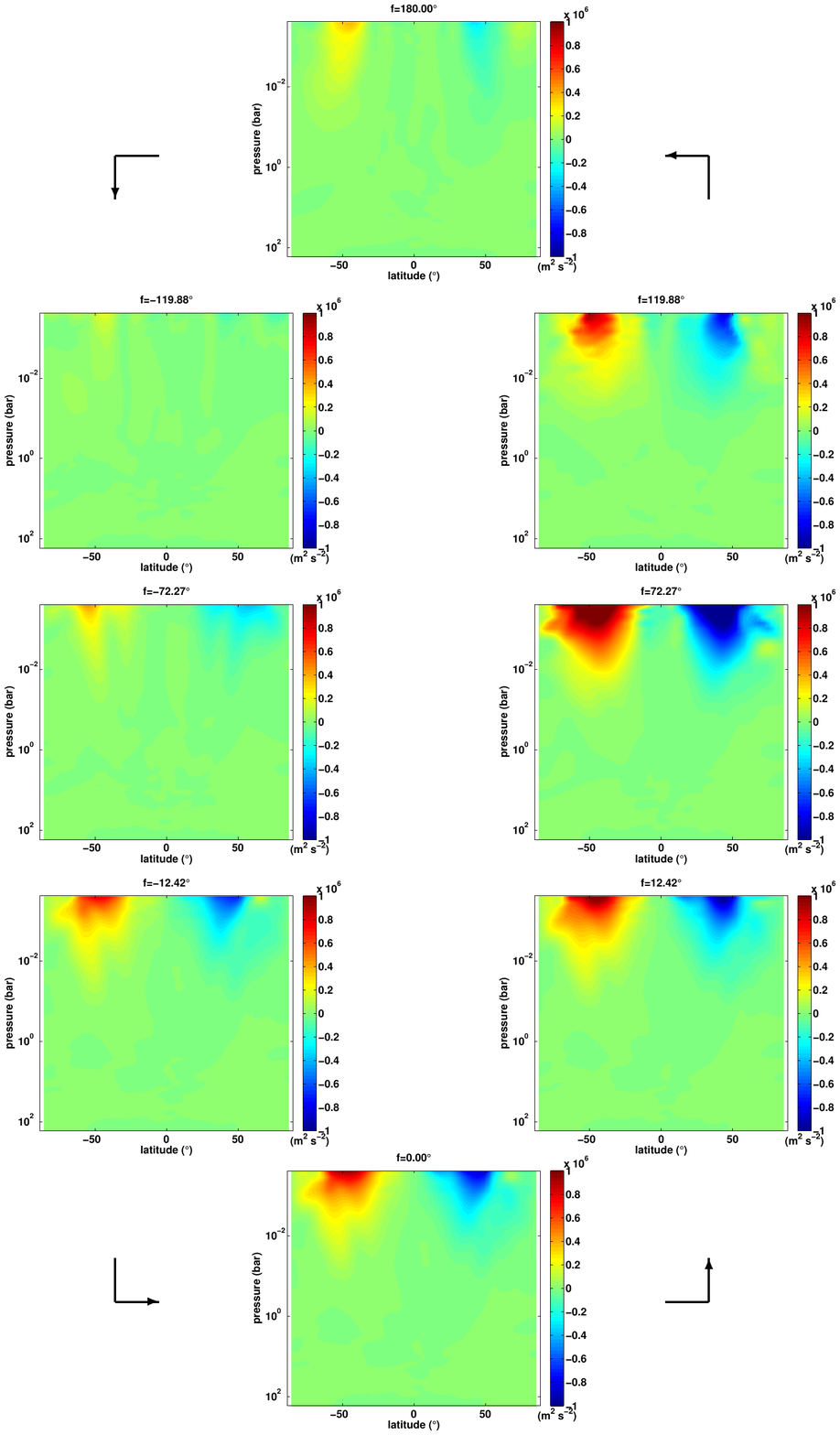}
\caption{Eddy momentum flux, $\overline{u'v'\ cos\phi}$, throughout the orbit of a model where $T_{eq}=951$ K, $e = 0.5$, and $a = 0.0534$ AU. Positive values indicate a northward flux of eastward eddy angular momentum, while negative values indicate a southward flux of eastward eddy angular momentum.  The maximum momentum flux occurs hours after periapse passage ($f=72^{\circ}$).  The snapshots correspond to model outputs from 6103-6108 Earth days.}
\label{momflux}
\end{figure*}

\subsection{Momentum budget and Equatorial superrotation}
In most of our simulations, the planet maintains an equatorial eastward jet throughout its orbit.  Superrotation is common in three-dimensional circulation models of hot Jupiters (e.g. Showman \& Guillot 2002, Showman et al. 2008, 2009, Menou and Rauscher 2009, Rauscher and Menou 2010, 2012, Cooper and Showman 2005,2006, Dobbs-Dixon \& Lin 2008, Dobbs-Dixon et al. 2010, Heng et al. 2011), and is also seen within our own solar system on Venus, Titan, Jupiter and Saturn. While superrotation on solar system planets has been extensively studied, only recently have the mechanisms for superrotation on synchronously rotating exoplanets been identified (Showman and Polvani 2011, hereafter SP2011).  As discussed in their paper, the generation of superrotation requires up-gradient momentum transport that pumps angular momentum from outside the jet to within it.  Only transport by waves and eddies can satisfy this requirement (Hide 1969).

SP2011 demonstrate that for synchronous hot Jupiters on near-circular orbits such as HD 189733b and HD 209458b, equatorial superrotation is generated by standing, planetary scale Kelvin and Rossby waves at the equator and midlatitudes, respectively, that are themselves the dynamical response to the planet's large day-night heating contrast.  Because the Kelvin waves propagate eastward and the Rossby waves propogate westward, a phase tilt of the wind vectors emerges from northwest to southeast in the northern hemisphere, and southwest to northeast in the southern hemisphere, with an overall shape resembling a chevron pointing eastward.  This pattern causes meridional eddy angular momentum fluxes that converge angular momentum onto the equator, generating equatorial superrotation.

Despite the fact that not all our model integrations are synchronous, our results suggest that similar mechanisms are still at play.  The chevron shape that SP2011 describe is exactly what we noted earlier in our plots of the wind and temperature profiles from Figure~\ref{snapshots}, suggesting that, as in SP2011, our models exhibit an equatorward flux of eddy angular momentum.  This is demonstrated quantitatively in Figure~\ref{momflux}, which shows the zonally averaged meridional flux of relative zonal eddy angular momentum, $\overline{u'v'\ cos\phi}$, at snapshots throughout the orbit (as in Figures~\ref{snapshots} and~\ref{snapshots_uz}).  Here, $u'$ and $v'$ represent the devation of the zonal and meridional winds, respectively, from their zonal averages.  Positive values indicate a northward flux of eastward eddy angular momentum, and negative values indicate a southward flux of eastward eddy angular momentum.  At apoapse, the small day-night forcing causes poleward transport of eddy angular momentum at latitudes greater than $50^{\circ}$ but equatorward transport of eddy angular momentum equator of $50^{\circ}$ latitude; the amplitudes are weak due to the low insolation at apoapse.   As the planet moves closer to the star (middle row, left column), the increase in day-night forcing strengthens the equatorward flux of eastward eddy angular momentum at latitudes between $0^{\circ}$ and $50^{\circ}$ in both hemispheres.  The momentum flux grows and persists until hours after periapse.  Angular momentum thus converges onto the equator, and maintains the equatorial jet against westward accelerations caused by advection, friction (if any), and Coriolis forces.

These mechanisms also explain the difference in wind speeds between the synchronous and pseudo-synchronous cases in Figure~\ref{synchcompare1}.  The synchronous case has a higher day-night heating contrast, hence a higher equatorward flux of eddy momentum, leading to a stronger, narrower equatorial jet.  Moreover, the equatorial jet in the pseudo-synchronous case extends to deeper pressures than that of the synchronous case; this is most likely due to a change in momentum budget.

\section{Observational Implications}
Observations of transiting exoplanets in circular orbits are known to be shaped by the atmospheric circulation.  As was first predicted for hot Jupiters by Showman \& Guillot (2002) and later observationally confirmed for hot Jupiter HD 189733b by \cite{knutson2007}, an equatorial superrotating jet causes an eastward displacement of the hot spot from the substellar point under appropriate conditions.  Hence, a lightcurve shows a peak in infrared flux minutes to hours before secondary eclipse.  We expect such a situation for eccentric exoplanets as well.  However, observations of eccentric planets can be complicated because the flux received by the planet varies throughout its orbit.  Previous studies explored such situations, but have not used three-dimensional models  (e.g. Langton and Laughlin 2008a and b, Cowan and Agol 2011).  We can expand on these studies and the three-dimensional study by Lewis et al. (2010) by using our SPARC model results.

\begin{figure}
\centering
%\epsscale{1.0}
%\plotone{fig13_viewing_geometries.pdf}
\includegraphics[trim = 0.8in 1.3in 0.3in 1.1in, clip, width=0.50\textwidth]{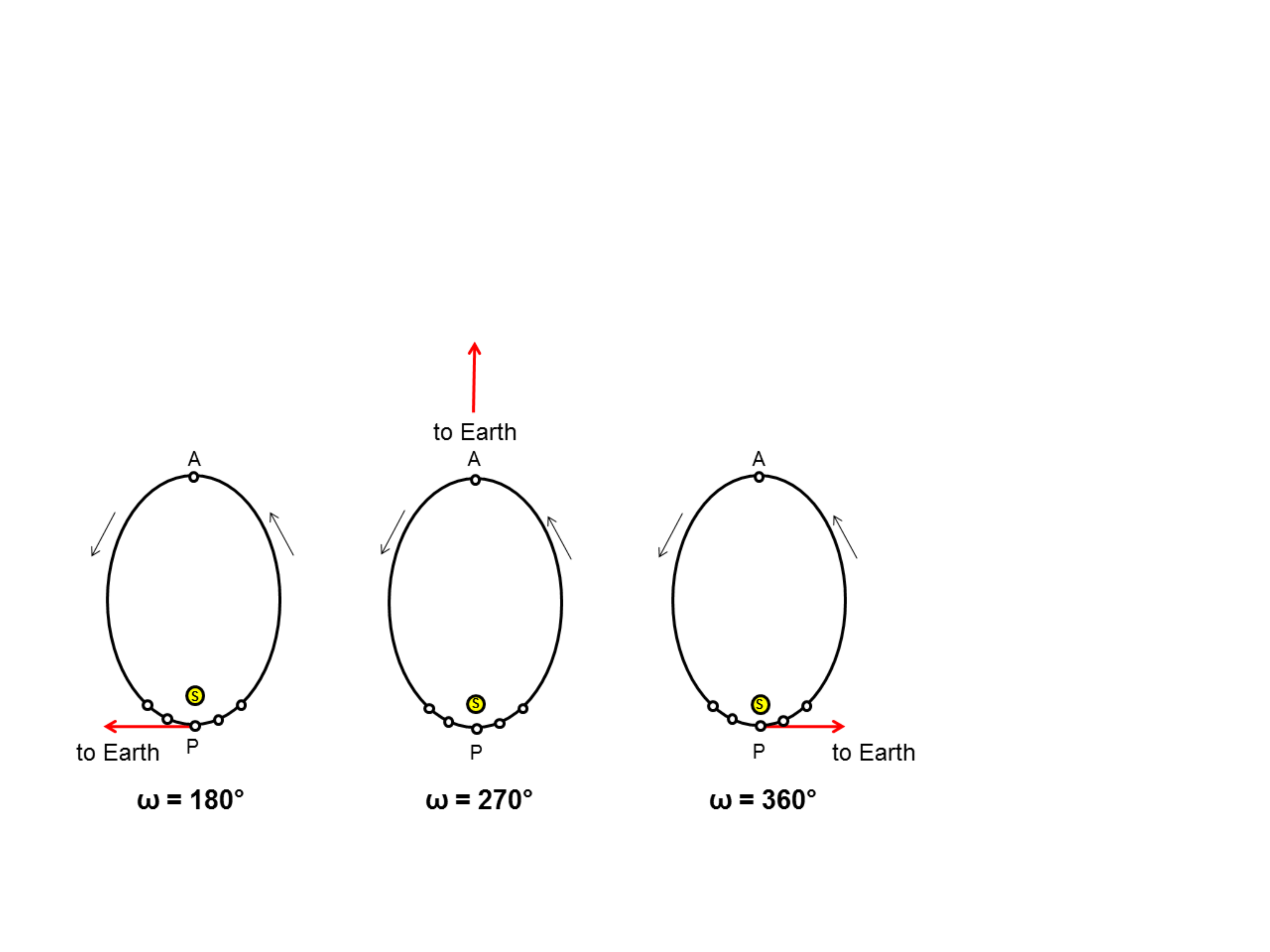}
\caption{Differing orbital viewing geometries explored in this study: $\omega=180^{\circ}$, $\omega=270^{\circ}$, $\omega=360^{\circ}$. }
\label{orbviews}
\end{figure}

\begin{figure}
%\epsscale{0.50}
\centering
\includegraphics[trim = 0.1in 0.1in 1.1in 6.1in, clip, width=0.50\textwidth]{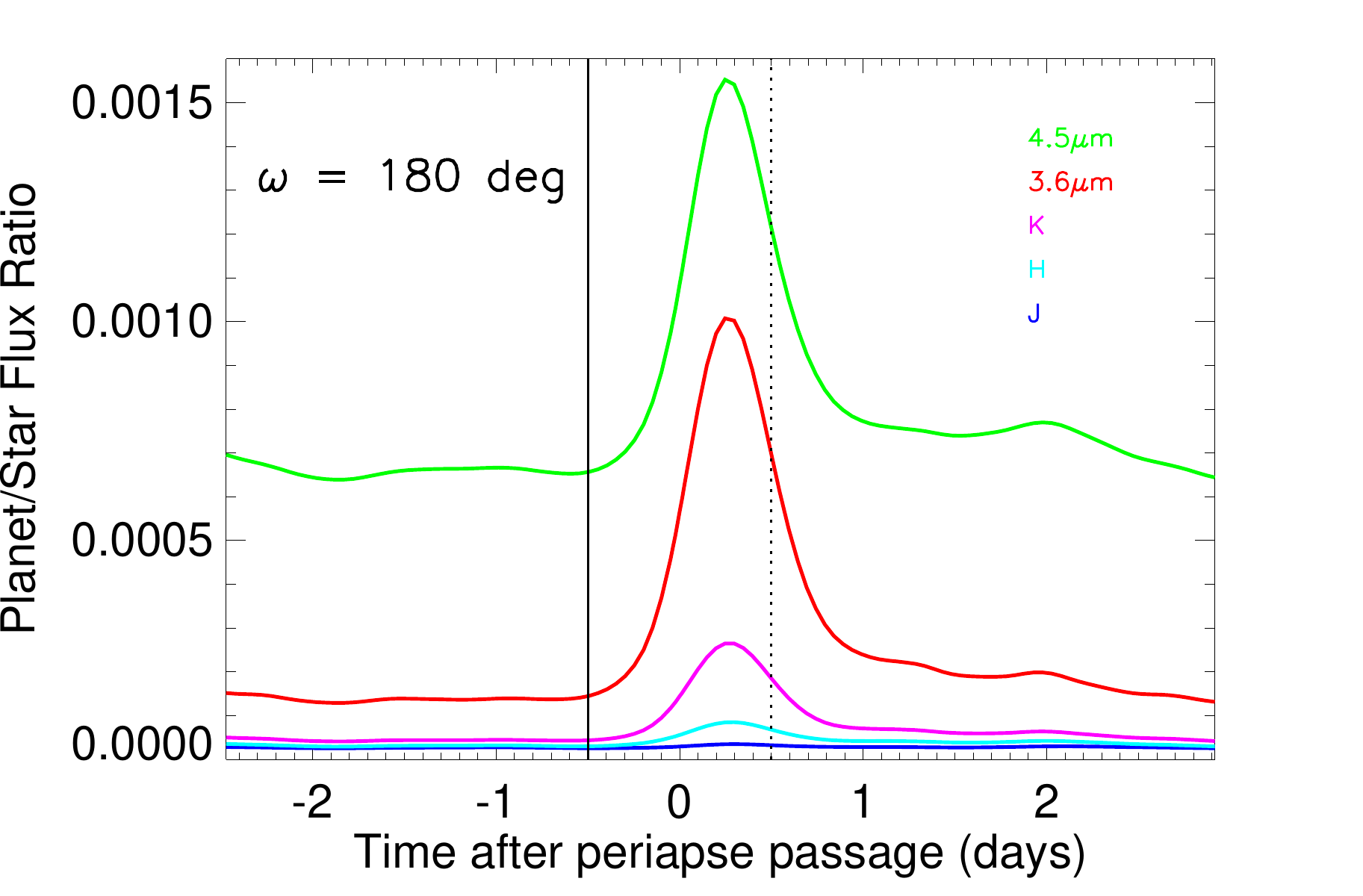}\\
\includegraphics[trim = 0.1in 0.1in 1.1in 6.1in, clip, width=0.50\textwidth]{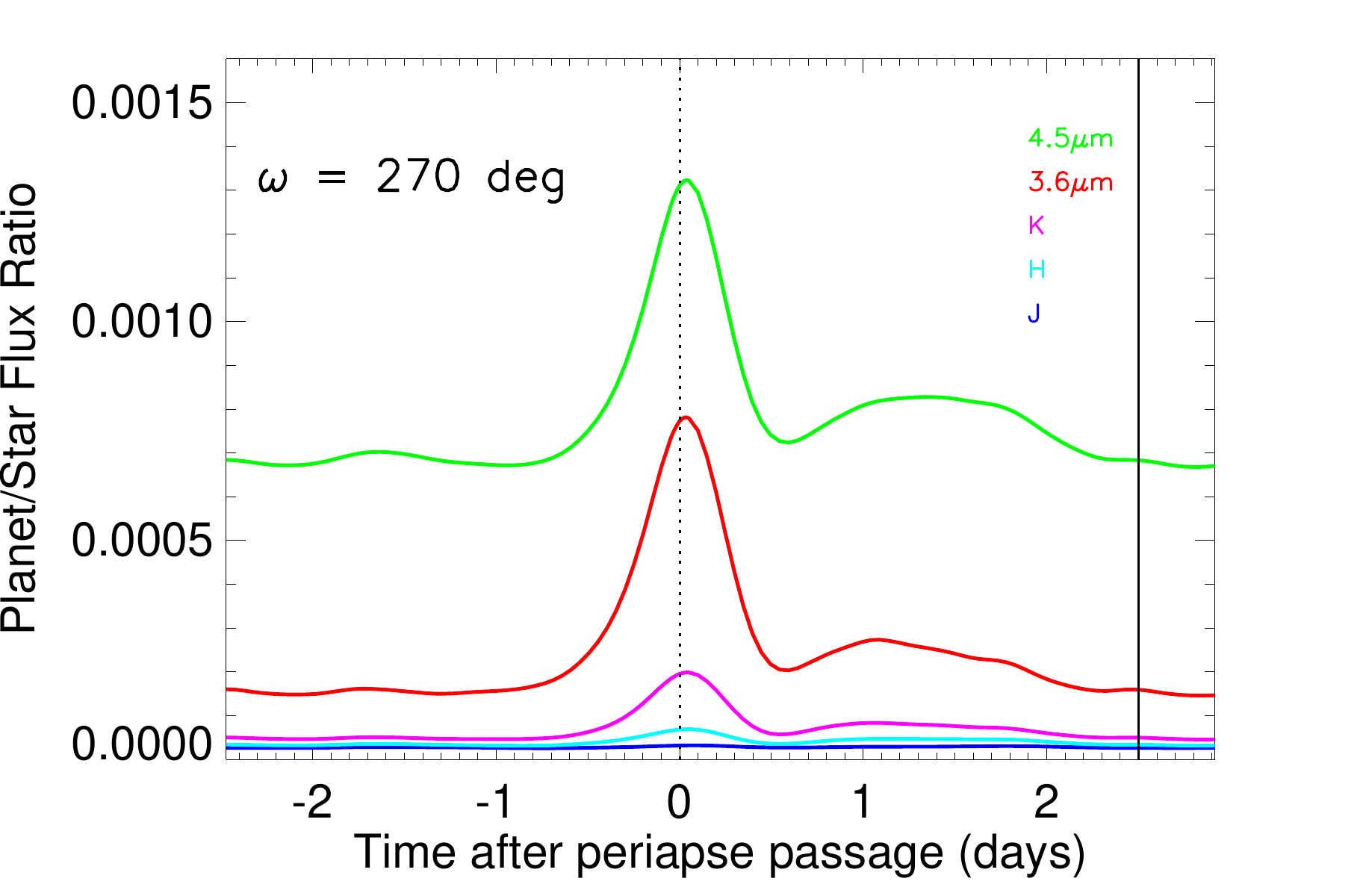}\\
\includegraphics[trim = 0.1in 0.1in 1.1in 6.1in, clip, width=0.50\textwidth]{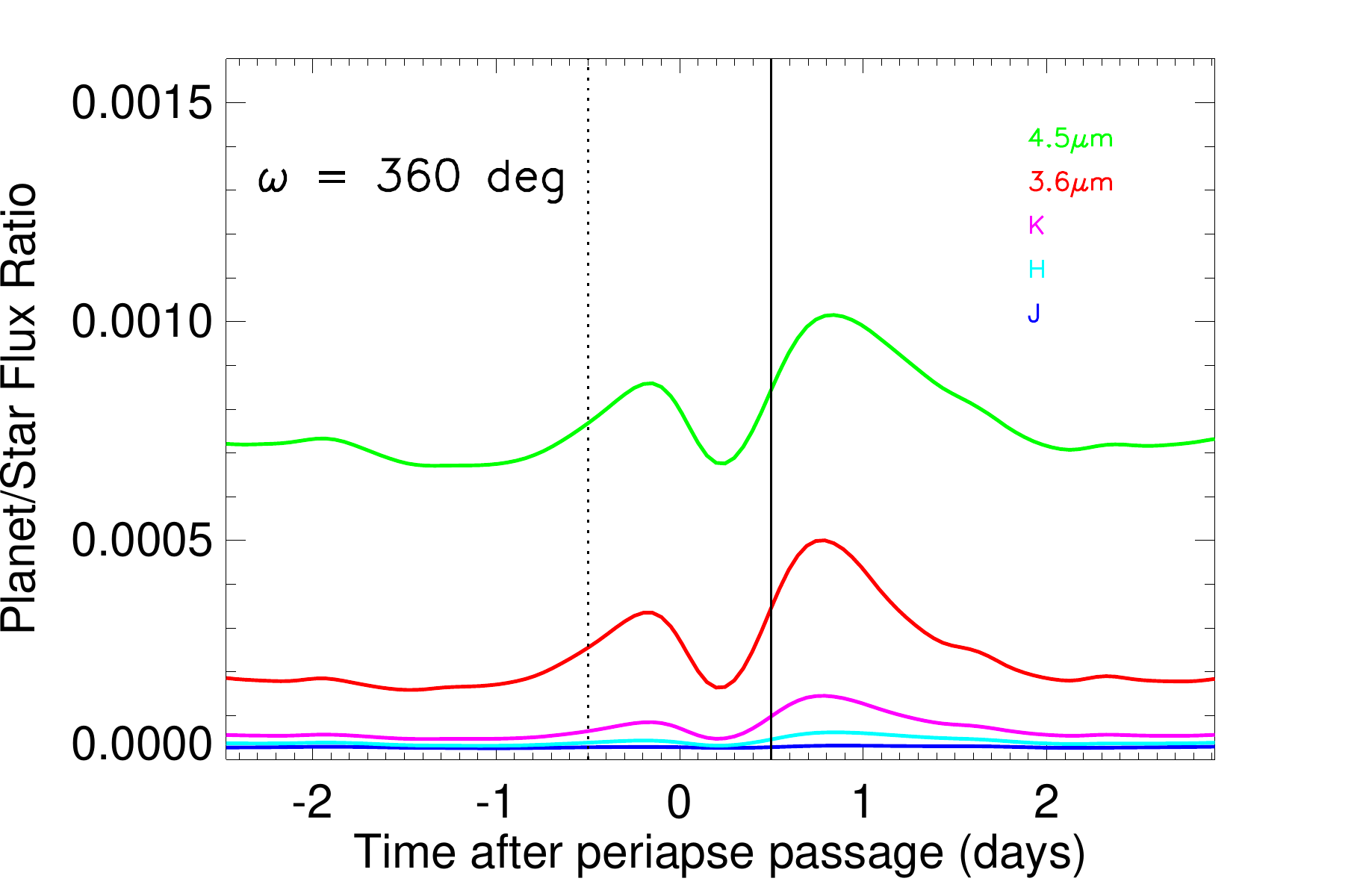}
%\plotone{fig14_lc_007_omega0.pdf}
\caption{Three lightcurves of the IR flux (expressed as a planet/star flux ratio) versus time throughout the full orbit for a single model integration where $T_{eq}=951$ K, $e = 0.50$, and $a = 0.0534$ AU.  The flux is plotted in warm Spitzer bandpasses, 3.6 and 4.5 $\mu m$, and JHK bands (1.26, 1.65, 2.20 $\mu m$, respectively).  These three cases only differ by their value of $\omega$, shown in Figure~\ref{orbviews}.  The solid lines denote transit, while the dotted lines denote secondary eclipse.  }
\label{lightcurve0.50}
\end{figure}

We choose to generate lightcurves following the procedures described in Fortney et al. (2006) for each model integration at three values of the argument of periastron ($\omega$), the angle between the radius vector to the ascending node and the periapse of the orbit.  In particular, we choose orientations where transit occurs before periapse (at $f=-90^{\circ}; \omega=180^{\circ}$), transit occurs at apoapse ($\omega=270^{\circ}$), and transit occurs after periapse (at $f=+90^{\circ}; \omega=360^{\circ}$)(Figure~\ref{orbviews}).  These values reflect the broad range of viewing geometries of transiting exoplanets: for example, HAT-P-2b ($e=0.52$), HAT-P-17b ($e=0.35$) and HD 97658b ($e=0.13$) have $\omega$ near $180^{\circ}$; WASP-8b ($e=0.31$), HAT-P-31b ($e=0.25$) and HD 80606b ($e=0.93$) have $\omega$ near $270^{\circ}$; and GJ 436b ($e=0.15$), XO-3b ($e=0.26$), and HAT-P-11b ($e=0.20$) have $\omega$ near $360^{\circ}$ (see {\it exoplanet.eu} for a full list of values).  Note that when transit occurs, the nightside is visible to Earth, while the dayside is visible at the time of secondary eclipse.

\begin{figure*}
\epsscale{0.80}
\centering
\includegraphics[trim = 1.8in 1.0in 0.9in 0.1in, clip, width=0.7\textwidth]{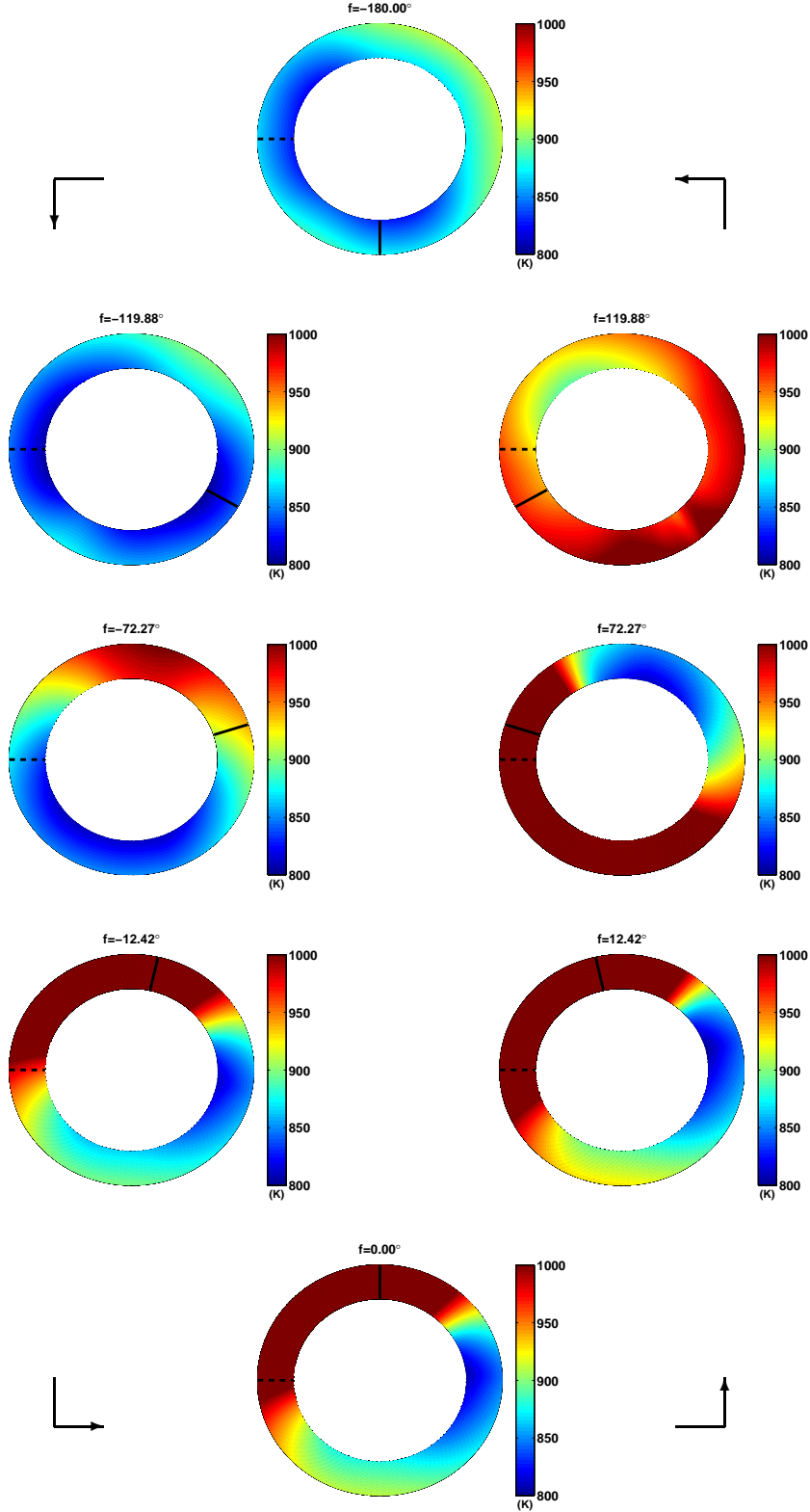}
\caption{Plots of temperature versus longitude and pressure in equatorial slices throughout one full orbit, for our nominal model with $T_{eq}=951$ K, $e=0.50$, and $a=0.0534$ AU. Each annulus shows temperature versus longitude and log-pressure for a snapshot at a particular orbital position (denoted by the value of $f$) as viewed from looking down from over the planet's north pole.  Pressures between 30 and 43 mbar are displayed in each annulus. The different annuli mark the time evolution throughout one complete orbit, starting from apoapse for the top annulus, proceeding to peripase, and then back to apoapse (follow arrows moving counterclockwise around the plot).  These snapshots correspond to model outputs from 6103-6108 Earth days.  The solid line in each snapshot indicates the longitude of the substellar point while the dashed line denotes the Earth-facing longitude for the particular case of $\omega=180^{\circ}$. For the cases of $\omega=270$ and 360 degrees, exactly the same plot is valid, except that the sub-Earth longitude would be aiming upward for $\omega=270^{\circ}$ and right for $\omega=360^{\circ}$.}
\label{cylinders}
\end{figure*}

\subsection{Observational effects of varying $\omega$}
Using the nominal case ($e=0.50, a=0.0534$ AU, $T_{eq}=951$ K), we illustrate the dependence of full-orbit lightcurves
on the value of $\omega$.  Figure~\ref{lightcurve0.50} shows three lightcurves of the \emph{same} model integration, but at the differing values of $\omega$ described in the previous section.  Hence, the differences in lightcurves results \emph{solely} from the differing viewing geometries.  The planet/star flux ratio is plotted as a function of Earth days from periapse in the warm Spitzer wavelength bands (3.6 and 4.5 $\mu$m), and JHK bands (1.26, 1.65, and 2.20 $\mu m$).  The solid and dashed lines correspond to transit and secondary eclipse, respectively.  For the lightcurve where $\omega=180^{\circ}$ (top panel), the peak in infrared (IR) flux occurs approximately 6 hours after periapse and before secondary eclipse.  The timing of the peak is due to a combination of the radiative lag of the planet and the fact that the hottest regions rotate into an Earth-facing orientation hours after periapse passage.  The dayside, which is most visible near secondary eclipse, is hottest after periapse.  Additionally, temperatures reach peak values at times close to when those hottest regions are facing Earth.  This helps explain the high amplitude of the lightcurve peak in the $\omega=180^{\circ}$ case relative to the other two cases.  These effects can be visualized by plotting the temperature near the photosphere along the equator in a polar projection (Figure~\ref{cylinders}).  Snapshots throughout the orbit plot the temperature in a polar projection as a function of pressure and orbital position at a latitudinal slice near the equator; the solid and dashed lines denote the substellar longitude and earth-facing longitude, respectively.  Based on the temperatures at the earth-facing longitude, it is apparent that the peak in IR flux should occur after periapse and before secondary eclipse (see $f=12^{\circ}$ and $72^{\circ}$); this is indeed the case.

For the same model but instead with $\omega=270^{\circ}$, the peak in IR flux occurs 1 hour \emph{after} eclipse and periapse.  In this case, the dayside is visible near periapse, and therefore we are seeing the planet increase in temperature (and IR flux) during its closest approach to the star.  However, the hottest regions are displaced eastward of the substellar point, and therefore the hottest regions face Earth significantly \emph{before} periapse passage (Figure~\ref{cylinders} at $f\approx72^{\circ}$).  Thus, by the time the peak temperatures are reached near and after periapse, the hottest regions are already rotating out of view from Earth.  This explains the lower amplitude of the lightcurve peak as compared to the peak of the $\omega=180^{\circ}$ case. The IR peak decreases as the planet moves further away from the star along our line of sight.  A second ramp-up in flux occurs approximately 1 day after periapse/eclipse; this is due to the hottest regions rotating back into view as the planet moves toward apoapse.

In the case where $\omega=360^{\circ}$, two distinct peaks are present -- a smaller peak that occurs after secondary eclipse and $\sim$3-4 hours before periapse, and a second higher peak that occurs after transit.  The first peak occurs as some of the dayside is visible as the planet is heated approaching periapse.  The peak decreases as the dayside rotates out of view.  For this geometry, the hottest regions are facing away from Earth when they reach their peak temperatures (Figure~\ref{cylinders}).  The second peak occurs after transit as the hottest regions again rotate into view.  At this point in the orbit, peak temperatures have already occurred, and therefore this peak is lower in amplitude than the other two lightcurves.  However, because the planet is hotter than it was after secondary eclipse, this second peak has a higher amplitude than the first one.

\subsection{High-$e$ model integrations at varying $\langle F \rangle$}
\begin{figure}
%\epsscale{0.80}
\centering
\includegraphics[trim = 0.0in 0.2in 1.1in 5.7in, clip, width=0.5\textwidth]{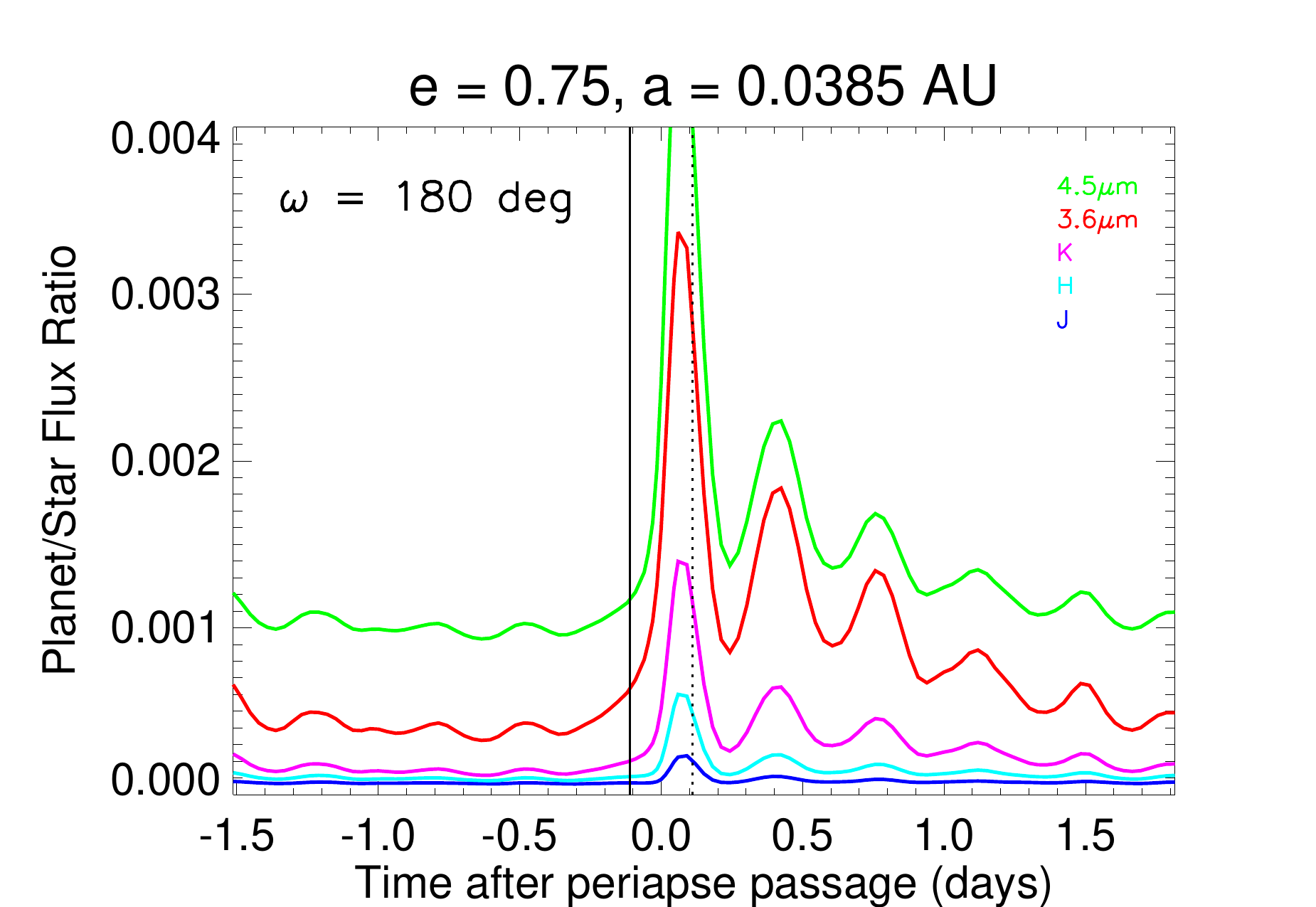}\\
\includegraphics[trim = 0.0in 0.2in 1.1in 5.7in, clip, width=0.5\textwidth]{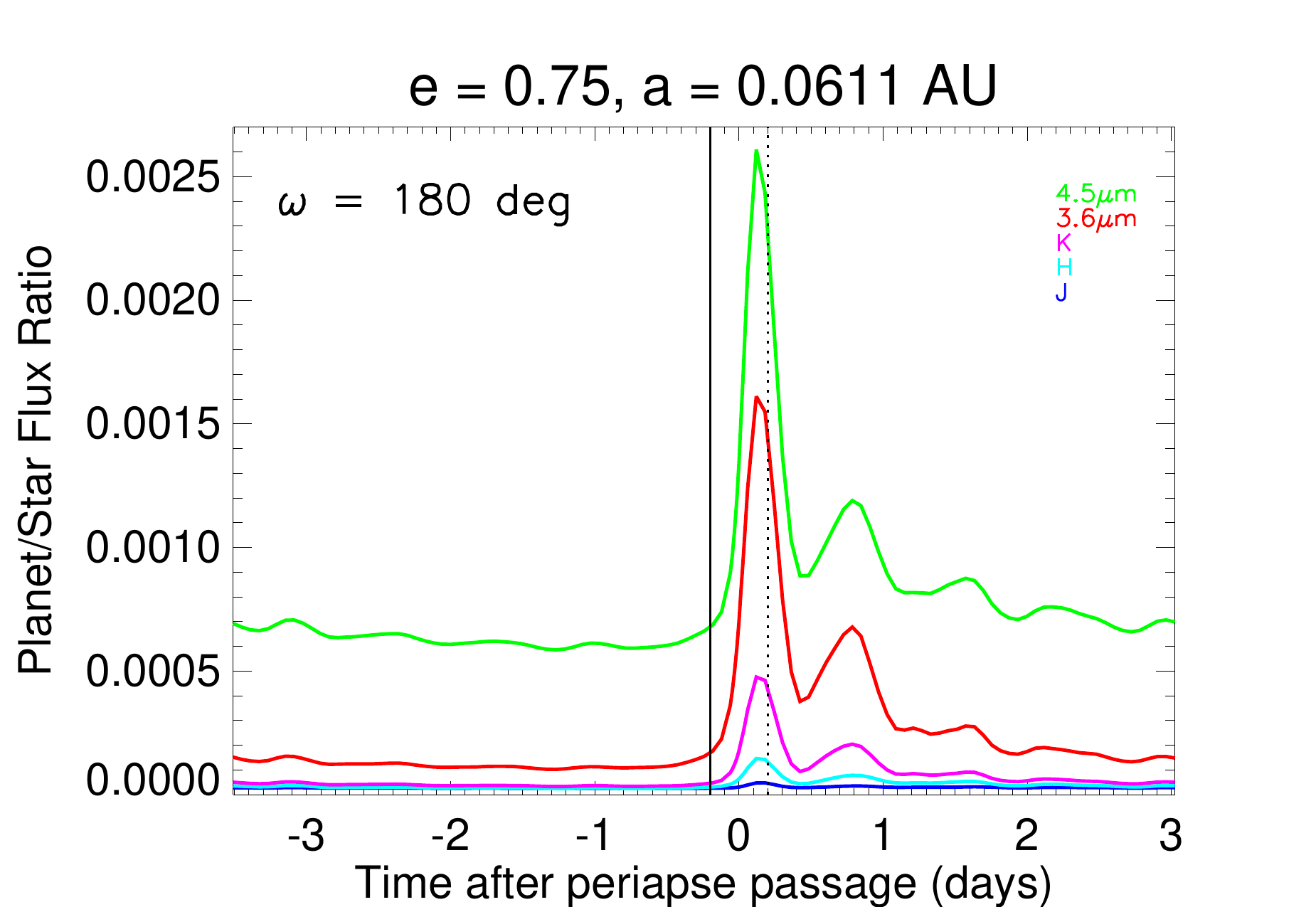}
\caption{Lightcurves for two different $e = 0.75$ model integrations: $T_{eq}=1199$ K, $a=0.0385$ AU (top), and $T_{eq}=951$ K, $a=0.0611$ AU (bottom), both with $\omega=180^{\circ}$.  The solid line denotes transit, while the dotted line denotes secondary eclipse.}
\label{lightcurve004_008}
\end{figure}

If an eccentric planet has a large enough day-night temperature difference and a sufficiently rapid rotation rate, its resultant lightcurve can show a periodic rise and fall in flux throughout its orbit.  The lightcurves shown in Figure~\ref{lightcurve004_008} are two $e=0.75$ model integrations that differ in $\langle F \rangle$; the top panel shows a case where $\langle F \rangle =468183~\mathrm{W~m^{-2}}$ ($T_{eq}=1199$ K) and $a=0.0385$ AU, while the bottom panel is a case where $\langle F \rangle =185691~\mathrm{W~m^{-2}}$ ($T{eq}=951$ K)and $a=0.0611$ AU.   For both cases, lightcurves are shown for the $\omega=180^{\circ}$ geometry (see Figure~\ref{orbviews}).  The differences in $\langle F \rangle$ lead to differences in lightcurve amplitude in the two models; the closer, hotter planet exhibits a higher flux amplitude.  Not surprisingly, the timing of peak IR flux for both models resembles the $\omega=180^{\circ}$ case in Figure~\ref{lightcurve0.50}.  The peak in each case occurs after periapse and before eclipse, due to the combined effects of the thermal lag of the planet and geometry of the system.  Additionally, both lightcurves exhibit a quasi-periodic rise and fall of IR flux days after periapse passage.  This phenomenon, called ``ringing", was also seen in models by Langton \& Laughlin (2008) and Cowan \& Agol (2011), caused by the hottest point of the planet rotating in and out of view from Earth.  For ringing to occur, the temperature difference from dayside to nightside must be large and it must survive for multiple planetary rotation periods.  In the models shown in Figure~\ref{lightcurve004_008}, this is aided by the rapid pseudo-synchronous rotation rate associated with an eccentricity of 0.75 (Table~\ref{simulationtable}).  For the top case, the day-night temperature difference is over $\sim~1300$ K, while the bottom case varies by $\sim750$ K (Figure~\ref{ecompare}).  The lightcurves from Figure~\ref{lightcurve0.50} do not exhibit ringing because the day-night temperature difference is low ($\sim300-400$ K).

Because the radiative and dynamical timescales in the models by Cowan \& Agol (2011) are calculated by a simple scaling with temperature, the ringing seen in their simulated lightcurves occurs with a period equal to the planet's assumed solid-body rotation period in the inertial frame of the winds.  In reality, an eccentric exoplanet approaching apoapse would become increasingly cold, the winds would weaken and decrease, and hence the ringing should be non-periodic.  Indeed, for our three dimensional models, the ringing in each case has varying periods -- the top case has a ringing period ranging from 0.3 to 0.4 days ($\mathrm{T_{rot}}=0.35$ days) while the bottom panel has a period ranging 0.7-0.8 days ($\mathrm{T_{rot}}=0.7$ days).

Despite these differences, this study and the studies by Langton \& Laughlin (2008) and Cowan \& Agol (2011) illustrate the influence of atmospheric dynamics on the observations of eccentric exoplanets.  The lightcurve features described here could be seen in future follow-up observations of eccentric exoplanets, and observers can use the conclusions drawn above to relate the observations to the planet's atmospheric structure.  For example, if ringing is present in an IR lightcurve, the planet must have a strong day-night temperature assymmetry, and hence we would expect the planet to have strong superrotation at the equator (Showman and Polvani 2011).  This is probably the case for highly eccentric ($e>0.5$), short-period exoplanets such as HD 17156b and HD 80606b.  If ringing were observed, it would also allow an estimate of the sum of the rotation speed and the superrotation speed, thereby placing constraints on both wind speeds and rotation rates of eccentric hot Jupiters.  The timing of peak IR flux relative to transit and secondary eclipse, and also relative to orbital position, can help to constrain the circulation further.

\section{Conclusions}
We present three-dimensional circulation models coupled with a two-stream, non-grey radiative transfer scheme for a number of theoretical eccentric hot Jupiters.  We have shown that as in published models with zero eccentricity, our high-eccentricity circulation models are dominated by eastward flow at photospheric levels which cause an eastward displacement of the hottest regions from the substellar point.  The rapid rotation rates associated with pseudo-synchronization at high eccentricity lead to a small Rossby deformation radius and in some cases multiple jets in the atmosphere.  Global-mean temperatures and day-night temperature differences peak not at periapse but several hours afterward due to finite radiative timescales in the planet's atmosphere.

Furthermore, we show that equatorial superrotation is generated and maintained by eddies formed by the strong day-night heating contrast, which induce a flux of momentum from midlatitudes to the equator.  The eddy magnitudes and momentum fluxes peak just after periapse passage leading to variations in the zonal-mean flow throughout the orbit.

Lastly, we have shown that the spatial and temporal variations of the wind and temperature structure, as well as the orbital viewing geometry of the system with respect to Earth, can affect the time and amplitude of peak IR flux seen in full-orbit lightcurves.  Depending on the viewing geometry of the orbit relative to Earth, we find that peaks in IR flux that either lead or lag periapse are possible; in all cases, a combination of temporal effects (temperatures changing over time) and geometric effects (hot spots rotating into or out of view) are important in controlling the timing and amplitude of the flux peaks.  In cases where the day-night temperature contrast is large and the rotational period is short, the lightcurve can also exhibit ``ringing" in flux as the hottest region of the planet rotates in and out of view.  This ringing is non-periodic, due to the variation in stellar heating as a function of distance.

\acknowledgments
This work was supported by NASA Origins and Planetary Atmospheres grants to APS.  T.K. also acknowledges support from the Harriet P. Jenkins Pre-Doctoral Fellowship Program (JPFP).  Resources supporting this work were provided by the NASA High-End Computing (HEC) Program through the NASA Advanced Supercomputing (NAS) Division at Ames Research Center.  The authors thank the anonymous referee for their helpful comments and suggestions.

\appendix
\section{Validation tests for updated radiative transfer scheme}
\begin{figure}
\centering
\includegraphics[trim = 0.2in 0.2in 1.0in 6.2in, clip, width=0.75\textwidth]{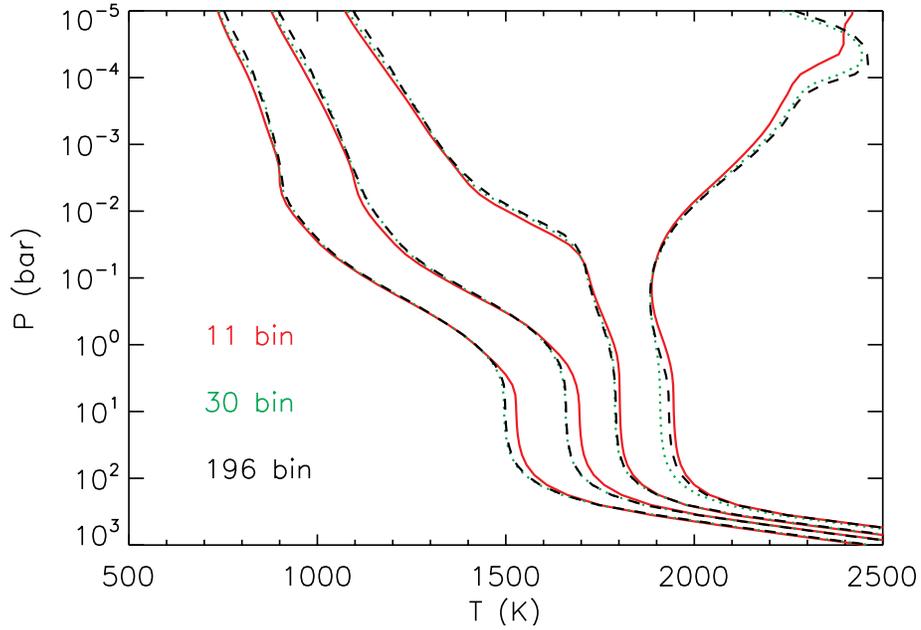}\\
\caption{Globally-averaged pressure temperature profiles from 1-D radiative-convective radiative transfer models of HD 189733b with 11 frequency bins (red profiles), 30 frequency bins (green profiles) and 196 frequency bins (black profiles).  From left to right, the profiles vary in stellar flux from the nominal value, as well as 2$\times$, 4$\times$ and 8$\times$ higher flux.   The 8$\times$ higher flux case has a temperature inversion.}
\label{rtcompare1d}
\end{figure}

\begin{figure}
\centering
%\epsscale{0.90}
\includegraphics[trim = 1.5in 3.0in 1.5in 2.7in, clip, width=0.4\textwidth]{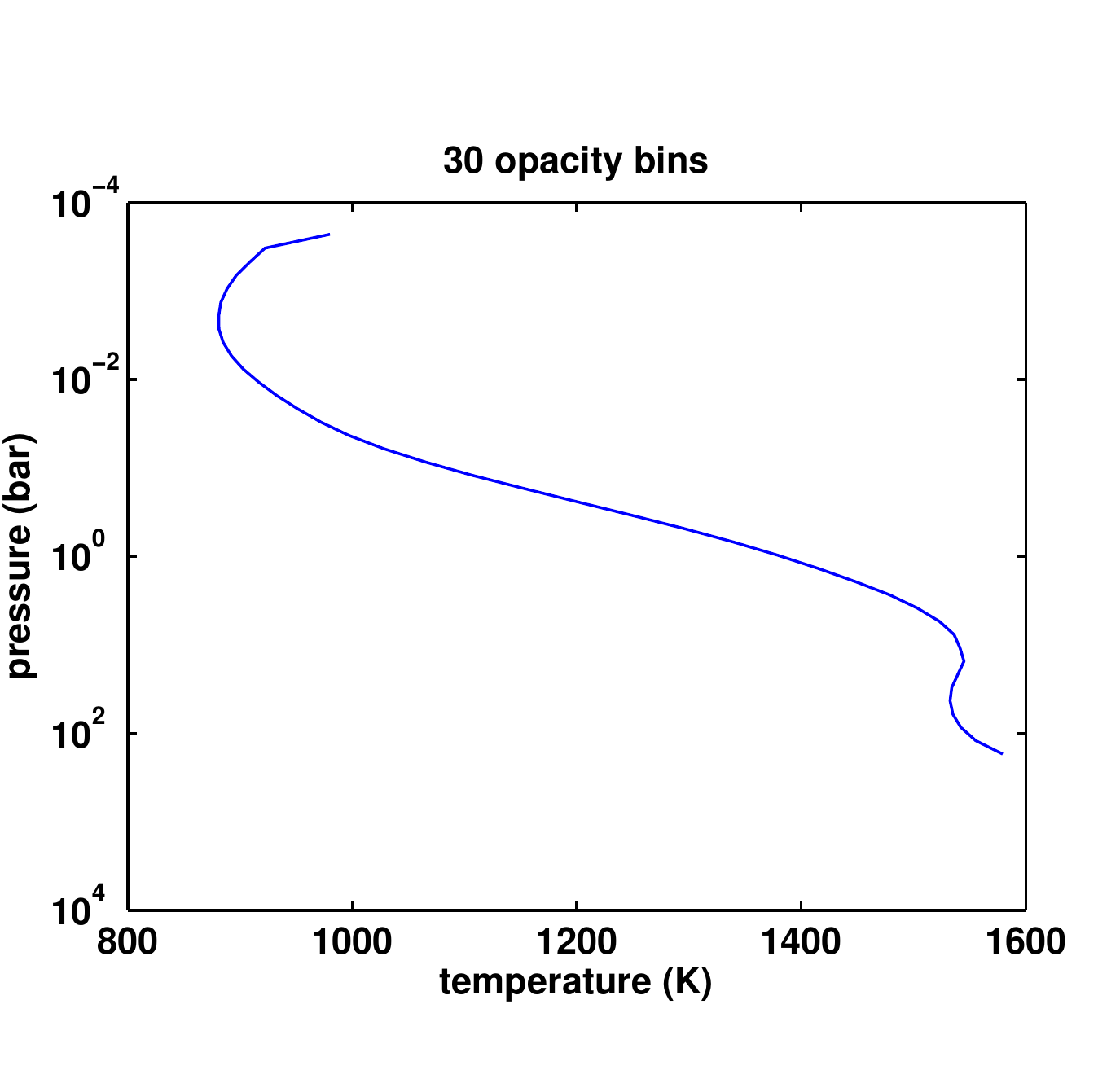}
\includegraphics[trim = 1.5in 3.0in 1.5in 2.7in, clip, width=0.4\textwidth]{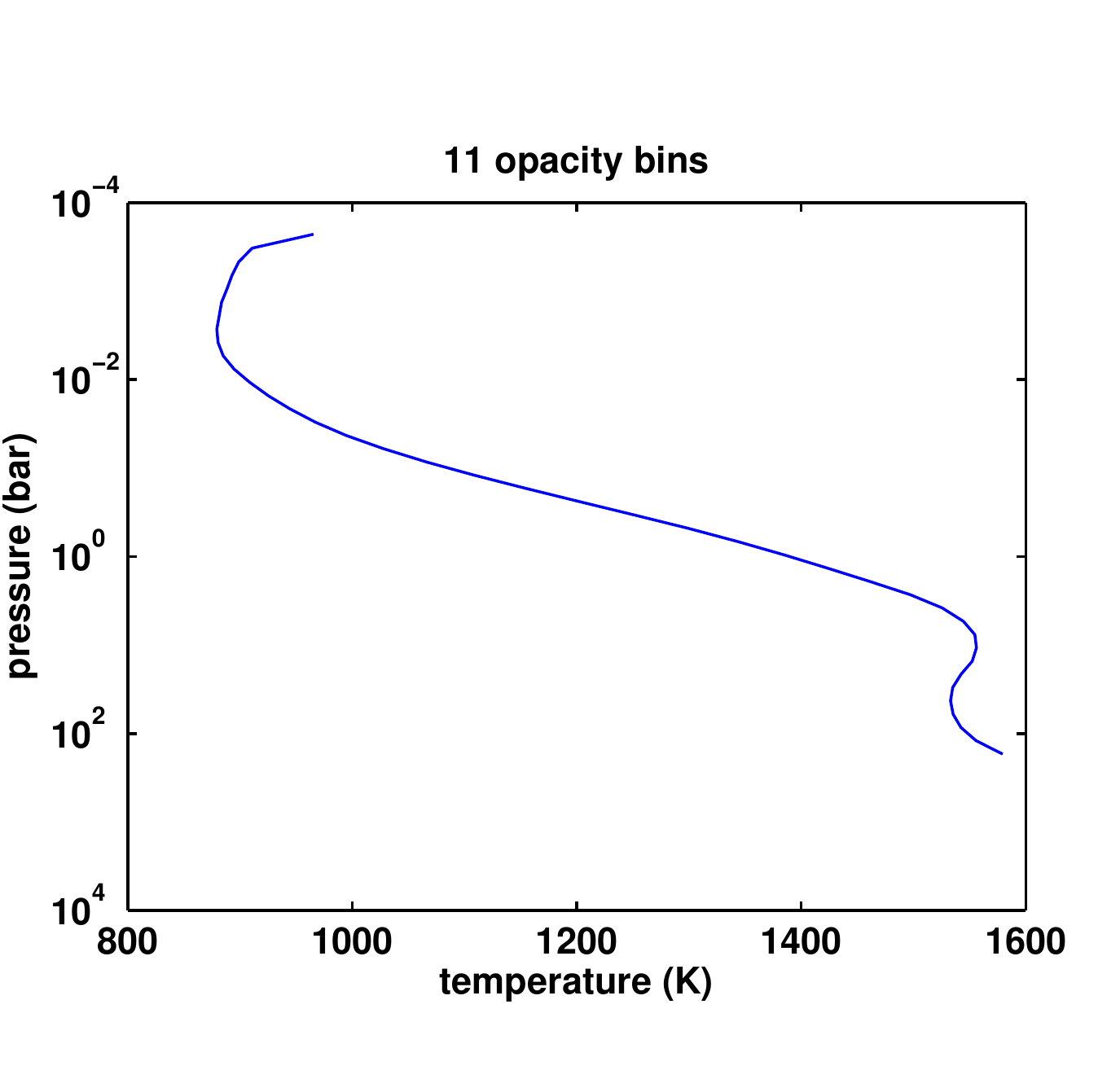} \\
\includegraphics[trim = 0.5in 2.6in 0.8in 2.6in, clip, width=0.42\textwidth]{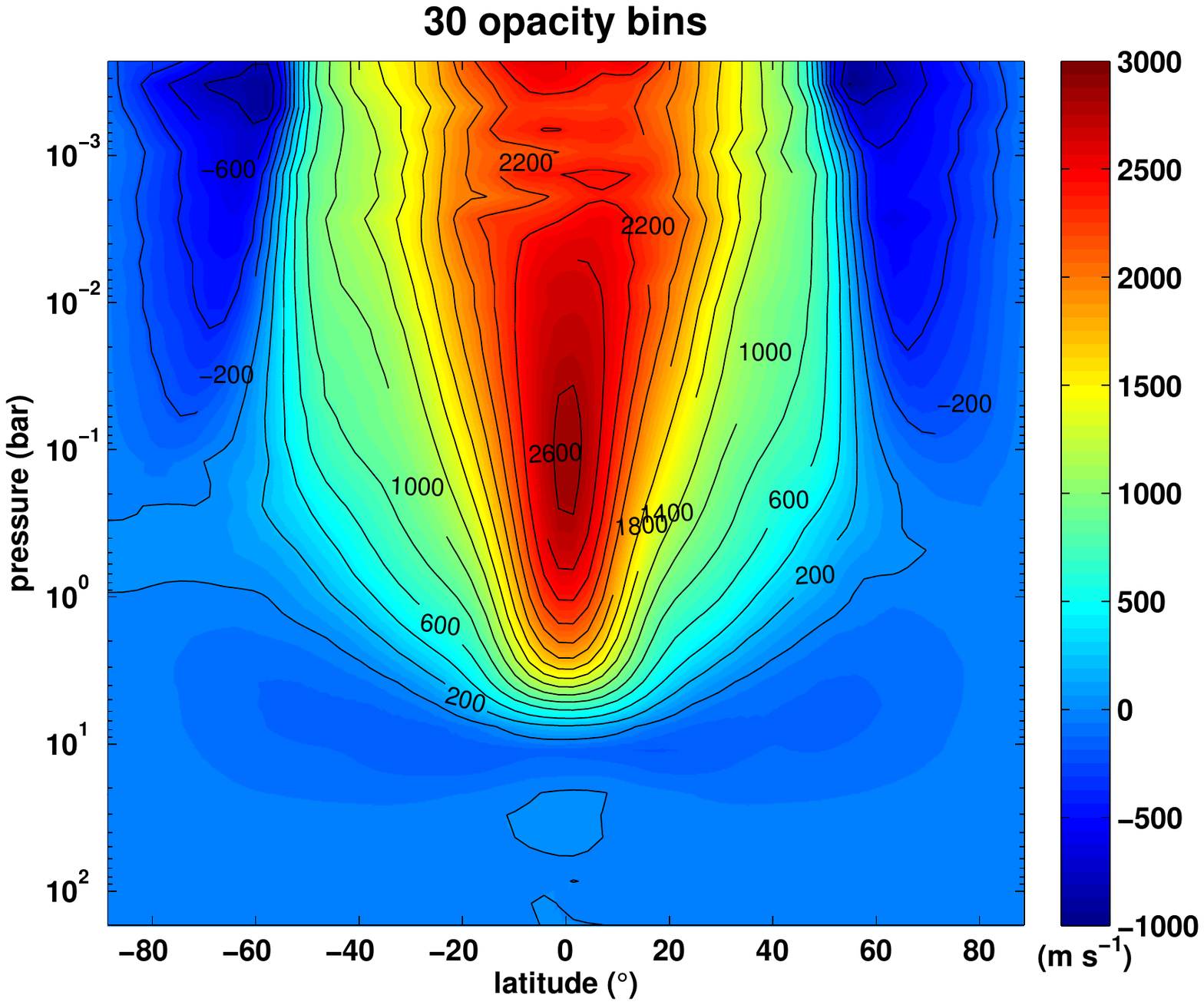}
\includegraphics[trim = 0.5in 2.6in 0.8in 2.6in, clip, width=0.42\textwidth]{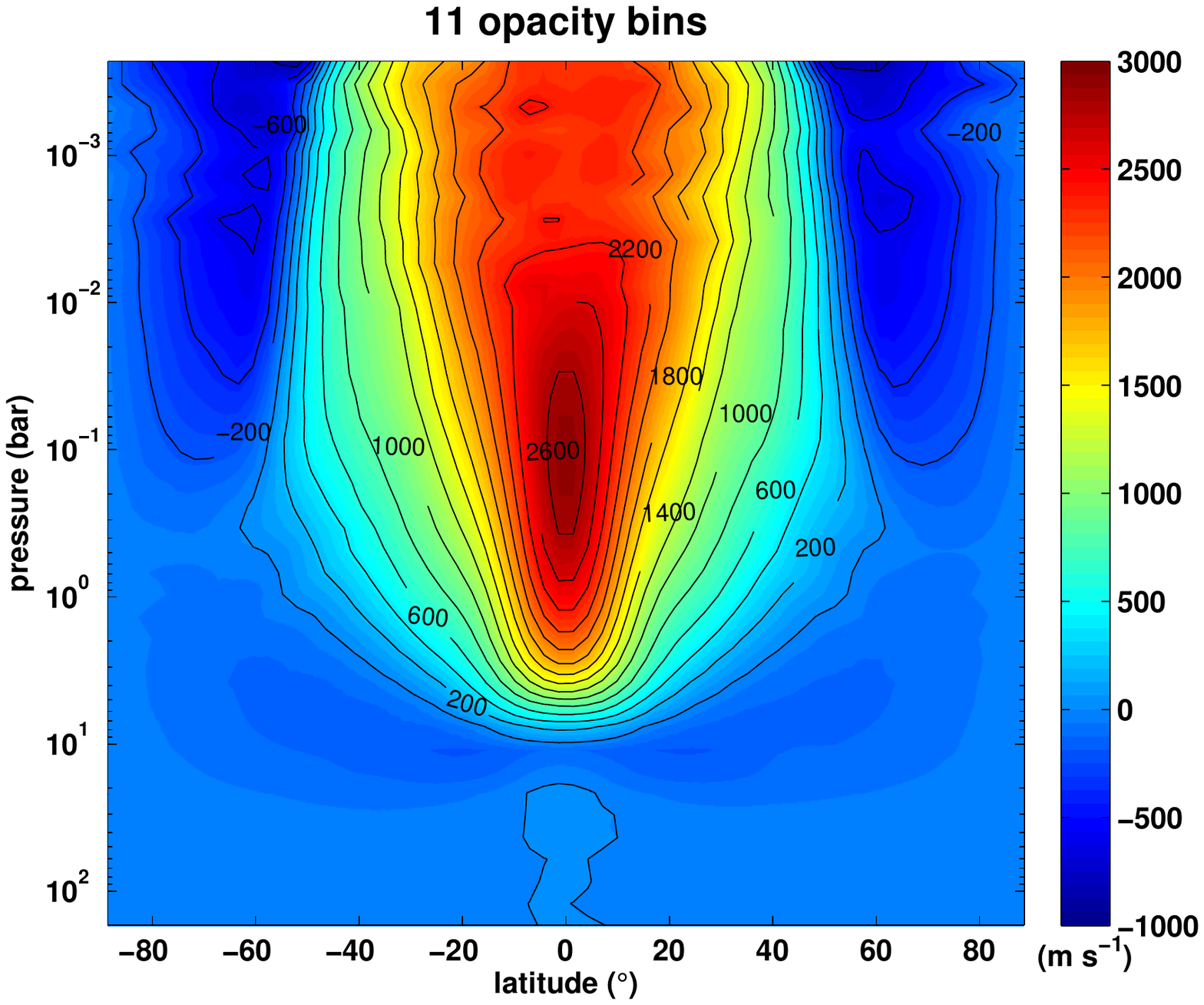} \\
\includegraphics[trim = 0.5in 2.9in 0.8in 2.8in, clip, width=0.42\textwidth]{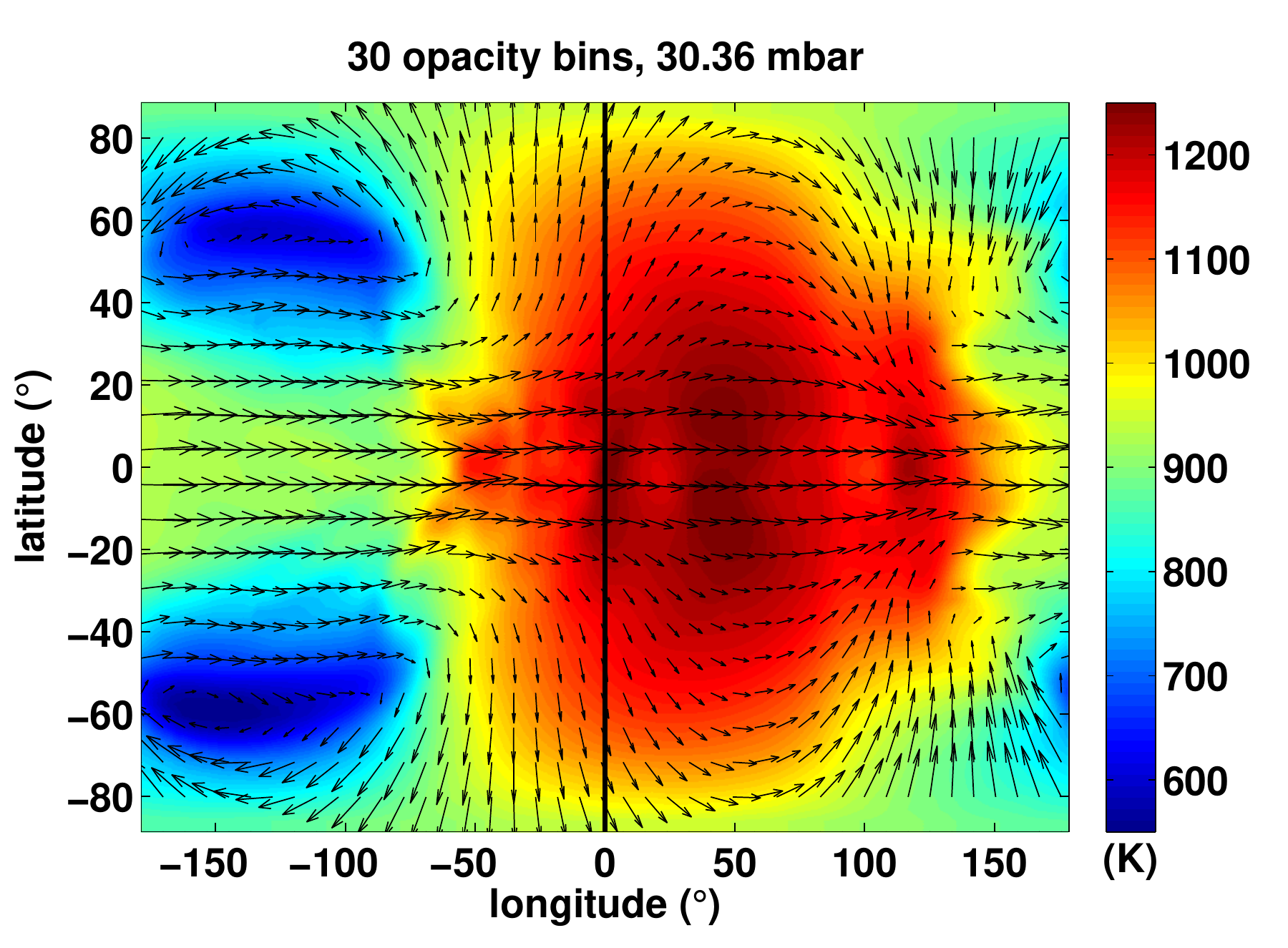}
\includegraphics[trim = 0.5in 2.9in 0.8in 2.8in, clip, width=0.42\textwidth]{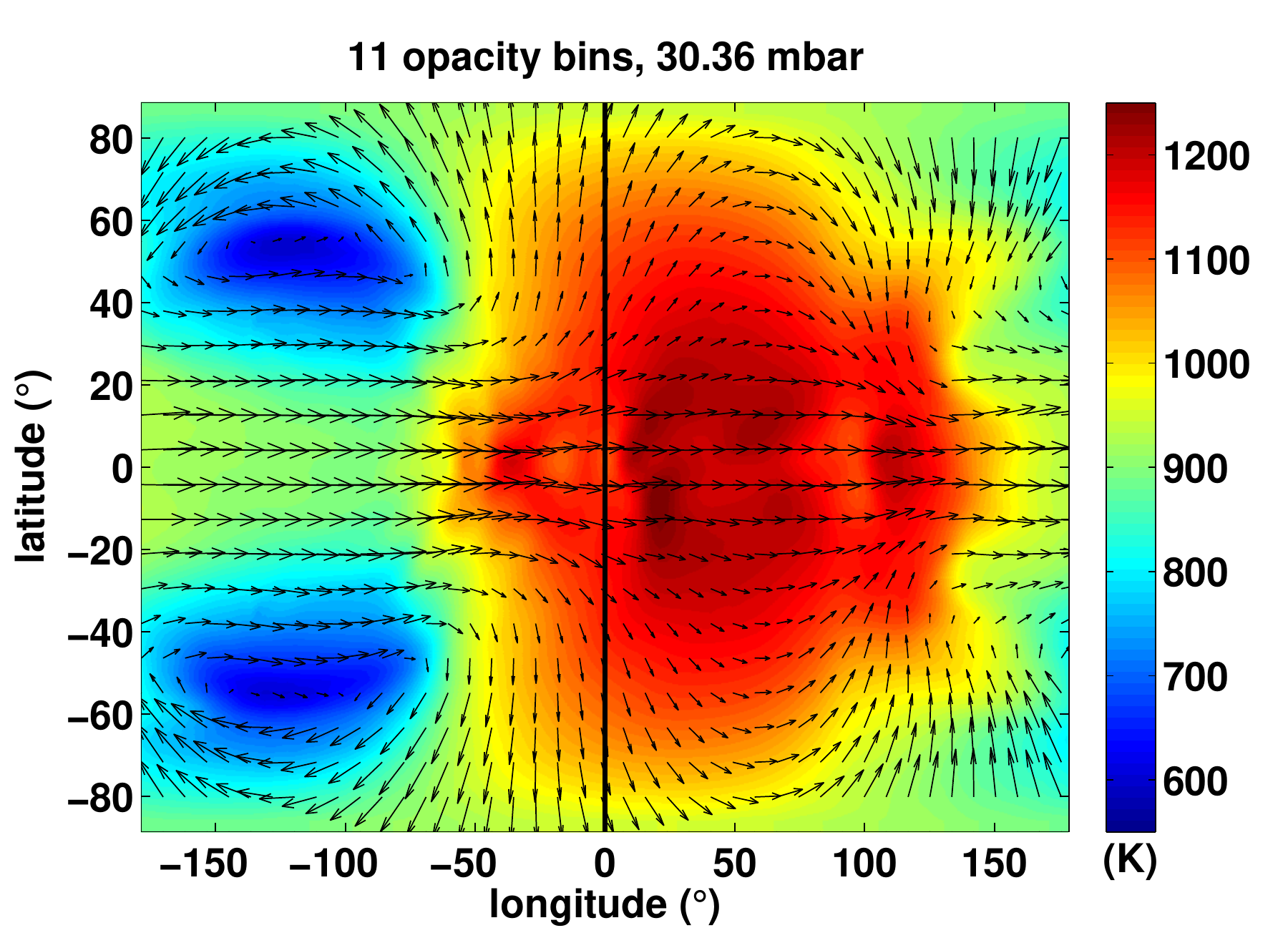} \\
\caption{Globally-averaged pressure-temperature profiles (top row),  zonal-mean zonal wind (middle row) and the wind/temperature profiles at 30 mbar (bottom row) for the 30-bin (left column) and 11-bin (right column) model integrations of HD 189733b.  Snapshots were output at 250 Earth days.}
\label{rtcompare3d}
\end{figure}

To validate our updated radiative transfer scheme using 11 frequency bins, we conducted the following tests:

\begin{enumerate}
	\item First, we tested the sensitivity of the number of frequency bins in a one-dimensional radiative-convective model of HD 189733b ($a=0.0313$ AU, $e=0.0$,  $R_P=8.2396 \times 10^7~\mathrm{m}$, $\mathrm{g=21.4~ms^{-2}}$).  Figure~\ref{rtcompare1d} compares the globally averaged pressure-temperature profiles of models run with 11 frequency bins, 30 frequency bins, and 196 frequency bins at (from left to right) 1$\times$, 2$\times$, 4$\times$, and 8$\times$ the average stellar flux of HD 189733b.  For each group of models, the 11-bin profile agrees well with the 30- and 196-bin cases from 1 $\mu$bar to 1 bar; the temperature varies by only tens of K.  Below 10 bars, the 30- and 11-bin profiles differ from the 196-bin cases by up to 50 K, due to the loss of resolution at low wavenumbers.  In the case of the highest average flux, the large amount of heating leads to a temperature inversion that is not fully captured by the 11-bin model, but can be improved with a better initial guess at the P-T profile.

	\item After testing the new RT scheme for one-dimensional models, we ran full three-dimensional simulations of HD 189733b at its nominal average stellar flux using the SPARC/MITgcm.  Figure~\ref{rtcompare3d} shows the globally-averaged PT profiles (top row), zonal-mean zonal wind (middle row) and wind/temperature profiles at 30 mbar (bottom row) for the 30-bin (left column) and 11-bin (right column) RT setups.  The globally-averaged profiles are nearly identical, with any temperature differences within a few K.  The zonal wind plots also show good agreement, with similar peak speeds ($>2600~\mathrm{ms^{-1}}$), equatorial jet width ($\sim60^{\circ}$), and jet level ($<100$ bar).  At 30 mbar, the 30- and 11-bin models share similar dayside-nightside temperature structure, with a hot spot (1100 K) eastward of the substellar point, and two colder regions on the nightside in the mid-latitudes.

\end{enumerate}

Both tests show that the bulk circulation and temperature structure is retained in the transition from 30 to 11 frequency bins; hence, we proceed to use the SPARC/MITgcm with the new scheme for these and future model simulations.

\clearpage

\clearpage

\clearpage

\end{document}